\documentclass[11pt]{article}
\usepackage{amsmath,amssymb,color,epsfig,cite}
\usepackage{xcolor}
\usepackage{mathrsfs}
\usepackage{slashed}
\usepackage{multirow}
\usepackage{graphicx}
\usepackage[hidelinks]{hyperref}
\usepackage{mathtools}

\usepackage{amsfonts}
\def\td{\tilde}

\makeatletter
\@addtoreset{equation}{section}
\makeatother

\newcommand{\be}{\begin{equation}}
\newcommand{\ee}{\end{equation}}
\newcommand{\bald} {\begin{aligned}}
\newcommand{\eald}{\end{aligned}}
\newcommand{\nn}{\nonumber}

\def\ft#1#2{{\textstyle{\frac{\scriptstyle #1}{\scriptstyle #2} } }}
\def\fft#1#2{{\frac{#1}{#2}}}
\def\CP{{{\mathbb C}{\mathbb P}}}
\def\0{{\sst{(0)}}}
\def\1{{\sst{(1)}}}
\def\2{{\sst{(2)}}}
\def\3{{\sst{(3)}}}
\def\4{{\sst{(4)}}}
\def\5{{\sst{(5)}}}
\def\6{{\sst{(6)}}}
\def\7{{\sst{(7)}}}
\def\8{{\sst{(8)}}}
\def\sst#1{{\scriptscriptstyle #1}}
\def\oneone{\rlap 1\mkern4mu{\rm l}}
\def\ep{{\epsilon}}
\def\del{{\partial}}
\def\ii{{\rm i}}
\def\RP{{{\mathbb R}{\mathbb P}}}
\def\nabslash{\slash \negthinspace \negthinspace \negthinspace
                \negthinspace \nabla}

\def\crampest{\medmuskip = 1mu plus 1mu minus 1mu}
\def\uncramp{\medmuskip = 4mu plus 2mu minus 4mu}

\def\dfft#1#2{{\displaystyle\fft{#1}{#2}}}

\def\phan{{{\phantom{\Sigma}}}}

\def\jm{{{\rm j\,}}}
\def\km{{{\rm k\,}}}

\def\cD{{{\cal D}}}

\def\im{{{\rm i\,}}}
\def\R{{\mathbb R}}

\def\Z{{\mathbb Z}}
\def\C{{\mathbb C}}

\def\cO{{{\cal O}}}

\def\wtd{\widetilde}

\def\ie{{i.e.~}}

\def\ii{{\rm i}}

\newcommand{\rt}{\right}
\newcommand{\lt}{\left}

\def\ben{\begin{equation}}
\def\bea{\begin{eqnarray}}
\def\een{\end{equation}}
\def\eea{\end{eqnarray}}

\def\ft#1#2{{\textstyle{\frac{\scriptstyle #1}{\scriptstyle #2} } }}
\def\fft#1#2{{\frac{#1}{#2}}}
\newcommand{\eq}[1]{(\ref{#1})}

\title{
    \vspace{-1.5cm}
    \begin{flushright}
        \small MI-HET-870 
    \end{flushright}
    \vspace{1cm}
$\text{Spin}^h$ Structure, Scalar and Charged Spinor Eigenfunctions on the $SU(3)/SO(3)$ Wu Manifold}
\bigskip\bigskip

\author{Cameron Gibson, Okan G\"unel, Gabriel Larios and C.N. Pope
\\ \\ \\ 
 Mitchell Institute for Fundamental Physics \& Astronomy,\\
         Texas A\&M University,
         College Station,
         TX 77843-4242,
         USA.
         \\ \\ 
}

\begin{document}

\maketitle

\begin{abstract}
Generalised spin structures are necessary for
placing fermions on manifolds that do not admit a standard spin structure. This is especially relevant in a dimensional 
reduction on such a
manifold,  which can then be compensated by using fermions
that are appropriately charged under some Maxwell or Yang-Mills field defined on the internal manifold.  A well known example in the physics literature is
$\CP^2$, which has four real dimensions and is the coset $SU(3)/U(2)$.
In this paper we focus on a five-dimensional coset space, namely the Wu manifold $SU(3)/SO(3)_{\rm max}$, where $SO(3)_{\rm max}$ is maximal in $SU(3)$.  Intriguingly, the Wu manifold does not admit a spin structure or spin$^c$ structure, it does admit a spin$^h$ structure.  We provide a physical interpretation of the spin$^h$ structure by considering spinors that are coupled to an $SO(3)$ Yang-Mills field defined on the Wu manifold, but which carry half-integer ``isospin,'' thereby canceling the minus sign in the holonomy for uncharged spinors that provides the original obstruction to an ordinary spin structure.  We also construct a gauge-covariantly constant spinor in the Wu manifold, and we show how this can be employed in order to construct spin$^h$ spinor harmonics from scalar harmonics.  We provide a very explicit construction of all the scalar and spin$^h$ harmonics.  In a follow-up paper, we shall employ the results we obtain here in order to discuss dimensional reductions and consistent reductions on the Wu manifold.

\end{abstract}

\bigskip 
\begin{center}
    \footnotesize{
camerongibson@tamu.edu, okan.gunel@tamu.edu, gabriellariosplaza@hotmail.com, pope@physics.tamu.edu
}
\end{center}
\thispagestyle{empty}
\newpage

\tableofcontents

\newpage
\section{Introduction}

   Not all manifolds allow ordinary fermions to be defined.  The situation can arise where there is a topological
obstruction that makes it inconsistent to define ordinary spinors on a manifold. Essentially, because the
spin group Spin$(n)$ on an $n$-dimensional manifold is the double cover of the tangent space group $SO(n)$, a
factor of $-1$ can arise in the holonomy for the parallel transport of spinors around a family of curves spanning a cycle in the manifold, implying that the spinors are not globally defined.  In some cases, generalised spinors may nevertheless be defined, by allowing them to couple minimally to an abelian or non-abelian gauge field defined on the manifold.  If the representation of the spinors under the gauge group is appropriately chosen, an additional $-1$ factor in the holonomy for gauge parallel transport of the spinors can arise, compensating the $-1$ factor that arose for uncharged spinors.  A classic example is the 
$\CP^2$ manifold, the complex projective plane, which has real dimension 4 and which does not admit an ordinary spin structure. In this case there exists a natural $U(1)$ connection whose field strength is rather like that of a Dirac monopole with magnetic charge $P$ in $\CP^2$, and by coupling fermions with an electric charge $e$ satisfying $2e P=\ft12+\, $integer, rather than the $2eP=\,$integer of the usual Dirac quantisation condition, an additional minus sign in the holonomy is introduced that compensates the $-1$ factor for the holonomy for uncharged spinors.  (See, for example, \cite{Hawking:1977ab}.)

   One reason why manifolds without a spin structure are of interest in physics is because one may, under suitable
circumstances, nevertheless use them when describing compactifications of higher-dimensional theories such as supergravities 
or string theories.  The study of coset space reductions of higher-dimensional supergravity
theories has a long history, dating back to the early 1980's
\cite{freundrubin,dufpop1982,birengdewnic,duffnilspope,dewitnicol}.
Most of the attention has been directed towards
coset reductions on spheres, with some of the most important examples
including the $S^7$ reduction of eleven-dimensional supergravity
\cite{dufpop1982,birengdewnic,duffnilspope,dewitnicol}, and
the $S^5$ reduction of ten-dimensional type IIB supergravity
\cite{kimromvan}.  These both provide examples of {\it consistent} reductions,
in which the subsector of Kaluza-Klein modes that form the massless
supergravity multiplet in the lower-dimensional theory can be (classically)
decoupled
completely, at the full non-linear order, from the remainder of the
Kaluza-Klein massive modes.  Put another way, any solution of the
lower-dimensional massless supergravity lifts to provide an exact solution
of the higher-dimensional theory.  The essence of the proof of
consistency of the $S^7$ reduction to ${\cal N}=8$ gauged $SO(8)$
supergravity in four dimensions was given in \cite{dewitnicol}.  The
analogous proof of the consistency of the $S^5$ reduction of type IIB
supergravity to give five-dimensional ${\cal N}=8$ gauged $SO(6)$
supergravity was given in \cite{baghohsam}.  One especially important
application of the $S^5$ reduction of type IIB supergravity is in
the context of holography and
the AdS$_5$/CFT$_4$ correspondence \cite{malda,gubklepol,withol}.

  The 5-sphere $S^5$ can be described as the coset $SO(6)/SO(5)$.  It
can also be described as the coset $SU(3)/SU(2)$.  There is another
five-dimensional coset space whose numerator group is $SU(3)$ that
is of considerable interest, and that is the coset $SU(3)/SO(3)$, where
$SO(3)$ is a maximal subgroup of $SU(3)$.  This coset space is commonly
referred to as the Wu manifold in the mathematical literature.  It has
many intriguing properties, including the fact that it does not admit
an ordinary spin structure, nor even a spin$^c$ structure which was first noticed by Landweber and Stong \cite{notspinc}. In 2017, it was proven that it does admit a spin$^h$ structure \cite{Chen:2017}. This result seems to have been proven in less standard language in an earlier physics paper \cite{balachandran}.

   The notion of manifolds that do not admit an ordinary spin structure
appeared in the physics literature in the late 1960s, with the work of
Geroch \cite{geroch1,geroch2}.  The focus in that work was on spacetime
manifolds with Lorentzian signature.  Subsequently, in the late 1970s,
physicists became interested in the subject in the context of manifolds
of positive-definite metric signature, when studying the Euclidean approach
to quantum gravity.  (See, for example, \cite{Hawking:1977ab}.)  The principal
example of interest in those investigations was the complex projective
plane $\CP^2$, which does not admit an ordinary spin structure but does
admit a spin$^c$ structure, as we mentioned above.
 It was also remarked in \cite{Hawking:1977ab} that more
general possibilities, referred to there as ``generalised spin structures,''
could arise, in which spinors in half-integer isospin representations of
a suitable Yang-Mills could also be considered, again with the goal of
cancelling the $-1$ holonomy for uncharged spinors.  Such non-abelian
generalisations of the notion of a spin$^c$ structure have subsequently
become known as spin$^h$ structures in the mathematical literature \cite{christian}.

The Wu manifold has been of interest to the mathematical community since the 1950s \cite{wu50,dold56,barden65}. It is a minimal example in the sense that $5$ is the lowest dimension in which an orientable manifold does not necessarily admit a spin$^c$ structure \cite{wu50}. Interestingly, all orientable manifolds of dimension $\leq5$ are spin$^h$ and all compact orientable manifolds of dimension $6$ and $7$ are spin$^h$ \cite{albanese23}.

The Wu manifold also plays an important role as a generator of the oriented cobordism group $\Omega_5^{SO}\cong \Z_2$ \cite{crowley11}. The Wu manifold was therefore used in \cite{debray22} to show the triviality of a potential nonperturbative anomaly of the $E_{7(7)}(\R)$ $U$-duality symmetry of $4d$ ${\cal N}=8$ supergravity. Spin$^h$ structures are useful for discussing global anomalies \cite{freed16,brennan23}. For example, in \cite{witten19}, the Dold manifold $(S^1\times \CP^2)/\Z_2$ (another five-dimensional spin$^h$ manifold that is not spin$^c$, with $\Z_2$ acting as antipodal identification on $S^1$ and complex conjugation on $\CP^2$) was equipped with a spin$^h$ structure to detect a new four-dimensional SU(2) anomaly. 

Our aim in the present paper will be to provide an understanding of some of the
key properties of the Wu manifold and its spin$^h$ structure, using where
possible the language and tools accessible to physicists. We shall take 
the discussion of spin$^c$ structures in\cite{Hawking:1977ab} as a guide, and
in fact we shall revisit some of the results for $\CP^2$ described in \cite{Hawking:1977ab}
using an approach and notation that we shall then develop further in order to
extend the understanding to the somewhat more subtle framework of the Wu
manifold.  In a companion paper, which is currently under construction, we shall be
making use of some of the results obtained in the present work in order to
discuss supergravity compactifications on the Wu manifold \cite{gigulapo2}.

The paper is structured as follows.  In section \ref{Wumanifoldsec} we give a basic introduction to the Wu manifold, showing how
its metric can be obtained by starting from the bi-invariant metric on the $SU(3)$ group manifold,
using a parameterisation of $SU(3)$ in which the maximal $SO(3)$ subgroup is manifest.  After factoring
by $SO(3)_{\rm max}$ the metric on the Wu manifold is obtained.  We give expressions for the metric, 
spin connection and curvature, and also for the $SO(3)$ Yang-Mills gauge connection which plays a
central role in the subsequent discussion of the spin$^h$ structure.  We also discuss briefly some global
aspects of the construction.  In section \ref{CP2sec}, by way of an introduction
to our subsequent discussion of the spin$^h$ structure in the Wu manifold, we review some of the key features of
the $\CP^2$ manifold and its spin$^c$ structure.  Our presentation follows the approach described in \cite{Hawking:1977ab},
but we elaborate on some of the details and introduce some formalism that we shall later apply to the case of the
Wu manifold.  We also develop further some of the observations in \cite{Hawking:1977ab} on generalised spin structures,
and in particular we show explicitly how $\CP^2$ can also support a spin$^h$ structure.  This discussion will
be closely paralleled in section \ref{Wuspinhsec}, where we turn to the Wu manifold and give an explicit demonstration of the absence of an ordinary spin
structure and of a spin$^c$ structure, and the existence of a spin$^h$ structure.  Included in our discussion is the construction
of a gauge-covariantly constant spinor in the Wu manifold.  We make use of this in section \ref{eigenfunctionssec} in order to construct the complete
spectrum of eigenfunctions of the gauge-covariant Dirac operator, in terms of the scalar eigenfunctions on the Wu manifold.
We also give an explicit derivation of the scalar eigenfunctions and eigenvalues.  In section \ref{noncompactdualssec}, we describe the construction of
the non-compact dual of the Wu manifold, defined on the coset $SL(3,\R)/SO(3)$.  In contrast to the standard compact Wu
manifold $SU(3)/SO(3)$, the non-compact dual is topologically trivial and there is no obstruction to the existence of an
ordinary spin structure.  In order to illustrate the parallels with $\CP^2$ we also discuss its non-compact dual on the
coset $SU(2,1)/U(2)$, which is again topologically trivial and admits an ordinary spin structure. 

After presenting some concluding remarks in section \ref{conclusionssec}, we give some further details and elaborations in a series of appendices.
Appendix \ref{Gbundlesec} contains a review of the description of a metric on the total space of the fibration over a base space, where the
fibre is a compact semi-simple Lie group.  Appendix \ref{paramsec} contains a detailed discussion of the identifications and periodicities
for the coordinates in the parameterisation of $SU(3)$ in eqn (\ref{SU3}), in order to establish the fundamental domain that 
gives precisely a single-fold cover of $SU(3)$.   In appendix \ref{SubmanSec} we give an explicit discussion of the totally-geodesic
submanifolds of the Wu manifold, some of which play a role in our description of the spin$^h$ structure.  Appendix \ref{2compsec} gives another
way to describe the construction of the gauge-covariantly constant spinor in the Wu manifold, by employing a two-component spinor
notation in the $SO(3)$ fibres.  Appendix \ref{eigenvaluessec} contains some of the detailed calculations needed for our explicit construction of the
scalar eigenfunctions in the Wu manifold in section \ref{eigenfunctionssec}.  Finally, in appendix \ref{sec:pin}  we discuss (s)pinors in $\RP^2$, demonstrating explicitly the
absence of an ordinary spin structure, and the existence of two Pin$^-$ structures. (Pinors in $n$ dimensions are representations of the 
double cover of $O(n)$, as opposed to spinors, which are representations of the double cover of $SO(n)$.  Pinors are needed in order
to define spinor-like objects in non-orientable manifolds, such as $\RP^2$ \cite{kirby}  Our discussion includes a fully explicit construction 
of the complete spectrum of eigen-pinors of the Dirac operator on $\RP^2$.

\bigskip
\noindent{\bf Notation and Conventions:}
\medskip

We conclude the introduction with some notation and conventions that we shall be using in the paper.

When discussing spinors in five dimensions, 
it will sometimes be convenient to choose an explicit representation for the Dirac matrices $\Gamma^a$.  We shall use
\bea
\Gamma^1 &=& \tau_2\otimes\tau_1\,,\qquad
\Gamma^2= \tau_2\otimes\tau_2\,,\qquad
\Gamma^3= \tau_2\otimes\tau_3\,,\nn\\
\Gamma^4 &=& -\tau_1\otimes\oneone_2\,,\qquad
\Gamma^5= \tau_3\otimes\oneone_2\,,\label{d5Gamma0}
\eea
where $\tau_i$ are the standard $2\times2$ Pauli matrices, and $\oneone_2$ is the $2\times 2$ identity matrix:
\bea
\tau_1=\begin{pmatrix} 0&1\\ 1&0\end{pmatrix}\,,\qquad
\tau_2=\begin{pmatrix} 0& -\im\\ \im&0\end{pmatrix}\,,\qquad
\tau_3=\begin{pmatrix} 1&0\\ 0& -1\end{pmatrix}\,.\label{Paulimatrices}
\eea
    
We shall also use the same Dirac matrices (\ref{d5Gamma0}) when we consider spinors in the four-dimensional $\CP^2$ manifold. Note that $\Gamma_5$ plays the role of the chirality operator in four dimensions, with $\Gamma_5=- \Gamma_1\Gamma_2\Gamma_3\Gamma_4=\hbox{diag}\,(1,1,-1,-1)$. Thus $\ft12(1\pm\Gamma_5)$ project onto the subspaces of right-handed and left-handed spinors respectively, in four dimensions.

We shall also occasionally make reference to the Gell-Mann matrices $\lambda_i$, which are a
standard basis for the (Hermitian) generators of $SU(3)$.  They are given by
{\small
\bea
\lambda_1 &=& \begin{pmatrix} 0&1&0 \\ 1&0&0 \\ 0&0&0 \end{pmatrix}\,,\qquad
\lambda_2 = \begin{pmatrix} 0&-\im &0 \\ \im &0&0 \\ 0&0&0 \end{pmatrix}\,,\qquad
\lambda_4 = \begin{pmatrix} 0&0&1 \\ 0&0&0 \\ 1&0&0 \end{pmatrix}\,,\nn\\
\lambda_5 &=& \begin{pmatrix} 0&0&-\im \\ 0&0&0 \\ \im &0&0 \end{pmatrix}\,,\qquad
\lambda_6 = \begin{pmatrix} 0&0 &0 \\ 0&0&1 \\ 0&1&0 \end{pmatrix}\,,\qquad
\lambda_7 = \begin{pmatrix} 0&0&0 \\ 0&0&-\im \\ 0&\im &0 \end{pmatrix}\,,\nn\\
\lambda_3 &=& \begin{pmatrix} 1&0&0 \\ 0&-1&0 \\ 0&0&0 \end{pmatrix}\,,\qquad
\lambda_8 = \fft1{\sqrt3}\begin{pmatrix} 1&0&0 \\ 0&1&0 \\ 0&0&-2 \end{pmatrix}\,.\label{su3gen}
\eea
}

\section{\label{Wumanifoldsec}The Wu manifold}

\subsection{A parameterisation of $SU(3)$}

Our construction of the metric on the $SU(3)/SO(3)_{\rm max}$ Wu manifold
begins by constructing the bi-invariant metric on $SU(3)$.  
We can parameterise $SU(3)$ matrices $U$ in the $3\times 3$ representation by writing
\bea
U= \cO_1\, B\,\cO_2^T\,,\label{SU3}
\eea
where locally, 
$\cO_1$ and $\cO_2$ are $SO(3)$ matrices and $B$ is a diagonal matrix in $U(1)\times U(1)$. We 
parameterise $\cO_1$ and $\cO_2$ as
\bea
\cO_1= e^{\Phi\, t_3}\, e^{\Theta\, t_2}\, e^{\Psi\, t_3}\,,\qquad
\cO_2= e^{\phi\, t_3}\, e^{\theta\, t_2}\, e^{\psi\, t_3}\,,\label{O1O2def}
\eea
where the $SO(3)$ generators are taken to be the $3\times3$ matrices 
$(t_i)_{jk}=-\ep_{ijk}$, that is
{\small
\bea
\label{tidef}
t_1=\begin{pmatrix} 0&0&0\cr 0&0&-1\cr 0&1&0\end{pmatrix}\,,\quad
t_2=\begin{pmatrix} 0&0&1\cr 0&0&0\cr -1&0&0\end{pmatrix}\,,\quad
t_3=\begin{pmatrix} 0&-1&0\cr 1&0&0\cr 0&0&0\end{pmatrix}\,.\label{t123def}
\eea
}
They obey the algebra $[t_i,t_j]= \ep_{ijk}\, t_k$.
The $3\times 3$ matrix $B$ is diagonal, given by
\bea
B={\rm diag}\, \Big(e^{\im (\mu+\ft13\nu)}, e^{\im (-\mu+\ft13\nu)},
e^{-\fft{2\im}{3}\, \nu}\Big)\,.\label{Bmatrixdef}
\eea
The matrix $B$ is a member of the Cartan subalgebra $H=U(1)\times U(1)$ of $SU(3)$.

   Defining the left-invariant 1-forms $\sigma_i$ and $\Sigma_i$ 
for the two $SO(3)$ algebras by
\bea
\cO_1^T\, d\cO_1 = \Sigma_i\, t_i\,,\qquad
\cO_2^T\, d\cO_2 = \sigma_i\, t_i\,,
\eea
we have
\crampest
\bea
\mkern-36mu
&&\sigma_1= \sin\psi\,d\theta - \cos\psi\,\sin\theta\,d\phi\,,\quad
\sigma_2= \cos\psi\,d\theta +\sin\psi\,\sin\theta\,d\phi\,,\quad
\sigma_3=d\psi+\cos\theta\,d\phi\,,\label{sigSig}\\
\mkern-36mu
&&\Sigma_1= \sin\Psi\,d\Theta - \cos\Psi\,\sin\Theta\,d\Phi\,,\
\Sigma_2= \cos\Psi\,d\Theta +\sin\Psi\,\sin\Theta\,d\Phi\,,\
\Sigma_3=d\Psi+\cos\Theta\,d\Phi\,.\nn
\eea
\uncramp
These obey $d\sigma_i=-\ft12 \ep_{ijk}\, \sigma_j\wedge\sigma_k$ and
$d\Sigma_i=-\ft12 \ep_{ijk}\, \Sigma_j\wedge\Sigma_k$. 

Due to the parameterisation of $SU(3)$ as in eqn \eq{SU3}, we 
have some redundancies concerning the coordinate ranges. From the definition of $B$ in eqn
\eq{Bmatrixdef}, we have the identifications 
$(\mu,\nu)\sim(\mu+2\pi,\nu)\sim (\mu+\pi,\nu+3\pi)$ 
on the central coordinates, thus restricting the ranges of $\mu$ and $\nu$ to lie within
a fundamental cell in a tiling of the $(\mu,\nu)$ plane by hexagons elongated in the $\nu$ direction. 
However, there are subtleties that result in further restrictions on 
the fundamental region of the coordinate ranges on $SU(3)$. These are discussed more 
fully in appendix \ref{ParametSec}, but we shall also summarise them here.  

Firstly, the four matrices
$\gamma_1=\hbox{diag}\,(1,1,1)$, $\gamma_2=\hbox{diag}\,(1,-1,-1)$, $\gamma_3=\hbox{diag}\,(-1,1,-1)$,
$\gamma_4=\hbox{diag}\,(-1,-1,1)$ that form the central elements in $SO(3)$ also lie in the Cartan
subgroup parameterised in terms of $\mu$ and $\nu$ in eqn (\ref{Bmatrixdef}). The matrices 
$\gamma$ form the dihedral group of order 4, which we denote as $D_4$.
Thus from
 \begin{equation}
    (\mathcal{O}_1)(B)(\mathcal{O}_2^T)=(\mathcal{O}_1 \gamma^{-1})(\gamma B ) (\mathcal{O}_2^T) \enspace \text{for} \enspace \gamma \in SO(3) \cap H \, , 
\end{equation}
it follows that in order to remove a redundancy in the parameterisation of $SU(3)$ in eqn (\ref{SU3}), we should
identify elements of $B$ under $B=\gamma_i\, B$ for each $i$. This implies a 2-fold 
reduction of the ranges of each of the central coordinates $\mu$ and $\nu$ that form the fundamental cell, which is 
therefore now defined
by the identifications (see appendix \ref{firstAct})
\bea(\mu,\nu)\sim(\mu+\pi,\nu)\sim (\mu+\ft12\pi,\nu+\ft32\pi)\,.\label{smallhex}
\eea

Secondly, we have
\begin{equation}
(\mathcal{O}_1)(B)(\mathcal{O}_2^T)=(\mathcal{O}_1 \omega^{-1})(\omega B \omega^{-1})(\omega \mathcal{O}_2^T) \enspace \text{for} \enspace  \omega \notin H \enspace \text{such that} \enspace \omega B \omega^{-1} \in H \enspace \forall B \in H \,.
\end{equation}
Here, $\omega$ corresponds to the Weyl transformations and as 
discussed in appendix \ref{secondAct}, the action of $\omega$ implies that the fundamental hexagonal cell
defined by (\ref{smallhex}) can actually be further subdivided into six wedges, with the true fundamental
domain being just one of the six wedges in Fig.~\ref{munu}, with the other five being filled out by the action of the elements $\omega$.
 
 Finally, we have 
\begin{equation}
(\mathcal{O}_1)(B)(\mathcal{O}_2^T)=(\mathcal{O}_1 \gamma')( B )(\gamma'^{-1} \mathcal{O}_2^T) \enspace \text{for} \enspace \gamma'  \enspace \text{such that} \enspace \gamma' B \gamma'^{-1} \in SO(3) \enspace \forall B \in H \,.
\end{equation}
In this case, $D_4$ acts on either left or right $SO(3)$, and it 
reduces the coordinate ranges from the usual coordinate ranges of 
$SO(3)$ (see appendices \ref{thirdAct} and \ref{su2Act}). However, 
we need this identification to hold also on the coset space since we 
expect the metric on the coset space to extend smoothly onto $\R^5$ 
in the neighbourhood of the points where the $U(1)$ Killing vector 
has a fixed point (see section \ref{StrucSubSec} for the details). 
Therefore, it will be the coordinates in $\mathcal{O}_2$, that will 
be affected by the action of $D_4$. (Yet, on the level of 
parameterising a unique element of $SU(3)$, $D_4$ can act on the 
coordinates in $\mathcal{O}_1$.)

\subsection{Local properties of the Wu metric}

   The bi-invariant metric on the $SU(3)$ group manifold is given by
\bea
ds_8^2 &=& \ft12{\rm tr}\,(dU^\dagger\,dU)\,,\label{su3met0}
\eea
which can be written in the form 
\bea
ds_8^2=ds_5^2 + \sum_i (\Sigma_i - A^i)^2\,,\label{su3met}
\eea
where 
\bea
\label{WuMetric}
ds_5^2 = d\mu^2+ \ft13 d\nu^2 + \sin^2(\mu-\nu)\,\sigma_1^2 +
  \sin^2(\mu+\nu)\, \sigma_2^2 + \sin^2 2\mu\, \sigma_3^2\,,
\eea
and 
\bea
A^1=\cos(\mu-\nu)\, \sigma_1\,,\qquad
A^2= \cos(\mu+\nu)\, \sigma_2\,,\qquad
A^3=\cos 2\mu\, \sigma_3\,.\label{so3YM}
\eea
Comparing with the general expression in eqn (\ref{Gbundlemet}) for the metric on a principal $G$ bundle over
a base space, it can be seen that eqn (\ref{su3met}) gives the metric on $SU(3)$ as a principal $SO(3)$ bundle 
over the base space $SU(3)/SO(3)_{\rm max}$. The coset $SU(3)/SO(3)_{\rm max}$ is also referred to as the
\textit{Wu manifold}.\footnote{Note that $SO(3)_{\rm max}$ is
the maximal real subgroup of $SU(3)$. In terms of the anti-Hermitian $SU(3)$ generators
$\tilde\lambda_i=\im \lambda_i$, where $\lambda_i$ are the standard Gell-Mann matrices given in eqns (\ref{su3gen}), 
$SO(3)_{\rm max}$ is generated by the three real matrices among the $\tilde\lambda_i$,
namely $(\tilde\lambda_2,\tilde\lambda_5,\tilde\lambda_7)$.  Up to signs, these coincide with
our $SO(3)$ generators $(t_3,t_2,t_1)$ defined in eqns (\ref{t123def}).  We shall frequently
omit the subscript ``max'' on $SO(3)$, but it will be understood that this is the one we are
always considering.} 

   The metric on the Wu manifold is thus given by eqn (\ref{WuMetric}), 
where the left-invariant 1-forms $\sigma_i$ are now defined on the 
lens space $S^3/D_8^* = SO(3)/D_4$ and $\mu\in [0,\pi/2]$, $\nu\in[-\mu,\mu]$ (see appendix \ref{ParametSec} for a detailed discussion). 
The connection $A^i$ on the $SO(3)$ fibres is given by eqns (\ref{so3YM}).
We shall use the natural vielbein basis $e^a$, with
\bea
e^1= \sin(\mu-\nu)\, \sigma_1\,,\quad
e^2= \sin(\mu+\nu)\, \sigma_2\,,\quad 
e^3=\sin2\mu\, \sigma_3\,,\quad
e^4=d\mu\,,\quad e^5=\fft{d\nu}{\sqrt3}\,.\label{wuviel}
\eea
    The connection 1-forms for the Wu space are given by
\bea
\omega_{12}&=& \cos2\mu\, \sigma_3\,,
\quad \omega_{13}= \cos(\mu+\nu)\, \sigma_2\,,\quad
\omega_{14}= \cos(\mu-\nu)\, \sigma_1\,,\nn\\
\omega_{15} &=& -\sqrt3\, \cos(\mu-\nu)\, \sigma_1\,,
\quad \omega_{23}= -\cos(\mu-\nu)\, \sigma_1\,,\quad
\omega_{24}= \cos(\mu+\nu)\, \sigma_2\,,\nn\\
\omega_{25}&=&\sqrt3\, \cos(\mu+\nu)\, \sigma_2\,,\quad
\omega_{34}=2\cos2\mu\, \sigma_3\,,\quad 
\omega_{35}= 0\,,\quad\omega_{45}=0\,.\label{wuspincon}
\eea
The curvature 2-forms $\Theta_{ab}= d\omega_{ab} + \omega_{ac}\wedge \omega_{cb}$ are
then given by
\bea
\label{spinconwu}
&&\Theta_{12}= e^1\wedge e^2 + 2 e^3\wedge e^4\,,\qquad
\Theta_{13}=e^1\wedge e^3 + e^2\wedge e^4 + \sqrt3\, e^2\wedge e^5\,,\nn\\
&&\Theta_{14}=e^1\wedge e^4 - e^2\wedge e^3 - \sqrt3\, e^1\wedge e^5\,,\qquad
\Theta_{15}=3 e^1\wedge e^5 -\sqrt3\,e^1\wedge e^4 +\sqrt3\,e^2\wedge e^3\,,
\nn\\
&& \Theta_{23}= e^2\wedge e^3 -e^1\wedge e^4 +\sqrt3\, e^1\wedge e^5\,,\qquad
\Theta_{24}=e^2\wedge e^4 + e^1\wedge e^3 +\sqrt3\, e^2\wedge e^5\,,\nn\\
&& \Theta_{25}=
3 e^2\wedge e^5 +\sqrt3\, e^1\wedge e^3 +\sqrt3\, e^2\wedge e^4\,,\qquad
\Theta_{34}= 4 e^3\wedge e^4 + 2 e^1\wedge e^2\,,\nn\\
&&\Theta_{35}=0\,,\qquad \Theta_{45}=0\,.
\eea
The metric is Einstein, with $R_{ab}=6\delta_{ab}$.

In terms of the vielbein $e^a$ on $SU(3)/SO(3)_{\rm max}$, defined in
eqns (\ref{wuviel}), the $SO(3)$ Yang-Mills field strengths 
$F^i=dA^i+\ft12\ep_{ijk}\, A^j\wedge A^k$ are given by
\bea
F^1 &=& e^1\wedge e^4-e^2\wedge e^3 - \sqrt3\, e^1\wedge e^5\,,\nn\\
F^2 &=& e^2\wedge e^4 + e^1\wedge e^3 +\sqrt3\, e^2\wedge e^5\,,\nn\\
F^3 &=& 2 e^3\wedge e^4 + e^1\wedge e^2\,.\label{SO3Fi}
\eea
The field strengths $F^i$ give the curvature of the principal $SO(3)$ bundle.
Note that they obey
\bea
F^i_{ac}\, F^i_{bc}= 6\delta_{ab}\,,\qquad
F^i_{ab}\, F^j_{ab}=10\delta_{ij}\,.\label{FFid}
\eea
(So the Yang-Mills field strengths span the entire five-dimensional base
space in an isotropic fashion.) It can also be verified that the
$F^i_{ab}$ satisfy
\bea
F^i_{ab}\, F^i_{cd} = F^i_{ac}\, F^i_{bd}- F^i_{ad}\, F^i_{bc}\,,
\eea
and
\bea
F^i_{[a|c}\, F^j_{c| b]} = \ft12 \epsilon_{ijk}\, F^k_{ab}\,.
\label{FFasym}
\eea

   Some properties of the $SO(3)$ Yang-Mills connection $A^i$ are as follows.
Firstly, it can be seen that the Hodge duals of the 2-form field strengths
$F^i$ in eqns (\ref{SO3Fi}) are
\bea
{*F}^1 &=& e^2\wedge e^3\wedge e^5 -e^1\wedge e^4\wedge e^5 +
    \sqrt3\, e^2\wedge e^3\wedge e^4\,,\nn\\
{*F}^2 &=& -e^1\wedge e^3\wedge e^5 - e^2\wedge e^4\wedge e^5 +
            \sqrt3\, e^1\wedge e^3\wedge e^4\,,\nn\\
{*F}^3 &=& 2 e^1\wedge e^2\wedge e^5 + e^3\wedge e^4\wedge e^5\,.
\label{starF}
\eea
It can now be verified that $F^i$ satisfies the Yang-Mills equations,
\bea
D{* F}^i \equiv d{*F}^i + \ep_{ijk}\, A^j\wedge F^k=0\,,
\eea
or in other words, $D_b F^i_{ab}= \nabla_b\, F^i_{ab} +
  \ep_{ijk}\, A^j_b\, F^k_{ab}=0$.  In fact, a stronger condition holds 
also, namely that $F^i_{ab}$ is gauge-covariantly constant,
\bea
D_a F^i_{bc}= \nabla_a\, F^i_{bc} +
  \ep_{ijk}\, A^j_a\, F^k_{bc}=0\,.
\eea

It is useful to note that the spin connection $\omega_{ab}$ in
eqns (\ref{wuspincon}) can be written simply in the form
\bea
\omega_{ab}= F^i_{ab}\, A^i\,.\label{omFA}
\eea
where $A^i$ denotes the $SO(3)$ Yang-Mills connection given in 
eqns (\ref{so3YM}). 
The curvature 2-forms $\Theta_{ab}= d\omega_{ab} + 
\omega_{ac}\wedge \omega_{cb}$ can then easily be seen, using 
eqns (\ref{FFasym}) and (\ref{omFA}), to be given by
\bea
\Theta_{ab}= F^i_{ab}\, F^i\,,\label{ThFF}
\eea
The components of the Riemann tensor $R_{abcd}$ can be read off from 
$\Theta_{ab}= \ft12 R_{abcd}\, e^c\wedge e^d$, which then implies that
\bea
R_{abcd} = F^i_{ab}\, F^i_{cd}\,.\label{RiemFF}
\eea

Since $R_{ab}=\Lambda \delta_{ab}$, with $\Lambda=6$, we get the invariant
\bea
\label{WuInv}
I_2\equiv \fft{R^{abcd} R_{abcd}}{\Lambda^2}= \fft{25}{3}\,.
\eea
This is a characteristic scale-invariant number for the $SU(3)/SO(3)_{\rm max}$ coset space, which can be seen to agree with that for the metric given in eqn (2.8) of ref. \cite{gilupo}.

   A more general family of $SU(3)$ metrics on the $SO(3)$ bundle over $SU(3)/SO(3)_{\rm max}$
can be written as 
\bea
ds_8^2 = ds_5^2 + c^2\, (\Sigma_i-A^i)^2\,,\label{SU3mets2}
\eea
using the base metric \eq{WuMetric} and the connection \eq{so3YM}, where $c$ is
a constant squashing parameter.
Plugging the above expressions into eqns (\ref{SU2Ricci}), we therefore
find that the vielbein components of the 
Ricci tensor on the $SU(3)$ metric (\ref{SU3mets2}) are given by
\bea
\wtd R_{ab} &=& (6-3c^2)\, \delta_{ab}\,,\qquad 
\hbox{$SU(3)/SO(3)_{\rm max}$ directions}\,,\nn\\
\wtd R_{ij} &=& \Big(\fft{5c^2}{2} + \fft{1}{2c^2}\Big)\, \delta_{ij}\,,
\qquad \hbox{$SO(3)$ fibre directions}\,.\label{RicSU3}
\eea
Requiring that the metric on $SU(3)$ be Einstein therefore implies
\bea
  c^2=1 \qquad \hbox{or} \qquad c^2=\fft1{11}\,.\label{cvals}
\eea
The first of these, where $c^2=1$, corresponds to the bi-invariant metric
on $SU(3)$ that we obtained in eqns (\ref{su3met0}) and (\ref{su3met}).  The second case, $c^2=\ft1{11}$,
corresponds to the squashed Einstein metric on $SU(3)$.  (See, for
example, \cite{gilupo}.)

\subsection{Global aspects of the Wu manifold}
\label{StrucSubSec}

The metric (\ref{WuMetric}) has an obvious set of
three Killing vectors of $SO(3)_2$ corresponding to the right-translations
of $\cO_2^T$.  Among these is the $U(1)$ Killing vector
\bea
K=\fft{\del}{\del\phi}\,.\label{U1KV}
\eea
Calculating the square of this Killing vector, we find
\bea
|K|^2 &\equiv& g_{mn}\, K^m\, K^n\,,\nn\\
&=&
\sin^2 2\mu\, \cos^2\theta + \Big[\cos^2\psi\, \sin^2(\mu-\nu) +
  \sin^2\psi\, \sin^2(\mu+\nu)\Big]\, \sin^2\theta\,.
\eea
This vanishes when both of the two terms on the second line vanish,
which therefore requires
$\mu=0$, and hence also $\nu=0$.  That is to say, the $U(1)$ generator
$K$ will have a fixed point at $\mu=\nu=0$.  To analyse the nature of this
fixed point, we may define polar coordinates $(\rho,\alpha)$ that
are adapted to describing the vicinity of $\mu=\nu=0$, by writing
\bea
\mu=\rho\, \sin\alpha\,,\qquad \nu=\sqrt3\, \rho \,\cos\alpha\,.
\eea

   The metric (\ref{WuMetric}) now takes the (exact) form
\bea
ds_5^2 &=& d\rho^2 + \rho^2\, d\alpha^2 +
   \sin^2\big[2\rho\, \sin(\alpha+\ft23\pi)]\, \sigma_1^2 +
 \sin^2\big[2\rho\, \sin(\alpha-\ft23\pi)]\, \sigma_2^2\nn\\
&& +
\sin^2\big[ 2\rho\,\sin\alpha\big]\,\sigma_3^2\,,\label{wumetrhoal}
\eea
where $\alpha \in [\frac{\pi}{3},\frac{2\pi}{3}]$ in order to cover $-\mu \leq \nu \leq \mu $ with $0 \leq \mu \leq \frac{\pi}{2}$ in the fundamental region of $(\mu,\nu)$ shown in Fig.~\ref{munu}.
In the vicinity of $\rho=0$, the metric (\ref{wumetrhoal}) takes the
approximate form
\bea
ds_5^2 \approx d\rho^2 + \rho^2\, ds_4^2 + {\cal O}(\rho^4)\,,
\label{nearrho=0}
\eea
where
\bea
ds_4^2= d\alpha^2 + 4\sin^2(\alpha+\ft23\pi)\, \sigma_1^2 +
  4\sin^2(\alpha-\ft23\pi)\, \sigma_2^2 +
  4\sin^2\alpha\, \sigma_3^2\,\label{triaxialS4}\,.
\eea
This metric is precisely of the form of the triaxial Bianchi IX form
of the metric on the unit 4-sphere described by Giulini in \cite{giulini} (with
his radial coordinate $t$ related to our coordinate $\alpha$ by $t=-\alpha$).
It can be straightforwardly verified that $ds_4^2$ is an Einstein metric,
satisfying $R_{ab}=3\, g_{ab}$. Thus the metric (\ref{nearrho=0}) does indeed
extend smoothly onto $\R^5$ in the neighbourhood of the point $\rho=0$.

As discussed in \cite{giulini}, the volume of the $S^4$
manifold on which the metric $ds_4^2$ is defined is
\bea
{\rm Vol}(S^4) &=& 8 \int_{\frac{\pi}{3}}^{\frac{2\pi}{3}} d\alpha\, \sin\alpha\, \sin\left( \alpha+\ft23\pi\right)\,
\sin\left(\alpha-\ft23\pi\right)\, \int_{S^3/D_8^*}\,
\sigma_1\wedge\sigma_2\wedge\sigma_3\nn\\
&=& 8\left[\frac{1}{12}\cos(3\alpha)\right]^{\frac{2\pi}{3}}_{\frac{\pi}{3}} (2\pi^2)\,\\
&=& \ft83\pi^2\,.
\label{volS4}
\eea
This is the correct volume for the unit 4-sphere.

  The fact that $ds_4^2$ is the metric on the unit 4-sphere is exactly what is
needed in order to demonstrate that the metric (\ref{nearrho=0}) in the
vicinity of $\rho=0$ indeed extends smoothly onto the point $\rho=0$ itself,
just like the origin of four-dimensional Cartesian coordinates when written in
terms of hyperspherical coordinates. Since the left-invariant 1-forms $\sigma_i$ must be interpreted as
being defined on $S^3/D_8^*$ in order to get a smooth degeneration of
$S^4$ orbits as $\rho$ approaches zero, this means that the same
identification under $D_8^*$ must hold everywhere in the five-dimensional
space $SU(3)/SO(3)$.




\section{\label{CP2sec}$\text{Spin}^c$ and spin$^h$ structures on $\CP^2$}

  Before continuing with our investigation of the properties of the Wu manifold it is instructive first to pause and look at some features of another manifold, namely the complex projective plane $\CP^2$.  This has complex dimension 2, and thus when viewed as a real manifold it is four dimensional.  One of the interesting features of the $\CP^2$ manifold is that, like the Wu manifold, it does not admit an ordinary spin structure.  However, in the case of $\CP^2$ it does have a spin$^c$ structure, whereas the Wu manifold does not.  The main purpose of this section, then, is to summarise some of the key features of $\CP^2$ that demonstrate the absence of an ordinary spin structure and the existence of a spin$^c$ structure, so that in the following section we shall be able to exhibit some of the parallels, and also the differences, between the $\CP^2$ and the Wu cases. 

\subsection{The Fubini-Study metric on $\CP^2$ in real coordinates}
  
  The standard complex Fubini-Study metric on $\CP^2$ can be recast in a very simple real form, by introducing real coordinates adapted to the symmetries as described in \cite{gibpop}:  
\bea
\label{cp2met2}
d\Sigma_2^2 =d\mu^2 + \ft14 \sin^2\mu\, (\sigma_1^2+\sigma_2^2) +
  \ft14 \sin^2\mu\, \cos^2\mu\, \sigma_3^2\,.
\eea
Here $\sigma_i$ are left-invariant 1-forms of $SU(2)$, and $0\le\mu\le\ft12\pi$.  We choose the
natural vielbein basis $e^a$ with
\bea
e^1= \ft12\sin\mu\, \sigma_1\,,\quad
e^2= \ft12\sin\mu\, \sigma_2\,,\quad
e^3= \ft12\sin\mu\,\cos\mu\, \sigma_3\,,\quad
e^4=d\mu\,.\label{cp2met}
\eea
The spin connection $\omega_{ab}$ and the curvature 2-forms $\Theta_{ab}$ are then given by
\bea
\omega_{12} &=& \ft14(\cos2\mu-3)\, \sigma_3\,,\qquad
\omega_{13}= \ft12 \cos\mu\, \sigma_2\,,\qquad
\omega_{14}= \ft12 \cos\mu\, \sigma_1\,,\nn\\
\omega_{23} &=& -\ft12\cos\mu\, \sigma_1\,,\qquad
\omega_{24} = \ft12 \cos\mu\, \sigma_2\,,\qquad
\omega_{34} = \ft12\cos2\mu\, \sigma_3\,,\label{CP2om}
\eea
and
\bea
\Theta_{12}&=& 4 e^1\wedge e^2 + 2 e^3\wedge e^4\,,\quad
\Theta_{13}= e^1\wedge e^3 + e^2\wedge e^4\,,\quad
\Theta_{14}= e^1\wedge e^4- e^2\wedge e^3\,,\nn\\
\Theta_{23}&=& e^2\wedge e^3 - e^1\wedge e^4\,,\quad
\Theta_{24}= e^2\wedge e^4+ e^1\wedge e^3\,,\quad
\Theta_{34}= 4 e^3\wedge e^4+ 2 e^1\wedge e^2\,.\qquad\label{CP2Theta}
\eea
The K\"ahler form $J$ is given by
\bea
J= \ft12 dA^0= e^1\wedge e^2 + e^3\wedge e^4\,,\qquad \hbox{where}\quad
A^0=-\ft12 \sin^2\mu\, \sigma_3\,.
\label{U1gf}
\eea
We choose conventions where $J$ is self-dual.  Note that the metric
(\ref{cp2met}) is Einstein with $R_{ab}=6\delta_{ab}$.

There is
also an $SU(2)$ Yang-Mills connection, given by
\bea
A^1= \cos\mu\, \sigma_1\,,\qquad
A^2= \cos\mu\, \sigma_2\,,\qquad
A^3= (1-\ft12 \sin^2\mu)\,\sigma_3\,.\label{cp2su2con}
\eea
The $SU(2)$ field strengths $F^i=dA^i+\ft12 \ep_{ijk}\, A^j\wedge A^k$ are given by
\bea
F^1=2(e^1\wedge e^4 - e^2\wedge e^3)\,,\quad
F^2=2(e^2\wedge e^4 - e^3\wedge e^1)\,,\quad
F^3=2(e^3\wedge e^4 - e^1\wedge e^2)\,.\label{cp2Fi}
\eea
In the convention where $J$ is self-dual, the Yang-Mills fields
$F^i$ are anti-self dual. Defining also $F^0\equiv 2J= dA^0$, these obey
\bea
F^i_{ac}\, F^i_{bc} &=& 12\delta_{ab}\,,\qquad 
F^i_{ab}\, F^j_{ab}=16\delta_{ij}\,,\qquad
F^i_{ac}\, F^j_{bc}= 4\delta_{ij}\, \delta_{ab} 
    -2\epsilon_{ijk}\, F^k_{ab}\,,\nn\\
F^0_{ab}\, F^j_{ab}&=&0\,,\hspace{40pt}
F^0_{ac}\, F^0_{bc}= 4\delta_{ab}\,.\label{FFidCP2}
\eea

The spin connection, the curvature 2-forms and the Riemann tensor can be written as
\bea
\omega_{ab} &=& \fft14 F^i_{ab}\, A^i + \fft34 F^0_{ab}\, A^0\,,
\label{omabCP2}\\
\Theta_{ab}&=& \fft14 F^i_{ab}\, F^i +\fft34 F^0_{ab}\, F^0\,,
\label{ThabCP2}\\
R_{abcd}&=&\frac{1}{4}F^i_{ab}F^i_{cd}+\frac{3}{4}F^0_{ab}F^0_{cd}\,.   \label{RiemCP2}
\eea

\subsection{\label{CP2spincsec}Spinors in $\CP^2$}

We now define the generators
\begin{equation}\widehat{F}^i\equiv \frac{\im}{16}F^i_{ab}\,\Gamma^{ab} , \qquad \widehat{F}^0 \equiv \frac{\im}{16} F^0_{ab}\,\Gamma^{ab}\,,
\label{Fgen}
\end{equation}
where $\Gamma_{ab}= \ft12(\Gamma_a\,\Gamma_b-\Gamma_b\,\Gamma_a)$ and $\Gamma_a$ are the Dirac matrices.
The $\widehat F^i$ generators obey the $SU(2)$ algebra
\begin{equation}
    [\widehat{F}^i,\widehat{F}^j]=\im\epsilon^{ijk}\widehat{F}^k\,,
\end{equation}
and they commute with the $U(1)$ generator $\widehat F^0$.

In the basis (\ref{d5Gamma0}) for the Dirac matrices $\Gamma^a$ that we are using in this paper, the chirality operator $\Gamma_5\equiv -\Gamma^1\Gamma^2\Gamma^3\Gamma^4$ takes the form
\bea
\Gamma_5=\hbox{diag}\, (1,1,-1,-1)\,,
\eea
and we have
\begin{equation}
\widehat{F}^0\widehat{F}^0=
     \text{diag}\left(\frac{1}{4},\frac{1}{4},0,0\right)\,,\qquad
    \widehat{F}^i \widehat{F}^i=
     \text{diag}\left(0,0,\frac{3}{4},\frac{3}{4}\right) \,.
     \label{OpSqr}
\end{equation}
In fact the generators $\widehat F^0$ and $\widehat F^i$ take the block forms
\bea
\widehat F^0= -\fft12 \begin{pmatrix}\tau_3 & 0 \cr
                           0 & 0\end{pmatrix}\,,\qquad
\widehat F^i= \fft12 \begin{pmatrix} 0 & 0\cr 0& \tau_i\end{pmatrix}
\,,\label{hF0hFi}
\eea
where each entry is a $2\times2$ matrix and $\tau_i$, as usual, denotes the $2\times 2$ Pauli matrices.

On the totally geodesic submanifold $S^2$ at $\mu=\frac{\pi}{2}$, the only nonzero gauge potentials are $A^3=\frac{1}{2}\sigma_3$ and $A^0=-\frac{1}{2}\sigma_3$. On this sphere, we therefore have
\bea
    \frac{1}{4}\omega_{ab}\Gamma^{ab}&=&\frac{1}{32}F^3_{ab}\Gamma^{ab}\sigma_3-\frac{3}{32}F^0_{ab}\Gamma^{ab}\sigma_3\,,\nn\\
&=& -\frac{i}{2}\widehat F^3
    \sigma_3+\frac{3i}{2}\widehat{F}^0\sigma_3\,,
\eea
and the holonomy around the $S^1$ equator of this $S^2$ is therefore
\begin{equation}
\begin{split}
  H=  \mathcal{P}\text{exp}\left(-\oint_{S^1}\frac{1}{4}\omega_{ab}\Gamma^{ab}\right)&=\text{exp}\left( \im\int_{S^1}\sigma_3\left(\frac{1}{2}\widehat{F}^3-\frac{3}{2}\widehat{F}^0\right) \right)\\
    &=\text{exp}\left( \im\int_{0}^{2\pi}(2d\phi)\left(\frac{1}{2
    }\widehat{F}^3-\frac{3}{2}\widehat{F}^0\right) \right)\\
    &=\text{diag}\left(e^{3 \pi i },e^{-3\pi i},e^{\pi i },e^{-\pi i }\right)\\
    &= -\oneone.
    \end{split}
\end{equation}

This obstruction can also be calculated from the curvature of the connection. Thus with the spinor-covariant exterior derivative being $\nabla=d+\ft14 \omega_{ab}\, \Gamma^{ab}$ it follows that $\nabla^2=\ft14 \Theta_{ab}\, \Gamma^{ab}$, and we have
the holonomy 
\bea
 H=\mathcal{P}\exp\left(-\int_{S^2}\frac{1}{4}\Theta_{ab}\Gamma^{ab}\right)\,,
 \eea
where $\Theta_{ab}= \ft14 F^i\, F^i_{ab} + \ft34 F^0\, F^0_{ab}$ (see eqn (\ref{ThabCP2})).  As can be seen from the definitions of the field strengths $F^i$ and $F^0$ we have 
\bea
\int_{S^2} F^3=-\ft12\int\sigma_1\wedge\sigma_2=-2\pi\,,\qquad
\int_{S^2} F^0= \ft12\int\sigma_1\wedge\sigma_2=2\pi\,,
\eea
with $F^1$ and $F^2$ integrating to give zero, and so
\bea
\int_{S^2}\frac{1}{4}\Theta_{ab}\Gamma^{ab} &=&
-2\pi \im\, \widehat F^3 + 6\pi \im\, \widehat F^0\,,\nn\\
&=& \im \pi\,\text{diag}(-3,3,-1,1)\,,
\eea
and hence the same conclusion that the integral over the
$S^2$ gives a holonomy factor $H=-\oneone$ for uncharged spinors. This implies that there is no spin structure. The $U(1)$ gauge-covariant spinor exterior derivative $D=d + \ft14 \omega_{ab}\, \Gamma^{ab}- \im e\, A^0\,\oneone$ has curvature $D^2= \ft14\Theta_{ab}\, \Gamma^{ab} -\im e F^0\,\oneone$, and so the holonomy over the $S^2$ for a spinor field of charge $e$ is given by exponentiating the integral of this curvature, which gives
\bea
\int_{S^2} \Big(\fft14\Theta_{ab}\, \Gamma^{ab} -\im\, e
F^0\,\oneone\Big)  &=&-2\pi \im\, \widehat F^3 + 6\pi \im\, 
\widehat F^0
 - 2\pi \im\,e\, \oneone\,,\nn\\
 &=& \im \pi\,\text{diag}(-2e-3,-2e+3,-2e-1,-2e+1)\,.
 \eea
Thus we get a holonomy factor $H=(-1)^{2e+1}$ over the $S^2$, for a spinor of charge $e$.  By choosing 
\bea
e=m+\fft12\,,\qquad m=\,\hbox{integer}\,,
\eea
we get spinors that are well defined.  This is the spin$^c$ structure on $\CP^2$ (see, for example, \cite{Hawking:1977ab}).

\subsection{$U(1)$ gauge-covariantly constant spinors on $\CP^2$}

As discussed in, for example,  \cite{Pope:1980ub}, and as we shall see below, there exist a pair of $U(1)$ gauge-covariantly
constant spinors $\eta$ with charges $e=\pm\ft32$ in $\CP^2$, obeying
\bea
D\eta\equiv \nabla\eta-\im e A^0\,\eta=0\,,\qquad \hbox{or}\qquad
D_a \eta = \nabla_a \,\eta -\im e\, A^0_a \,\eta=0\,,\label{gaugecc2}
\eea
where $\nabla=d + \fft14 \omega_{ab}\, \Gamma^{ab}$ is the spinor-covariant exterior derivative.  $D$ is the $U(1)$ gauge-covariant spinor exterior derivative. 

The integrability condition $D^2\eta=0$, or $[D_a,D_b]\,\eta=0$, for (\ref{gaugecc2}) implies 
\bea
\ft14 \Theta_{ab}\, \Gamma^{ab}\eta -\im e F^0\,\eta=0\,,
\qquad\hbox{or}\qquad \ft14 R_{abcd}\, \Gamma^{cd}\,\eta
 - \im e\, F^0_{ab}\, \eta=0.
\eea
Using eqn (\ref{ThabCP2}) gives
\bea
\ft1{16} F^i\,F^i_{ab}\,\Gamma^{ab}\eta + 
\ft3{16} F^0\,F^0_{ab}\,\Gamma^{ab}\eta -\im e F^0\,\eta=0\,.
\eea
The self-dual and anti-self dual projections of this equation, implemented using $P_\pm=\ft12(1\pm *)$, must separately vanish. Since $F^0$ is self-dual, while the $F^i$ are anti-self dual, it then follows, after using also the definitions (\ref{Fgen}), that
\bea
\widehat F^i\,\eta=0 \qquad\hbox{and}\qquad  3\widehat F^0\,\eta + e\,\eta=0\,.
\label{genseta}
\eea
The spinor $\eta$ must therefore be right-handed, $\Gamma_5\,\eta=+\eta$, with $e=\pm\ft32$.  Specifically, we will have one solution for $e=+\ft32$ and one for $e=-\ft32$ (see eqns (\ref{hF0hFi}):
\bea
e=+\ft32:\quad \eta=f_{\sst+}\,\begin{pmatrix} 1\cr 0\cr 0\cr 0\end{pmatrix}\,,
\qquad\qquad e=-\ft32:\quad 
\eta=f_{\sst-}\,\begin{pmatrix} 0 \cr 1 \cr  0 \cr 0\end{pmatrix}\,,
\label{etapm}
\eea
where $f_{\sst\pm}$ are functions to be determined by solving $D\eta=0$.

  The condition $D\eta=0$ for a gauge-covariantly constant spinor
gives, after using eqn (\ref{omabCP2}) and eqns (\ref{Fgen}),
\bea
0=D\eta &=&d\eta + \ft14\omega_{ab}\,\Gamma^{ab}\eta 
-\im e\, F^0\,\eta\,,\nn\\
&=& d\eta -\im A^i\, \widehat F^i\,\eta 
-3\im\, A^0\, \widehat F^0\,\eta -\im e\, A_0\,\eta\,,\nn\\
&=& d\eta -\im A_0\, (3 \widehat F^0 \,\eta + e\,\eta)\,,
\eea
(since $\widehat F^i\,\eta=0$ as in eqn (\ref{genseta})). Thus
from the second equation in (\ref{genseta}) it follows that
the two spinors in eqns (\ref{etapm}) will obey $D\eta=0$,
provided that the prefactor functions $f_{\sst\pm}$ are chosen to be just constants.

The existence of this $U(1)$ gauge-covariantly constant spinor ensures 
that $\CP^2$ is at least a spin$^c$ manifold. It also plays an important role in 
building the spectrum of (gauged) Dirac operators using the prescription given 
in section \ref{SpecDiracSubsec}.

\subsection{$SU(2)$ gauge-covariantly constant spinors on $\CP^2$}

If we instead use the $SU(2)$ connections \eq{cp2su2con} to define an 
$SU(2)$ gauge-covariant derivative, with
\bea
D\eta= d\eta +\ft14\omega_{ab}\, \Gamma^{ab}\, \eta  -\im A^i\,T_i\,\eta\,,\qquad \hbox{or}\qquad
D_a \eta = \nabla_a \,\eta -\im \, A^i_a T^i\,\eta=0\,,
\label{SU2gaugeD}
\eea
where $T^i$ are the (Hermitian) generators of $SU(2)$ in the spin-$j$ representation with $j$ as yet arbitrary, obeying $[T^i,T^j]=\im\epsilon_{ijk}\, T^k$, then the
integrability condition $D^2\eta=0$, or 
$[D_a,D_b]\,\eta=0$, now implies 
\bea
F^i\, \widehat{F}^i\,\eta + 
3  {F}^0 \widehat{F}^0 \,\eta
 +F^i \,T^i \, \eta=0 \,,
\eea
The self-dual and anti-self dual projections must separately vanish and so we therefore have
\bea
 \widehat F^0\, \eta=0\qquad \hbox{and}\qquad 
 (\widehat{F}^i + T^i)\,\eta=0\,. \label{Fihateigv}
\eea
The first equation here implies that $\eta$ must be left-handed, $\Gamma_5\,\eta=-\eta$, (see eqns (\ref{hF0hFi})).  Multiplying the second equation by
$\widehat F^i$, and comparing with the result of multiplying instead by $T^i$, shows that 
\bea
\widehat F^i\,\widehat F^i\, \eta= T^i T^i\,\eta\,.
\eea
Since from eqns (\ref{OpSqr}) $\widehat F^i\,\widehat F^i$ is
equal to $\fft34$ when acting on left-handed spinors, it follows
that in order to satisfy the integrability condition, the generators $T^i$ should be in the $j=\ft12$ representation of $SU(2)$. One can, for example, take $T^i=\ft12\tau_i$ where $\tau_i$ are the Pauli matrices, in which case the second equation in (\ref{Fihateigv}) can be written as
\bea
\Big[ \begin{pmatrix} 0 & 0\cr 0 & \tau_i \end{pmatrix} \otimes \oneone_2 
 + \oneone_4\otimes \tau_i \Big] \,\eta=0\,.
 \eea
There is a unique solution for $\eta$, up to scaling.

Returning to the gauge-covariant constancy equation $D\eta=0$ itself, we have
\bea 
0=D\eta &=& d\eta + \ft14\omega_{ab}\,\Gamma^{ab}\,\eta - 
  \im A^i\, T^i\,\eta\,,\nn\\
&=& d\eta +\ft1{16}\, A^i\, F^i_{ab}\,\Gamma^{ab}\, \eta
+ \ft3{16} \, A^0\, F^0_{ab}\, \Gamma^{ab}\, \eta -
\im A^i\, T^i\, \eta\nn\\
&=& d\eta -\im\, A^i\, \widehat F^i\,\eta -
3\im A^0\, \widehat F^0\, \eta -\im A^i\, T^i\,\eta\nn\\
&=& d\eta -\im A^i\,(\widehat F^i + T^i)\, \eta\,,\nn\\
&=& d\eta\,,
\eea
where the penultimate line follows from the first equation in (\ref{Fihateigv}) and the final line follows from the second equation in (\ref{Fihateigv}).  Thus, we conclude that if the spinor $\eta$ carries the $j=\ft12$ representation of the $SU(2)$ gauge group, and if it obeys the conditions (\ref{Fihateigv}) and its components are constant, then it is $SU(2)$ gauge-covariantly constant. In \cite{Dolan:1984ve}, the $SU(2)$ connection \eq{cp2su2con} is used to calculate the Atiyah-Singer index of $\CP^2$, and the $SU(2)$ gauge-covariantly constant spinor in this paper corresponds to the zero mode that is discussed there. The spectra of the $U(1)$ and $SU(2)$ gauged-Dirac operators are also calculated in \cite{Dolan:2003bj}.

\subsection{Spin$^h$ spinors on $\CP^2$}

The fact that we were able to construct a gauge-covariantly constant spinor in the $j=\ft12$ representation on the $SU(2)$ Yang-Mills group on $\CP^2$ suggests that it should be possible to find larger classes of Yang-Mills charged spinors on $\CP^2$ that are also well-defined.  Indeed this can be done, as we now explicitly show.

Motivated by the discussion we gave in section \ref{CP2spincsec} for spin$^c$ spinors on $\CP^2$, we may consider parallel propagating a spinor in a general spin-$j$ representation of the $SU(2)$ Yang-Mills group, with the gauge-covariant spinor exterior derivative $D=d+ \ft14\omega_{ab}\, \Gamma^{ab} - \im A^i\, T^i$ as in eqn (\ref{SU2gaugeD}).  When such a spinor is parallel propagated around a family of closed loops spanning the totally-geodesic $S^2$ at $\mu=\ft12\pi$ in the $\CP^2$ metric we therefore pick up a holonomy factor
\bea
H={\cal P}\, \exp\int_{S^2}\Big(\ft14\Theta_{ab}\, \Gamma^{ab} -
\im F^i\, T^i\Big)\,.\label{spinhCP2}
\eea
Integrated over the $S^2$ we have 
\bea
\int_{S^2} F^0= 2\pi\,,\qquad \int_{S^2} F^3=-2\pi\,,
\eea
with the integrals of $F^1$ and $F^2$ giving zero.
Thus we find that the holonomy factor in eqn (\ref{spinhCP2}) gives
\bea
H&=&\exp \left(\im\pi \begin{pmatrix} 3\tau_3 &0\cr 0& \tau_3 \end{pmatrix}\right) \, \exp(2\pi\im\, T^3)\,,\nn\\
&=& -\exp(2\pi\im\, T^3)\,.\label{holoCP2}
\eea
Now, in the spin-$j$ representation we can choose a basis for the $SU(2)$ generators in which 
\bea
T^3 = \hbox{diag}\, (j\,, j-1\,, j-2\,,\cdots \,,
-j+1\,, -j)\,,
\eea
and so $\exp(2\pi\,\im T^3)= (-1)^{2j}$.  It then follows from eqn (\ref{holoCP2}) that the holonomy factor for spinors in the spin-$j$ representation of $SU(2)$ will be
\bea
H=(-1)^{2j+1}\,,
\eea
and consequently, it will be $+1$ for any spinor in a spin-$j$ representation of $SU(2)$ with $j$ being a half odd integer.

In summary, we have shown explicitly that spinors in half odd integer spin representations of $SU(2)$ are well defined on $\CP^2$.  This provides an alternative to the usual spin$^c$ construction of well-defined spinors on $\CP^2$. It was observed in \cite{Hawking:1977ab} that such a non-abelian generalisation of spin$^c$ structures could be carried out, and our discussion here gives a concrete illustration of this.  Constructions of this type have since become known as spin$^h$ structures. Note that a spin$^c$ structure always implies the existence of a spin$^h$ structure by embedding the spin$^c$ connection in a spin$^h$ connection, as discussed in, for example, \cite{witten19}.

\section{\label{Wuspinhsec}Spin\texorpdfstring{$^h$}{ h} structure on the Wu manifold}

The original argument in \cite{Hawking:1977ab} can be summarised as follows. Consider an $S^2$ submanifold within an orientable Riemannian $n \geq 3$-dimensional manifold $M$ as a map from $S^2$ into $M$. Choose a point $p$ on this $S^2$ submanifold. Then consider a family of loops $\lambda_u$ with basepoint $p$ that continuously span the entire $S^2$ with $\lambda_0=\lambda_1$ being just the point $p$ and $\lambda_\frac{1}{2}$ being an equator of $S^2$ through p.  Considering the spin connection as an $SO(n)$ gauge field, each loop $\lambda_u$ gives an $SO(n)$ holonomy factor $h(u)$. Since $\lambda_0=\lambda_1$, $h(0)=h(1)$ and so $h(u)$ traces out a closed loop in $SO(n)$ between $h(0)$ and $h(1)$. This closed loop defines an element of $\pi_1(SO(n))=\Z_2$.  

Now, consider parallel transporting a spinor around the same set of loops to get a set of Spin($n$) holonomy factors $H(u)$, where Spin($n$) is the double cover of $SO(n)$. Whether a spinor is consistent over the sphere can be determined by using two coordinate patches on each half of $S^2$ and calculating the multiplicative difference in the spin holonomy at the equator as $+1$ or $-1$, which corresponds to the two elements of $\pi_1(SO(n))=\mathbb{Z}_2$. If $M$ is a spin manifold, $H(0)=H(1)$ since $\lambda_0=\lambda_1$, which is a point. However, if $H(0)=-H(1)$, spinors cannot be globally defined and so $M$ cannot be a spin manifold. 

The totally geodesic $S^2$ within $\CP^2$ gives a result of $-1$ that can then be cancelled by charging fermions with a $U(1)$ or $SO(3)$ gauge field that gives an additional $-1$ factor when going from $H(0)$ to $H(1)$. 

This exact argument will be used for an $S^2$ submanifold of the Wu manifold, where it will be seen that an $SO(3)$ gauge field is needed to provide an additional $-1$ factor.

\subsection{Description using Stiefel-Whitney classes}

For the Wu manifold, the holonomy argument can be interpreted in the following way. The mathematical background for this interpretation can be found in, for example, \cite{classes,hatcher}. Any totally geodesic submanifold $M_1$ of a manifold $M_2$ is given by an immersion, $f:M_1 \rightarrow M_2$. Let $f^\ast TM_2$ denote the pullback of the tangent bundle on $M_2$ to a vector bundle on $M_1$ via the map $f$. Then there is a Whitney sum decomposition into tangent and normal bundles $f^\ast TM_2 \cong TS^2+NS^2$ (see Corollary 3.5 of \cite{classes}). Specifically, consider the immersion of the totally geodesic $S^2$ in the Wu manifold $f:S^2\rightarrow \text{Wu}$. Then, the Stiefel-Whitney classes obey \cite{classes}
\begin{equation}
    w_1(f^\ast T\text{Wu})=w_1(TS^2)+w_1(NS^2)=0+0=0.
\end{equation}
Note that $w_1(TS^2)=0$ since $S^2$ is orientable.
\begin{equation}
\begin{split}
w_2(f^\ast T\text{Wu})=&w_2(TS^2)+w_1(TS^2)w_1(NS^2)+w_2(NS^2)\\
=&0+0+w_2(NS^2)\\
=&w_2(NS^2).
\end{split}
\end{equation}
Note that $w_2(TS^2)=0$ since $S^2$ is a spin manifold.

This matching of Stiefel-Whitney classes means that any failure of Wu to be spin on this $S^2$ is detected by its normal bundle. Since $NS^2$ is a rank 3 orientable bundle over $S^2$, it is an $SO(3)$ bundle. $SO(3)$ bundles over $S^2$ are determined by a map from the equator $S^1$ to $SO(3)$ and therefore classified by $\pi_1(SO(3))=\mathbb{Z}_2$. When we calculate the difference of holonomies of a spinor around the equator of $S^2$ using two coordinate patches, a factor of $+1$ implies it is the trivial $SO(3)$ bundle that can lift to an $SU(2)$ bundle (which has $\pi_1(SU(2))=0$). On the other hand, a factor of $-1$ implies it is the nontrivial $SO(3)$ bundle that does not lift to an $SU(2)$ bundle. 

Introducing an $SO(3)$ bundle $Q'$ to the Wu manifold with $w_1(Q')=0$ and $w_2(Q')=1$ means that the combined holonomy factor from the normal bundle and $Q'$ is $(-1)(-1)=+1$. When the holonomy calculation is carried out explicitly on $S^2$ in Wu, it will be seen that the normal components of the spin connection give the $-1$ factor which is then cancelled by the contribution from an $SO(3)$ connection.

\begin{table}[b]
\centering
\scriptsize
\setlength{\tabcolsep}{4pt}
\renewcommand{\arraystretch}{1.15}
\begin{tabular}{|c||c|c|c|c||c|c|c|c|}
\hline
\multicolumn{1}{|c||}{\multirow{2}{*}{$k$}}
& \multicolumn{4}{c||}{$W=SU(3)/SO(3)$}
& \multicolumn{4}{c|}{$\mathbb{CP}^2$} \\
\cline{2-9}
\multicolumn{1}{|c||}{}
& $\pi_k(W)$
& $H_k(W;\mathbb Z)$
& $H^k(W;\mathbb Z)$
& $H^k(W;\mathbb Z_2)$
& $\pi_k(\mathbb{CP}^2)$
& $H_k(\mathbb{CP}^2;\mathbb Z)$
& $H^k(\mathbb{CP}^2;\mathbb Z)$
& $H^k(\mathbb{CP}^2;\mathbb Z_2)$ \\
\hline\hline
0 & 0 & $\mathbb Z$ & $\mathbb Z$ & $\mathbb Z_2$ & 0 & $\mathbb Z$ & $\mathbb Z$ & $\mathbb Z_2$ \\
\hline
1 & 0 & 0 & 0 & 0 & 0 & 0 & 0 & 0 \\
\hline
2 & $\mathbb Z_2$ & $\mathbb Z_2$ & 0 & $\mathbb Z_2$ & $\mathbb Z$ & $\mathbb Z$ & $\mathbb Z$ & $\mathbb Z_2$ \\
\hline
3 & $\mathbb Z_4$ & 0 & $\mathbb Z_2$ & $\mathbb Z_2$ & 0 & 0 & 0 & 0 \\
\hline
4 & 0 & 0 & 0 & 0 & 0 & $\mathbb Z$ & $\mathbb Z$ & $\mathbb Z_2$ \\
\hline
5 & $\mathbb Z \oplus \mathbb Z_2$ & $\mathbb Z$ & $\mathbb Z$ & $\mathbb Z_2$
  & $\mathbb Z$ & 0 & 0 & 0 \\
\hline
\end{tabular}
\caption{Homotopy, integral (co)homology and $\mathbb Z_2$ cohomology groups of the Wu manifold and $\CP^2$.} \label{homology}
\end{table}

\subsection{Description using characteristic classes}

The property that $\mathbb{CP}^2$ is a spin$^c$ manifold can also be described in the following way (see, for example, \cite{freedwitten}). The first Stiefel-Whitney class $w_1(\mathbb{CP}^2)=0$ is trivial since $\mathbb{CP}^2$ is orientable. However, it has nontrivial second Stiefel-Whitney class $w_2(\mathbb{CP}^2)=1$ which lies in $H^2(\mathbb{CP}^2,\mathbb{Z}_2)=\mathbb{Z}_2$. Since $H^3(\mathbb{CP}^2,\mathbb{Z})=0$, there exists a  $U(1)$ bundle $Q$ such that 
$w_2(Q)=w_2(\mathbb{CP}^2)$.  Here $w_2(Q)$ is the mod 2 reduction of the first Chern class of the $U(1)$ bundle. For $S^2$, which is a generator of $H_2(\mathbb{CP}^2,\mathbb{Z})$, the flux quantisation condition for spin$^c$ is
\begin{equation}
   \frac{1}{2}\int_{S^2} w_2(\mathbb{CP}^2)= \int_{S^2} \frac{F}{2\pi}=\frac{1}{2} \mod \mathbb{Z}.
\end{equation}

The Wu manifold $SU(3)/SO(3)$ is also orientable and not spin, so $w_1(\text{Wu})=0$ and $w_2(\text{Wu})=1$. However, $H^2(\text{Wu},\mathbb{Z})=0$, so there are no nontrivial $U(1)$ bundles to provide a spin$^c$ structure. However, we can introduce a principal $SO(3)$ bundle $Q'$ such that $w_2(Q')=w_2(\text{Wu})=1$ \cite{Chen:2017}. Since $H_2(\text{Wu},\mathbb{Z})=\mathbb{Z}_2$ and $H_2(\text{Wu},\mathbb{Z}_2)=\mathbb{Z}_2$, we can consider a nontrivial $S^2$ or $\mathbb{RP}^2$ generator, denoted by $L$. Then they can be paired (i.e. through a holonomy calculation) to get
\begin{equation}
    (w_2(Q'),L)=(w_2(\text{Wu}),L)=1 \mod 2 .
\end{equation}

For reference, the homotopy, integral (co)homology and $\Z_2$ cohomology groups of $\CP^2$ and the Wu manifold are listed in Table~\ref{homology}.

\subsection{Holonomy}

Note that the identity $R_{abcd}=F^i_{ab}F^i_{cd}$ in eqn (\ref{RiemFF}) 
will be used repeatedly, together with the identity (\ref{FFasym}), which
can be rewritten as $F^i_{ac}\, F^j_{cb}- F^i_{bc}\, F^j_{ca}
\equiv [F^i,F^j]_{ab}= \epsilon_{ijk}\, F^k_{ab}$.  Thus
the $F^i_{ab}$, viewed as $5\times 5$ antisymmetric
matrices, are the generators of $SO(3)$ in the $j=2$ representation, as 
is confirmed by noting from eqns (\ref{FFid}) that $F^i_{ab}F^{i\,bc}=-2(2+1)\delta_a^c$.  When we need an explicit representation for the Dirac matrices we shall use the expressions given in eqns (\ref{d5Gamma0}).

\subsubsection{Vectors}
Consider the local holonomy of a vector $V^a$ around a loop defined by the differential 2-form $\delta\Sigma^{cd}$:
\begin{equation}
        \delta V^a=R^a {}_{bcd}\, V^b \,\delta \Sigma^{cd}=(F^i_{cd}\,\delta\Sigma^{cd})\,F^{i\,a} {}_{ b} \,V^b\equiv c^i\, 
        F^{i\,a}{}_b\,V^b \, ,
\end{equation}
for constants $c^i \equiv F^i_{cd}\,\delta \Sigma^{cd}$. This shows that vectors transform in the $j=2$ irrep of $SO(3)$.

\subsubsection{Uncharged spinors}

Consider the holonomy for an uncharged spinor $\psi$ around the same loop defined by $\delta\Sigma^{cd}$:
\begin{equation}
    \delta \psi=\frac{1}{4}R_{abcd} \,\Gamma^{ab} \psi \, \delta \Sigma^{cd}=(-\im F^i_{cd}\,\delta\Sigma^{cd})\,\frac{\im}{4}F^{i}_{a b} \,\Gamma^{ab}\psi=-\im c^i\, \widehat F^i\,\psi \, , 
\end{equation}
where the generators $\widehat F^i$ are defined as 
\ben
\widehat F^i\equiv \frac{\im}{4} F^i_{ab} \Gamma^{ab} \,.
\een
They satisfy
\be
\bald
\widehat F^i \widehat F^i=&-\frac{1}{16}R_{abcd}\,\Gamma^{ab}\,\Gamma^{cd} 
=\frac{R}{8} \oneone =\frac{15}{4} \oneone \, ,
\eald
\ee
using \eq{RiemFF}. The generators $\widehat F^i$ satisfy the commutator $[\widehat F^i,\widehat F^j]=
\im \epsilon^{ijk}\widehat F^k $. Thus, uncharged spinors transform in the $j=\frac{3}{2}$ irrep of $SU(2)$.

\subsubsection{Charged spinors}

Finally, consider the holonomy for a charged spinor $\psi$ around the same loop defined by $\delta\Sigma^{cd}$:
\begin{equation}
\begin{split}
    \delta \psi&=\frac{1}{4}R_{abcd} \,\Gamma^{ab} \psi \,\delta \Sigma^{cd}-\im F_{cd}^i\,T^i \psi \,\delta \Sigma^{cd}\\
    &=\frac{1}{4}F_{ab}^i \,F_{cd}^i\,\Gamma^{ab} \psi\,\delta \Sigma^{cd}-\im F_{cd}^iT^i \psi \,\delta \Sigma^{cd}\\
    &=(-\im F^i_{cd}\,\delta \Sigma^{cd})\left(\frac{\im}{4}F^i_{ab}\,\Gamma^{ab}+T^i\right)\psi\\
    &=-\im c^i\,(\widehat F^i+T^i) \psi \, ,
    \end{split}
\end{equation}
where $T^i$ are the (Hermitian) generators of $SU(2)$ in the spin$-j$ representation. Therefore, we have that $J^i \equiv \widehat F^i \otimes \oneone + \oneone \otimes T^i$ is the generator of the holonomy of charged spinors. In other words, $J^i$ generates the tensor product of two irreps, one generated by $\widehat F^i$ and one generated by $T^i$. 

$\widehat F^i$ has already been shown to generate the 
$j=\frac{3}{2}$ irrep of $SU(2)$, denoted $V_\frac{3}{2}$. Let $T^i$ generate a spin $j$ irrep of $SU(2)$, denoted $V_j$, i.e. $T^i T^i = j(j+1) \oneone $. Then, by the standard Clebsch-Gordan decomposition, $J^i$ generates 
\be
V_{\frac{3}{2}} \otimes V_j = V_{|j-\frac{3}{2}|} \oplus ...\oplus V_{j+\frac{3}{2}} \, .
\ee
This will give a set of states $\{|J,M\rangle\}$ for $|j-\frac{3}{2}|\leq J \leq j+\frac{3}{2}$ and $-J \leq M \leq J $.

The gauge-covariantly constant spinor $|0,0\rangle$ corresponds to $V_0$, the $j=0$ irrep of $SU(2)$. Since $V_0$ only appears in the direct sum above when $j=\frac{3}{2}$, we have to choose the $j=\frac{3}{2}$ irrep of $SU(2)$ for $T^i$ to get a gauge-covariantly constant spinor. For this choice, we get the decomposition $V_\frac{3}{2} \otimes V_\frac{3}{2}=V_0 \oplus V_1 \oplus V_2 \oplus V_3$. All these states (and states for any other choice of $j$) can be read off from Clebsch-Gordan coefficients.

For example, the Clebsch-Gordan coefficients imply the gauge-covariantly constant spinor is 
\ben
\label{GCovConst}
\Big|0,0\Big\rangle=\frac{1}{2}\left(\Big|\frac{3}{2},-\frac{3}{2}\Big\rangle- \Big|\frac{1}{2},-\frac{1}{2}\Big\rangle + \Big|-\frac{1}{2},\frac{1}{2}\Big\rangle- \Big|-\frac{3}{2},\frac{3}{2}\Big\rangle\right) \, , 
\een
which is written in 2-component spinor notation in eqn \eq{psires}. Note that 
\begin{equation}
    \widehat F^3 \equiv 
\frac{\im}{4}F^3_{ab}\Gamma^{ab} = \text{diag}\left(-\frac{3}{2},\frac{3}{2},\frac{1}{2},-\frac{1}{2}\right)
\end{equation}
in our conventions for the gamma matrices \eq{d5Gamma0} and $T^3 = \text{diag}(\frac{3}{2},\frac{1}{2},-\frac{1}{2},-\frac{3}{2})$. This also makes it easy to write down a triplet of states $V_1=\{|1,-1\rangle,\,|1,0\rangle,\,|1,1\rangle\}$ etc.

\subsection{Charged spinors on the Wu manifold}
\label{sec:sphereholonomy}

The metric on the Wu manifold is (\ref{WuMetric})
\begin{equation}
    ds_5^2=d\mu^2+\frac{1}{3} d\nu^2+\sin^2(\mu+\nu)\sigma_1^2+\sin^2(\mu-\nu)\sigma_2^2+\sin^2 2\mu\,\sigma_3^2 \, .
\end{equation}
where $0 \leq \mu \leq \frac{\pi}{2}$ and $-\mu \leq \nu \leq \mu$. $\sigma_i$ are the left-invariant one-forms defined on the lens space $S^3/D_8^\ast=SO(3)/D_4$.

Consider $d\theta=d\phi=0$ and $\nu=0$ (so $0 \leq \mu \leq \frac{\pi}{2}$). Note that these choices imply $d\nu=0$, $\sigma_1=\sigma_2=0$, $\sigma_3=d\psi$. Then, the metric can be written as
\begin{equation}
\label{ParTransMet}
    ds_2^2=d\mu^2+\sin^2 2\mu\, d\psi^2 \, ,
\end{equation}
with $\mu \in [0,\frac{\pi}{2}]$ and $\psi \in [0,\pi)$. We can rewrite the metric as
\begin{equation}
    ds_2^2=\frac{1}{4}\left(d \hat \theta^2+\sin^2\hat \theta\,d\hat\psi\right) \, ,
\end{equation}
where $\hat \theta\equiv 2\mu$, $\hat \psi \equiv 2\psi$. Since $\hat \theta \in [0,\pi]$ and $\hat \psi \in [0,2\pi)$, this is the metric of a 2-sphere with equator at $\hat \theta=\frac{\pi}{2}$. This 2-sphere is a totally geodesic submanifold of the Wu manifold (see appendix~\ref{SubmanSec} for a full list of totally geodesic submanifolds of the Wu manifold which matches \cite{klein2009}). Therefore, the equator is a geodesic of both $S^2$ and the Wu manifold.

Parallel transporting a charged spinor around the equator $\gamma$ at $\hat \theta=\frac{\pi}{2}$ multiplies it by the factor
\begin{equation}
    U(\gamma) \equiv\mathcal{P} \text{exp}\left(\oint _\gamma \im A^i (\widehat F^i+T^i)\right)=  \text{exp}\left(\oint _\gamma \im A^i J^i\right)=\text{exp}\left(\oint _\gamma \im A^3 J^3\right),
\end{equation}
where $A^3=\cos2\mu\, \sigma_3=\frac{1}{2}\cos\hat \theta\, d\hat\psi$. In order to globally define a gauge field, we introduce $A^+\equiv \frac{1}{2}(-1+\cos\hat \theta)\, d\hat\psi$ on the top half of the sphere and $A^- \equiv \frac{1}{2}(1+\cos\hat \theta)\, d\hat\psi$ on the bottom half of the sphere. Then 
\begin{equation}
    U(\gamma) = \text{exp}\left(\oint _\gamma \im (A^--A^+) J^3\right)=\text{exp}\left(\oint _\gamma \im d\hat \psi J^3\right).
\end{equation}
This is then equal to 
\bea
    U(\gamma)&=&
    \text{diag}(e^{2 \pi i  J},e^{2\pi i (J-1)},...,e^{2 \pi i (-J+1)},e^{-2\pi i J})\\
   & =&(-1)^{2J} \oneone\\
   & =&(-1)^{(2j+1)} \oneone \, , \label{quants2}
\eea
where we used the fact that $J$ takes the values $|j-\frac{3}{2}|,...,(j+\frac{3}{2})$. For the holonomy factor to be consistent, $j$ must be a half-integer. 

Note that the same calculation with uncharged spinors would give the holonomy factor
\begin{equation}
    \text{diag}(e^{-3\pi \im},e^{3\pi \im},e^{\pi \im},e^{-\pi \im})=-\oneone \, .
\end{equation}
Note also that this calculation equivalently gives a flux quantisation condition for the field strength $F^3 \equiv dA^3$ integrated over the sphere of radius $\frac{1}{2}$ defined by $\hat{\theta} \in [0,\pi],\hat{\psi} \in [0,2\pi)$ which we denote $\Sigma$.
\begin{equation}
\int_{\Sigma} F^3   =\frac{2\pi n}{M} \, ,\quad \quad  F^3= -\frac{1}{2}\sin \hat{\theta} d \hat{\psi} \, . \label{s2quant2}
\end{equation}
where $n$ is any integer and $M$ is the $J^3$-eigenvalue (i.e. $M=|j-\frac{3}{2}|,...,(j+\frac{3}{2})$).

A comment on the significance of the above calculations is in order.  The Wu manifold $SU(3)/SO(3)$ is endowed with an $SO(3)$ connection $A^i$, whose field strength $F^i$ gives the curvature of the principal $SO(3)$ fibre in the description of $SU(3)$ 
as an $SO(3)$ bundle over the Wu manifold.  Starting from a well-defined field on the Wu manifold,
for example an uncharged scalar field, we could then consider a generalisation to scalar fields that were charged
with respect to the $SO(3)$ Yang-Mills gauge group. Such charged scalars would also be well-defined, provided they were in any integer-spin representation of $SO(3)$.  It would not be permissible to take the scalars to be in a half-integer spin representation, since only integer spin is allowed within $SO(3)$.  Concretely, one would find that in a holonomy calculation along the lines of the ones we performed earlier for spinors in the Wu manifold, the scalar in a half-integer spin representation of the Yang-Mills group would acquire a $-1$ holonomy and hence would not be globally well-defined.  By contrast, when we instead consider the holonomy for spinors in the Wu manifold, as we did above, there is already a holonomy factor of $-1$ arising for uncharged spinors.  Therefore, the way to cancel this is precisely to consider instead charged spinors that are in a half-integer spin representation of the Yang-Mills group, so that the further $-1$ holonomy factor thereby introduced will cancel the $-1$ holonomy for uncharged spinors.

In this respect, the situation is closely analogous to that for spinors in $\CP^2$.  In that case, the way to cancel the $-1$ holonomy factor for uncharged spinors was to take them to be charged and coupled to the $U(1)$ connection in $\CP^2$, but with a charge $e$ that violates the standard Dirac quantisation consistency condition $2eP=\,$integer that would be necessary when considering charged generalisations of neutral fields that were themselves already well-defined, and that instead satisfies the modified quantisation condition $2eP=\ft12 +\,$integer, so as to introduce a $-1$ holonomy factor to cancel the $-1$ holonomy for uncharged spinors.

\subsection{Flux quantisation on \texorpdfstring{$\RP^2$}{RP2}}
There is a totally geodesic $\RP^2$ within the Wu manifold with $\mu=\frac{\pi}{2}$ and $\nu=0$:
\begin{equation}
ds^2=d\theta^2+\sin ^2 \theta d \phi^2 \, ,
\end{equation}
with $\theta \in [0,\pi]$, $\phi \in [0,2\pi)$ and an antipodal identification $(\phi,\theta) \sim (\phi+\pi,\pi-\theta)$. $\RP^2$ does not admit a spin structure but instead admits two $\text{Pin}^-$ structures, as discussed in appendix \ref{sec:pin}. This means that, instead of  antipodally identifying spinors on the double cover $S^2$, spinors are identified up to a factor of $\pm \im \tau_2$.  (Taking $\tau_1$ and $\tau_2$ as the Dirac matrices in two dimensions, the matrix 
$\im\tau_2$ has the interpretation of being the charge conjugation matrix.)

Furthermore, there are two $U(1)$ bundles over $\RP^2$, determined by the sign of the transition function on the orientation-reversing loop. On $\RP^2$, $A^1=A^2=0$ and the gauge field $A^3$ corresponds to the nontrivial $U(1)$ bundle, since under the antipodal identification $(\theta,\varphi)\longrightarrow (\pi-\theta,\varphi+\pi)$ on the double cover $S^2$, we have
\begin{equation}
    A^3(\pi-\theta,\phi+\pi)=-A^3(\theta,\phi) \, .
\end{equation}
Equivalently, it can be said that, on $\RP^2$, $U(1)$ gauge fields are classified up to the sign of the holonomy around the orientation-reversing loop (see, for example, \cite{Metlitski}).

The holonomy of a charged spinor $|J,M\rangle$ around orientation-reversing loop $\gamma'$ in $\RP^2$ is 
\bea
    H&=&\mathcal{P} \text{exp} \left(-\im\oint_{\gamma'} A^3 J^3 \right) |J,M\rangle\\
    &=&\exp\left(-\im \left(\int _0^\pi 2 d\phi\right)J^3\right)|J,M\rangle\\
    &=&e^{-2\pi \im J^3}|J,M\rangle\, . 
\eea
This gives the same conclusion as obtained on $S^2$ in eqn (\ref{quants2}).

Note that the field strength $F^3=\sin \theta d\theta \wedge d \phi $ is also odd under the antipodal map. The flux quantisation condition on $\RP^2$ can be written as half the integral over the double cover \cite{Tachikawa2019}.
\begin{equation}
\frac{1}{2}\int_{S^2} F^3 = \frac{2\pi n} {M} \, , \quad \quad F^3 \equiv \sin \theta d \theta \wedge  d \phi \,,
\end{equation}
where $n$ is any integer and $M $ is the $J^3$-eigenvalue. This precisely matches eqn (\ref{s2quant2}).

\subsection{Gauge-covariantly constant spinor on the Wu manifold}

Similarly to the case of $\CP^2$, we may seek gauge-covariantly constant spinors on the Wu manifold. Their existence proves that the Wu manifold admits a spin$^h$ structure. Gauge-covariantly constant spinors $\eta$ obey
\bea
D_a\,\eta\equiv \nabla_a\,\eta -\im A_a^i\, T^i\, \eta=0\,,
\label{gcc}
\eea
where $\eta$ is a spinor in some representation of the $SU(2)$ gauge
group, and $T^i$ are the (Hermitian) $SU(2)$ generators in this representation, 
obeying
\bea
[T^i,T^j] = \im \ep_{ijk}\, T^k\,.
\eea
As we shall see, we should choose the 4-dimensional
representation of $SU(2)$.

   The integrability condition $[D_a,D_b]\,\eta=0$ implies
\bea
\fft14 R_{abcd}\, \Gamma^{cd}\, \eta -\im F^i_{ab}\, T^i\, \eta=0\,,
\label{spinint}
\eea
where $F^i=dA^i + \ft12\ep_{ijk}\, A^j\wedge A^k$. 
Multiplying eqn (\ref{spinint}) on the left by $\Gamma^{ab}$ gives
\bea
-\fft12 R\,\eta - \im  F^i_{ab}\, \Gamma^{ab}\, T^i\, \eta=0\,,
\eea
with $R=30$ being the Ricci scalar, and hence
\bea
F^i_{ab}\, \Gamma^{ab}\, T^i\, \eta= 15\im \, \eta\,. \label{Fev1}
\eea
On the other hand, using eqn (\ref{RiemFF}), eqn (\ref{spinint}) implies
\bea
\ft14 F^j_{ab}\, F^j_{cd}\, \Gamma^{cd}\,\eta -\im  F^j_{ab}\,T^j\eta=0\,.
\eea
Multiplying by $F^i_{ab}$ and using eqn (\ref{FFid}) gives
\bea
\ft14 F^i_{ab}\,\Gamma^{ab}\, \eta -\im\, T^i\,\eta=0\,.\label{Fev11}
\eea
Multiplying this by $T^i$ then gives
\bea
F^i_{ab}\, \Gamma^{ab}\, T^i\, \eta = 4\im \, C_2\, \eta\,,\label{Fev2}
\eea
where $C_2=T^i T^i$ is the quadratic Casimir of the $SU(2)$ representation.
Comparing eqns (\ref{Fev1}) and (\ref{Fev2}) implies that a necessary 
requirement for the integrability condition $[D_a,D_b]\,\eta=0$ to be
satisfied is that the quadratic Casimir $C_2$ must be be given by
\bea
\label{Casimir}
C_2 = \fft{15}{4}\,.
\eea
This implies that we should take the spinor $\eta$ to be in the
4-dimensional representation of $SU(2)$, which is consistent with the discussion in section \ref{sec:sphereholonomy}.

  It may be verified that there is just one eigenspinor $\eta$ of eqn 
(\ref{Fev1}).  It may furthermore be verified that this spinor satisfies
the full second-order integrability condition (\ref{spinint}):
\bea
\ft14 R_{abcd}\, \Gamma^{cd}\, \eta -\im F^i_{ab}\, T^i\, \eta&=&
 \ft14 F^i_{ab}\, F^i_{cd}\, \Gamma^{cd}\, \eta
  -F^i_{ab}\, T^i\,\eta\,,\nn\\
  &=& F^i_{ab}\, \Big( \ft14  F^i_{cd}\, \Gamma^{cd}\, \eta -T^i\,\eta\Big)
\nn\\
&=& 0\,,
\eea
where the last line follows from eqn (\ref{Fev11})\,.

It can furthermore be verified that if the components of $\eta$ 
are taken to be constants, independent of the coordinates on the Wu 
manifold, then $\eta$ 
satisfies eqn (\ref{gcc}), and thus it is gauge-covariantly
constant:
\bea
D\,\eta &=& d\eta + \ft14\omega_{ab}\, \Gamma^{ab}\,\eta -
         \im A^i\, T^i\,\eta\nn\\
&=& d\eta +\ft14 F^i_{ab} \,A^i\,\Gamma^{ab}\,\eta -
      \im A^i\, T^i\,\eta\nn\\
&=& d\eta +A^i\,\Big( \ft14  F^i_{ab}\, \Gamma^{ab}\, \eta -\im T^i\,\eta\Big)
\nn\\
&=& d\eta =0\,.
\eea
Note that we used eqn (\ref{omFA}) in obtaining the second line, and then
eqn (\ref{Fev11}) in obtaining the last line.  Finally, we used that
$d\eta=0$ if $\eta$ is assumed to be a constant spinor. 
We can write down the $SU(2)$ gauge-covariantly constant spinor $\eta$ explicitly 
using the gamma matrices in eqns \eq{d5Gamma0} and $T^3 = \text{diag}(\frac{3}{2},\frac{1}{2},-\frac{1}{2},-\frac{3}{2})$. The spinor $\eta$ has now two indices $\eta^{\bar m} {}_{\alpha}$, where $\alpha$ is the spinor index and ${\bar m}$ are matrix indices of each generator of $SU(2)$, written as $(T^i)^{\bar m} {}_{ \bar n}$. Then, we have
\bea
\eta^{\bar 1} {}_\alpha =  \begin{pmatrix}-1\cr0\cr 0\cr0\end{pmatrix}_{\!\!\!\!\alpha} \,,\quad
\eta^{\bar 2} {}_\alpha =  \begin{pmatrix}0\cr0\cr 0\cr1\end{pmatrix}_{\!\!\!\!\alpha} \,,\quad
\eta^{\bar 3} {}_\alpha =  \begin{pmatrix}0\cr0\cr -1\cr0\end{pmatrix}_{\!\!\!\!\alpha} \,,\quad
\eta^{\bar 4} {}_\alpha =  \begin{pmatrix}0\cr1\cr 0\cr0\end{pmatrix}_{\!\!\!\!\alpha} \,.\quad
\eea
Note that this is the same result as eqn \ref{GCovConst} up to an overall normalising constant, with each nonzero entry corresponding to each of the four Clebsch-Gordan coefficients. This spinor can also be written using 2-component spinors as in eqn \eq{psires} (see appendix \ref{2compsec}).

\section{\label{eigenfunctionssec}Scalar and Dirac spinor spectrum of the Wu manifold}

Here, we first show how a gauge-covariantly constant spinor $\eta$ can be used in order to construct modes of the gauged Dirac
operator, by starting from scalar eigenfunctions $\phi$ of the scalar wave equation.  Then, we show how the scalar eigenfunctions on a group manifold or on a symmetric coset space may be constructed explicitly.  Our principal interest is in the case of the
Wu coset manifold $SU(3)/SO(3)_{\rm max}$ (there is a discussion of scalar eigenfunctions on $SU(n)/SO(n)$ in, for example, \cite{gudmundsson}), since, as we have seen, this does indeed admit a gauge-covariantly constant spinor.

\subsection{Spectrum of the (gauged) Dirac operator}
\label{SpecDiracSubsec}

Let $\psi$ be a spinor composed as
\be
\psi=\phi \, \eta + \ii \alpha \nabla_a \phi \, \Gamma^a \eta \, ,
\ee
where $\eta$ is a gauge-covariantly constant spinor, i.e. $D_a \eta = 0$,
and $\phi$ is an eigenfunction of the scalar Laplacian with eigenvalue $-\lambda$, i.e. $\Box \phi= -\lambda \,\phi$. We have in mind here a case where $A_a$ is a $U(1)$ gauge field.  Now applying the Dirac operator to $\psi$, a value for the
constant $\alpha$ such that $\psi$ becomes an eigenspinor can be determined:
\bea
\nn
\ii \Gamma^a  \nabla_a \psi & = & \ii  \nabla_a \phi \, \Gamma^a \eta + \ii \phi \, \Gamma^a \nabla_a \eta 
-\alpha \Box \phi \, \eta - \alpha \nabla_b \phi \, \Gamma^a \Gamma^b \nabla_a \eta \\ \nn
& = &  \ii  \nabla_a \phi \, \Gamma^a \eta 
- e  \phi \, \Gamma^a A_a \eta + \lambda \alpha \phi\, \eta 
- \ii e \alpha \nabla_b \phi \, \Gamma^a \Gamma^b A_a \eta  \\ \nn
& = &   \ii  \nabla_a \phi \, \Gamma^a \eta 
+ \lambda \alpha \phi\, \eta - e \Gamma^a A_a \lt( \phi\, \eta 
+ \ii \alpha \nabla_b \phi \,\Gamma^b\eta \rt) \\
& = & \lambda \alpha \phi\, \eta + \ii \nabla_a \phi\, \Gamma^a \eta - e \Gamma^a A_a \psi \, ,
\eea
where in the second line we used that $\eta$ is a gauge-covariantly constant spinor. If we move the last term to the left, we get the gauged Dirac operator on 
the left hand side
\be
\ii \Gamma^a \lt( \nabla_a - \ii e A_a \rt) \psi = \lambda \alpha \lt( \phi\, \eta
+ \frac{\ii}{\alpha \lambda} \nabla_a \phi\, \Gamma^a \eta \rt) \, .
\ee
For this to be an eigenspinor, we need
\be
\alpha^2 = \frac{1}{\lambda} \, ,
\ee
which gives us the eigenvalue $1/\alpha$, or $\sqrt{\lambda}$,
\be
\ii \Gamma^a \lt( \nabla_a - \ii e A_a \rt) \psi =\sqrt{\lambda} \, \psi \, .
\ee
Thus, one can generate higher modes of the gauged Dirac operator using the 
gauge-covariantly constant spinor and the scalar eigenfunctions. The 
resulting spinor has the square root of the eigenvalue of the scalar as the eigenvalue. Note that in a case where there is 
covariantly-constant spinor, i.e. a \textit{parallel spinor}, then since $A_a$ is zero, the construction just gives uncharged spinor harmonics, starting from uncharged scalar harmonics.  (Examples of this kind were discussed in \cite{Hawking:1978ghb}.)  More general possibilities work too, where $\eta$ is a gauge-covariantly constant spinor carrying a a charge under a non-abelian group, such as the example of the gauge-covariantly constant spinor in the Wu manifold that we discussed earlier.  In such a case, the $U(1)$ gauge potential $A_a$ is replaced by $A_a^i T^i$.  Note also that one can consider instead starting from scalar harmonics of a gauged scalar Laplacian, in which case one can arrive at spinor harmonics in other representations of the gauge group.

\subsection{Spectrum of the scalar eigenfunctions}
  Much of the discussion below generalises in the obvious way from $SU(3)$ to
any other compact Lie group, and to other cosets that are Riemannian symmetric 
spaces.

\subsubsection{Scalar eigenfunctions on \texorpdfstring{$SU(3)$}{SU(3)}}
\label{ScalarEigSubsec}
  Let $g$ be a general $SU(3)$ group element, obtained by exponentiating
the (anti-Hermitian) algebra generators $T_i$ in some representation $R$ of $SU(3)$, with
the coordinates on $SU(3)$ as parameters.  We then have 
\bea
g^{-1}\, dg = T_i\, \sigma_i\,,\label{LI1forms}
\eea
where in this section we are using $\sigma_i$ to denote the left-invariant 1-forms on $SU(3)$.  We shall use
the bi-invariant metric $ds^2 = \sigma_i^2$ on $SU(3)$.  From eqn 
(\ref{LI1forms}) we have $dg= g\, T_i\,  \sigma_i$, and so 
\bea
(\nabla_i g)\, \sigma_i = g\, T_i\, \sigma_i\,,
\eea
where $\nabla_i$ are the components of the gradient operator in the basis of
the vielbein $\sigma_i$.  (Note that the generators here are anti-Hermitian.)
Thus we have
\bea
\nabla_i\, g = g\, T_i\,.\label{nabg}
\eea

   Acting with another derivative, and using (\ref{nabg}) again, we therefore
have
\bea
\nabla_i\nabla_i\, g = g\, T_i T_i = -C_2\, g\,,
\eea
where $C_2=-T_iT_i$ is the quadratic Casimir in the representation $R$. In 
other words, we have shown that
\bea
-\square \,g= C_2\, g\,,
\eea
where $\square$ is the scalar Laplacian on $SU(3)$.  That is to say, each component of the $SU(3)$ matrix $g$ is a scalar eigenfunction, with eigenvalue given by $C_2$.
  
  It should be noted that we can take $\nabla_i$ to be the covariant 
derivative here.  This can be seen as follows:
Acting with $d$ on eqn (\ref{LI1forms}) gives $-g^{-1}dg \wedge g^{-1} dg
= T_i\, d\sigma_i$; that is, $-T_j T_k\, \sigma_j\wedge \sigma_k = 
T_i\, d\sigma_i$, implying $-\ft12 [T_j,T_k]\, \sigma_j\wedge \sigma_k = 
T_i\, d\sigma_i$.  Now we have
\bea
[T_j,T_k]= f_{ijk}\, T_i\,,\label{algebra}
\eea
where the structure constants are totally antisymmetric, and indices are raised
and lowered using the Kronecker delta.  Thus we have
\bea
d\sigma_i= -\ft12 f_{ijk}\, \sigma_j\wedge \sigma_k\,.
\eea
From this, it can be seen that the torsion-free spin connection, defined by
$d\sigma_i=-\omega_{ij}\wedge \sigma_j$ and $\omega_{ij}=-\omega_{ji}$,
is given by
\bea
\omega_{ij}= -\ft12 f_{ijk}\, \sigma_k\,.
\eea
Its components $\omega_{k\, ij}$, defined by $\omega_{ij}= 
 \omega_{k\, ij}\, \sigma_k$, are therefore given by
\bea
\omega_{k\, ij}= -\ft12 f_{ijk}\,.
\eea
Therefore, the covariant divergence of a vector field $V_i$ is given by
\bea
\nabla_i\, V_i &=& \del_i\, V_i + \omega_{i\, ij}\, V_j\,,\nn\\
&=& \del_i\, V_i -\ft12 f_{iji}\, V_j = \del_i\, V_i\,,
\eea
where $\del_i$ are the vielbein components of the partial derivative
operator; that is, acting on a scalar $\phi$ we have 
$d\phi= (\del_i\phi)\, \sigma_i$.

\subsubsection{Construction of the scalar eigenfunctions on $SU(3)/SO(3)$}
\label{ConsScalEig}
  The bi-invariant metric $ds_8^2$ on $SU(3)$ can be written in terms of
the metric $ds_5^2$ on the $SU(3)/SO(3)$ base as in eqn (\ref{Gbundlemet}):
\bea
d\tilde s_8^2 = ds_5^2 + c^2\, (\Sigma^i - A^i)^2\,,
\eea
where $\Sigma^i$ are the left-invariant 1-forms on the $SO(3)$ denominator
group, and $A^i$ denotes the $SO(3)$ gauge potential, given in eqn 
(\ref{so3YM}).  The constant $c=1$ here, for the bi-invariant metric on $SU(3)$. It can be verified from formulae in appendix \ref{Gbundlesec}
that the scalar Laplacian on $SU(3)$ can be written as
\bea
\tilde \square_8 = D^\mu\, D_\mu + \fft1{c^2}\, (\nabla^{SO(3)})^2 \,,
\label{Lap8to5}
\eea
where 
\bea
D_\mu= \nabla_\mu + A^i_\mu\, \nabla^{SO(3)}_i\,,
\eea
and $\nabla^{SO(3)}_i$ is the covariant derivative on the $SO(3)$ fibre.  
Here $\nabla_\mu$ denotes the covariant derivative on the 5-dimensional 
$SU(3)/SO(3)$ base (the Wu manifold).  Observe from (\ref{Lap8to5}) that
if $f$ is a function on $SU(3)$ that is independent of the coordinates
of the $SO(3)$ denominator, then $\tilde\square_8\, f= \square_5\, f$,
where $\square_5=\nabla^\mu\nabla_\mu$ is the scalar Laplacian on the
Wu manifold.  Thus we have that if $f$ is an eigenfunction of the scalar
Laplacian on $SU(3)$ and {\it if it is a singlet under $SO(3)$}, then it
will also be an eigenfunction of the scalar Laplacian on the Wu manifold,
with the same eigenvalue:
\bea
-\tilde\square_8\, f =\lambda\, f\qquad \hbox{implies} \qquad 
-\square_5\, f = \lambda\, f\,.\label{su3towu}
\eea
The set of all the $SU(3)$ scalar harmonics that are singlets 
under the $SO(3)$ denominator group can be constructed as follows:

  In terms of the (Hermitian) $3\times 3$ 
Gell-Mann matrices $\lambda_i$, which are the
generators of $SU(3)$, the expression (\ref{nabg}) becomes
\bea
\nabla_i\, g = \im g \, \lambda_i \,,\label{nabg2}
\eea
where $g$ is in $SU(3)$.  Denoting the components of the Gell-Mann matrices 
by $\lambda_i^\alpha{}_\beta$, and the components of the $SU(3)$ matrix $g$ by 
$g^\alpha{}_\beta$, eqn (\ref{nabg2}) and its conjugate give
\bea
\nabla_i\, g^\alpha{}_\beta = \im g^\alpha{}_\gamma\, 
\lambda_i^\gamma{}_\beta\,,\qquad
\nabla_i\, \bar g^\alpha{}_\beta = -\im \bar g^\alpha{}_\gamma\,
 \bar\lambda_i^\gamma{}_\beta
\,.\label{nabg3}
\eea

  Our goal is to construct the scalar eigenfunctions on $SU(3)/SO(3)$.   These
will comprise the subset of scalar harmonics on $SU(3)$ that are singlets
under the $SO(3)$ denominator group.  Since we construct the 
$SU(3)$ representative as $g={\cal O}_1\, K$, where ${\cal O}_1$ is the 
$SO(3)$ denominator-group representative and $K$ is the $SU(3)/SO(3)$
coset representative (see eqn (\ref{SU3})), it can be seen that $g^T g$ is independent of 
${\cal O}_1$.  In terms of the $SU(3)$ group indices, this is equivalent to
\bea
(g^T g)_{\alpha\beta} \equiv G_{\alpha\beta} 
 = g^\gamma{}_\alpha\, g^\gamma{}_\beta\,.
\eea
(Note that $(g^T)^\alpha{}_\beta=  g^\beta{}_\alpha$.)  

   We can now write down the general scalar eigenfunctions on $SU(3)/SO(3)$.
They are given by
\bea
f_{p,q} = S^{\alpha^\phan_1\cdots\alpha^\phan_{2p}}{}_{\beta_1\cdots\beta_{2q}}\,
  G_{\alpha^\phan_1\alpha^\phan_2}\cdots 
      G_{\alpha^\phan_{2p-1}\alpha^\phan_{2p}}\,
\bar G^{\beta_1\beta_2}\cdots \bar G^{\beta_{2q-1}\beta_{2q}}\,,\label{fpqdef}
\eea
where $S^{\alpha^\phan_1\cdots\alpha^\phan_{2p}}{}_{\beta_1\cdots\beta_{2q}}$
is an arbitrary constant tensor subject to the symmetry and tracelessness
conditions
\bea
S^{\alpha^\phan_1\cdots\alpha^\phan_{2p}}{}_{\beta_1\cdots\beta_{2q}} &=&
S^{(\alpha^\phan_1\cdots\alpha^\phan_{2p})}{}_{\beta_1\cdots\beta_{2q}}=
S^{\alpha^\phan_1\cdots\alpha^\phan_{2p}}{}_{(\beta_1\cdots\beta_{2q})}
\,,\label{Ssym}\\
S^{\alpha^\phan_1\cdots\alpha^\phan_{2p}}{}_{\alpha^\phan_1\beta_2\cdots
                                    \beta_{2q}}&=&0\,,\label{Straceless}
\eea
and $\bar G^{\alpha\beta}= \bar g^\gamma{}_\alpha\, \bar g^\gamma{}_\beta$.
Note that because of the symmetry properties in eqns (\ref{Ssym}), 
the condition (\ref{Straceless}) implies tracelessness for any contraction 
of an upper index with a lower index. 
These scalar eigenfunctions of the scalar Laplacian on $SU(3)/SO(3)$ obey
$-\square f_{p,q} = \lambda_{p,q} f_{p,q}$, with the eigenvalues
\bea
\lambda_{p,q}= \fft83\, \Big( 2p^2 + 2 q^2 + 3p + 3q + 2 p\, q\Big)\,.
\label{lampq1}
\eea
The eigenvalues can be calculated by using the Fierz-like identities \eq{su3ids} that the Gell-Mann matrices satisfy (see appendix \ref{CalEigDeg} for details). 
The degeneracies follow from calculating the number of independent
components in the constant tensor 
$S^{\alpha^\phan_1\cdots\alpha^\phan_{2p}}{}_{\beta_1\cdots\beta_{2q}}$,
subject to the conditions in eqns (\ref{Ssym}) and (\ref{Straceless}).  These
can easily be seen to imply the degeneracies
\bea
d_{p,q}&=& \ft14 (2p+1)(2p+2)(2q+1)(2q+2)-\ft14(2p)(2p+1)(2q)(2q+1) \,,\nn\\
&=& (2p+1)(2q+1)(p+q+1)\,.
\eea

Consider the subset of these scalar eigenfunctions \eq{fpqdef} that has been constructed using 
\bea
S^{\alpha^\phan_1\cdots\alpha^\phan_{2p}}{}_{\beta_1\cdots\beta_{2q}}= a^{\alpha^\phan_1} \cdots a^{\alpha^\phan_{2p}} \, b_{\beta_1} \cdots b_{\beta_{2q}} \, ,
\eea
where $a$ and $b$ are arbitrary constant vectors in $\C^3$ and are orthogonal to each other, i.e. $a^i b_i=0$, such that the tracelessness conditions \eq{Straceless} are still satisfied. These scalar eigenfunctions satisfy the so called {\it conformality condition},
\bea
(\nabla^i f)\, (\nabla_i f) = -\kappa\, f^2\,,\label{conf11}
\eea
where $\kappa$ is a constant. Any power of an eigenfunction $f$ of the scalar Laplacian that also satisfies eqn \eq{conf11} will be an eigenfunction of the scalar Laplacian (see appendix \ref{ConfScal} for details)
\bea
-\square h = \hat{\lambda} \, h \, ,
\eea
where $h=f^n$. If one starts with the eigenfunction $f=f_{1,0}$, then the eigenvalues are
\bea
\hat{\lambda}= \frac{8n (2n+3)}{3} \, ,
\eea
which coincide with eqn \eq{lampq1} with $p=n$ and $q=0$.

\section{\label{noncompactdualssec}Non-compact duals of Wu and $\CP^2$ manifolds}

\subsection{Non-compact dual of the Wu manifold}

  The simplest and most useful way to construct the metric on the non-compact
dual of the Wu manifold, \ie on the coset $SL(3,\R)/SO(3)$, is as follows.
The essential idea is that we just send
\bea
\mu\longrightarrow \im \mu\,,\qquad \nu\longrightarrow \im \nu\label{cnc}
\eea
in the Wu metric (\ref{WuMetric}), at the same time sending $ds_5^2\longrightarrow 
d\tilde s_5^2= -ds_5^2$.
Thus from the Wu metric (\ref{WuMetric}) we now get the non-compact
dual metric
\bea
d\tilde s_5^2 = \mu^2 + \ft13 d\nu^2 + \sinh^2(\mu-\nu)\, \sigma_1^2 +
  \sinh^2(\mu+\nu)\, \sigma_2^2 + \sinh^2 2\mu\, \sigma_3^2\,.\label{ncwumet}
\eea
Taking the vielbein now to be
\bea
\td e^1= \sinh(\mu-\nu)\, \sigma_1\,,\quad
\td e^2= \sinh(\mu+\nu)\, \sigma_2\,,\quad
\td e^3=\sinh2\mu\, \sigma_3\,,\quad
\td e^4=d\mu\,,\quad \td e^5=\fft{d\nu}{\sqrt3}\,,\label{ncwuviel}
\eea
all the spin connection 1-forms $\td\omega_{ab}$ will be exactly the same as
in eqns (\ref{wuspincon}), except that $\cos$ is replaced by $\cosh$
everywhere. For the curvature 2-forms $\wtd\Theta_{ab}$, the components
will all be exactly the negatives of those given for the Wu metric in
eqns (\ref{spinconwu}).  In particular, it follows from
this that the Ricci tensor
is given by
\bea
\wtd R_{ab}=-6\delta_{ab}\,,
\eea
rather than $R_{ab}=6\delta_{ab}$ for the standard compact Wu metric.

 Note that the coordinates $\mu$ and $\nu$ will now range over the 
 intervals
\bea
\mu\ge0 \,,\qquad  -\mu\le \nu \le \mu\,.
\eea
It can still be seen that the discussion given in appendix \ref{StrucSubSec} for the
compact Wu manifold will still apply when studying the behaviour of the
non-compact metric near the point $\mu=\nu=0$, showing that, as in eqn
(\ref{wumetrhoal}) the left-invariant 1-forms $\sigma_i$
must be subject to the identification under $SO(3)/D_4$.  However, unlike the Wu manifold, $SL(3,\R)/SO(3)$ is contractible and therefore possess a unique spin structure \cite{contractible}. 

  The non-compact Wu metric in eqn (\ref{ncwumet}) can also be seen to
come from a coset construction that generalises in a very natural way
the one we gave in eqns (\ref{SU3}) to (\ref{su3met}) for the compact Wu manifold.  
Thus we now
make use of the fact that a general element $S$ of $SL(3,\R)$ can be
parameterised as
\bea
S  = {\cal O}_1\, D\, {\cal O}_2^T\,,
\eea
where ${\cal O}_1$ and ${\cal O}_2$ are $SO(3)$ matrices given, as previously,
by eqns (\ref{O1O2def}), and now we have introduced the real diagonal matrix
\bea
D= \hbox{diag}\, \Big( e^{-\mu-\ft13\nu}, e^{\mu-\ft13\nu}, e^{\ft23\nu}\Big)
\,.\label{Dmatrix}
\eea
Note that $D$ is precisely the same as the $B$ matrix defined in
eqn (\ref{Bmatrixdef}) that was used in the $SU(3)$ parameterisation,
after sending
$\mu\longrightarrow \im \mu$ and $\nu\longrightarrow \im\nu$.  We will now
have the eight-dimensional metric $d\td s_8^2$ on the $SL(3,\R)$ group
manifold, given by
\bea
d\td s_8^2 =-\fft12 \hbox{tr}\,
(dS^{-1}\, dS) = d\td s_5^2 -(\Sigma^i - \wtd A^i)^2\,,
\eea
where $d\td s_5^2$ is the non-compact Wu metric (\ref{ncwumet}), and
the $SO(3)$ Yang-Mills potentials are given by
\bea
\wtd A^1=\cosh(\mu-\nu)\, \sigma_1\,,\qquad
\wtd A^2= \cosh(\mu+\nu)\, \sigma_2\,,\qquad
\wtd A^3= \cosh 2\mu\, \sigma_3\,.
\eea
The corresponding field strengths $\wtd F^i=d\wtd A^i + 
\ft12 \epsilon_{ijk}\,\wtd A^j\wedge \wtd A^k$ are just given by the
negatives of the expressions $F^i$ in eqns (\ref{SO3Fi}, so
\bea
\wtd F^1 &=& -\td e^1\wedge \td e^4+ \td e^2\wedge \td e^3 + \sqrt3\, \td e^1\wedge \td e^5\,,\nn\\
\wtd F^2 &=& -\td e^2\wedge \td e^4 - \td e^1\wedge \td e^3 -\sqrt3\, \td e^2\wedge \td e^5\,,\nn\\
\wtd F^3 &=& -2 \td e^3\wedge \td e^4 - \td e^1\wedge \td e^2\,.\label{SO3Finc}
\eea
The quadratic identities for the $F^i$ fields given in eqns (\ref{FFid}) and
(\ref{FFasym}) hold also for the $\wtd F^i$ fields we are considering here. Note
that the expressions now for the spin connection and curvature in terms of
the non-compact dual will be like those in eqns (\ref{omFA}), (\ref{ThFF}) and (\ref{RiemFF}) for
the compact Wu manifold, but with an overall minus sign in each case:
\bea
\td \omega_{ab}= -\wtd F^i_{ab}\, \wtd A^i\,,\qquad
\wtd \Theta_{ab}= - \wtd F^i_{ab}\, \wtd F^i\,,\qquad
\wtd R_{abcd}= -\wtd F^i_{ab}\, \wtd F^i_{ab}\,.
\eea

\subsection{Non-compact dual of the $\CP^2$ manifold}

In a similar vein, one can easily construct the non-compact dual of the $\CP^2$ manifold.
Starting from the real form of the Fubini-Study metric, given in eqn (\ref{cp2met2}), we
can just make the replacement $\mu\longrightarrow \im \mu$, and then send $ds_4^2\longrightarrow d\td s_4^2
= -ds_4^2$, yielding the non-compact dual metric
\bea
d\td s_4^2 = d\mu^2 + \ft14\sinh^2\, (\sigma_1^2+\sigma_2^2) +\ft14\sinh^2\mu\, \cosh^2\mu\, \sigma_3^2\,,
\label{ncCP2met}
\eea
with $\mu\ge0$ and $\sigma_i$ being the left-invariant 1-forms of $SU(2)$.  The level surfaces $\mu=\,$constant
are squashed 3-spheres, viewed as $U(1)$ bundles over $S^2$, with the squashing (actually a stretching) of the $U(1)$ fibres becoming more pronounced as $\mu$ increases.  As $\mu$ goes to zero the $S^3$ level surfaces approach round 3-spheres,
and the metric extends smoothly onto $\mu=0$ just like the origin of Euclidean 4-space described using hyperspherical polar coordinates:
\bea
\mu <<1 :&& d\td s_4^2 \sim d\mu^2 + \ft14 \mu^2\, (\sigma_1^2+\sigma_2^2+\sigma_3^2)= d\mu^2 +
  \mu^2\, d\Omega_3^2\,,\nn\\
  \mu>>1: && d\td s_4^2 \sim d\mu^2 + \ft1{16} e^{2\mu}\, \Big(\sigma_1^2+\sigma_2^2 + 
             \ft14 e^{2\mu}\, \sigma_3^2\Big)\,.
\eea
It is evident, therefore, that the non-compact $\CP^2$ manifold is just $\R^4$.  Being topologically trivial, 
there is therefore no obstruction to the existence of an ordinary spin structure on the non-compact
$\CP^2$ manifold.

We shall use the natural vielbein frame
\bea
\td e^1=\ft12\sinh\mu\, \sigma_1\,,\qquad \td e^2=\ft12\sinh\mu\, \sigma_2\,,\qquad
\td e^3=\ft12\sinh\mu\, \cosh\mu\, \sigma_3\,,\qquad \td e^4=d\mu\,,
\eea
The $U(1)$ and $SU(2)$ gauge potentials $\wtd A^0$ and $\wtd A^i$ are given by
\bea
\wtd A^0= \ft12 \sinh^2\mu\, \sigma_3\,,\quad
\wtd A^1=\cosh\mu\, \sigma_1\,,\quad  \wtd A^2=\cosh\mu\, \sigma_2\,,\quad
\wtd A^3 = \ft12 (1+\cosh^2\mu)\, \sigma_3\,.
\eea
The field strengths $\wtd F^0=d\wtd A^0$ and 
$\wtd F^i=d\wtd A^i+\ft12 \ep_{ijk}\, \wtd A^j\wedge \wtd A^k$ are given by
\bea
\wtd F^0 &=& -2(\td e^1\wedge \td e^2 + \td e^3\wedge \td e^4)\,,\label{cp2ncF0}\\
&&\nn\\
\wtd F^1 &=& -2(\td e^1\wedge \td e^4 - \td e^2\wedge \td e^3)\,,\quad
\wtd F^2=-2(\td e^2\wedge \td e^4 - \td e^3\wedge \td e^1)\,,\nn\\
\wtd F^3 &=& -2(\td e^3\wedge \td e^4 - \td e^1\wedge \td e^2)\,.\label{cp2ncFi}
\eea
These are all the negatives of the analogous field strengths $F^0$ and $F^i$
in the compact $\CP^2$ metric.  They obey the same quadratic identities 
(\ref{FFidCP2}) as in the compact $\CP^2$.  

The spin connection, the curvature 2-forms and the Riemann tensor can be written as
\bea
\td\omega_{ab} &=& -\fft14 \wtd F^i_{ab}\, \wtd A^i - \fft34 \wtd F^0_{ab}\, \wtd A^0\,,
\label{omabCP2nc}\\
\wtd \Theta_{ab}&=& -\fft14 \wtd F^i_{ab}\, \wtd F^i -\fft34 \wtd F^0_{ab}\, \wtd F^0\,,
\label{ThabCP2nc}\\
\wtd R_{abcd}&=&-\frac{1}{4}\wtd F^i_{ab}\wtd F^i_{cd}-\frac{3}{4}\wtd F^0_{ab}\wtd F^0_{cd}\,.   
\label{RiemCP2nc}
\eea

Note that the $SU(2,1)/U(2)$ coset structure of the non-compact $\CP^2$ can be seen by
making a small modification of the usual $\CP^2$ construction, as follows (see, for example,
section 3.1 of \cite{popscu}).  Introducing complex coordinates $Z^A$ on $\C^3$ and defining 
$\eta_{AB}=\hbox{diag}\,(-1,1,1)$, with $A,B,\cdots$ ranging over 0, 1 and 2, we consider the 
flat metric $d\hat s^2= \eta_{AB} dZ^A d\bar Z^B$, subject
to the restriction $\eta_{AB} Z^A\,\bar Z^B=-1$.  This defines the ``unit'' metric on AdS$_5$.
Both the metric and the constraint are invariant under $SU(2,1)$. 
Introducing inhomogeneous complex coordinates $\zeta^\alpha=Z^\alpha$, where $\alpha=1,2$,
the AdS$_5$ metric can be rewritten as
\bea
d\hat s^2 = -\Big|\fft{dZ^0}{Z^0} -|Z^0|^2\, \bar d\zeta^\alpha\, \zeta^\alpha\Big|^2 +
d\td s_4^2\,,\label{AdS5met}
\eea
where 
\bea
d\td s_4^2 = |Z^0|^2 \, d\zeta^\alpha\, d\bar\zeta^\alpha + 
|Z^0|^4\, |\bar\zeta^\alpha\,d\zeta^\alpha|^2\,.
\eea
From the constraint $\eta_{AB} Z^A\,\bar Z^B=-1$ and the definition $\eta_{AB} Z^A\,\bar Z^B=-1$ it
follows that $|Z^0|^2 = (1-|\zeta|^2)^{-1}$ where $|\zeta|^2 = \zeta^\alpha\,\bar\zeta^\alpha$, 
and so if we write $Z^0=|Z^0|\, e^{\ft{\im}{2}\,\tau}$ and project the AdS$_5$ metric (\ref{AdS5met})
orthogonally to the $\dfft{\del}{\del\tau}$ fibres, we see that 
\bea
d\td s_4^2 = \fft{\bar d\zeta^\alpha\, d\zeta^\alpha}{1-|\zeta|^2} +
    \fft{|\bar\zeta^\alpha\, d\zeta^\alpha|^2}{(1-|\zeta|^2)^2}
\eea
must be the metric on the non-compact $\CP^2$ base.  Rewriting using real coordinates adapted to
the $U(2)$ symmetry of the denominator group,
\bea
\zeta^1= \tanh\mu\, \cos\ft12\theta\, e^{\ft{\im}{2} (\psi+\phi)}\,,\qquad
\zeta^2= \tanh\mu\, \sin\ft12\theta\, e^{\ft{\im}{2} (\psi-\phi)}\,,
\eea
we recover the non-compact $\CP^2$ metric in the form (\ref{ncCP2met}) that we obtained by 
analytic continuation.

\subsection{Non-compact Wu and $\CP^2$ and toroidal supergravity reductions}

  Finally, it is interesting to note that both the non-compact Wu manifold $SL(3,\R)/SO(3)$ and the non-compact $\CP^2$
  manifold $SU(2,1)/U(2)$ arise also in the toroidal Kaluza-Klein reductions of certain supergravity theories.  
  
  If one reduces eleven-dimensional supergravity on a 3-torus then the resulting eight-dimensional theory has an
  $SL(3,\R)\times SL(2,\R)$ global symmetry.  In the scalar sector, the Lagrangian in eight dimensions has the form
  of a non-linear sigma model, with the scalar fields defined on the product manifold $\dfft{SL(3,\R)}{SO(3)} \times \dfft{SL(2,\R)}{SO(2)}$.
  The $\dfft{SL(3,\R)}{SO(3)}$ factor is precisely the non-compact Wu manifold. One way in which this can be seen in the supergravity reduction involves constructing the $SL(3,\R)/SO(3)$ scalar coset by exponentiating the positive-root generators of $SL(3,\R)$ with axionic scalars as coefficients, and exponentiating the Cartan subalgebra generators with dilatonic scalars as coefficients; see \cite{cjlp1} for a detailed discussion.  The outcome is that the metric on the non-compact Wu manifold can be written in the form
  \bea
  ds_5^2 = d\phi_1^2 + d\phi_2^2 + e^{2\phi_1-2\sqrt3\, \phi_2}\, d x^2 +
e^{4\phi_1}\, (dy -2x \,d z)^2
 +e^{2\phi_1+2\sqrt3\, \phi_2}\, d z^2\,.
\label{sl3so3met}
\eea
(The coordinates $\phi_1$ and $\phi_2$ correspond to the dilatonic scalars, and the coordinates $x$, $y$ and $z$ correspond to the axionic scalars.)  A rather complicated coordinate transformation maps the metric (\ref{sl3so3met}) into the metric (\ref{ncwumet}).

In the case of the non-compact $\CP^2$ manifold $SU(2,1)/U(2)$, this arises as the scalar manifold when the four-dimensional Einstein-Maxwell system is dimensionally reduced to $D=3$ on a circle. In this case two vector gauge fields must be dualised to axionic scalars in $D=3$ in order for the $SU(2,1)$ global symmetry to become locally realised in the scalar field non-linear sigma model.  The procedures have been described in \cite{brgima} and \cite{cjlp3}.  The outcome is a metric on the non-compact $SU(2,1)/U(2)$ manifold that takes the form
\bea
ds_4^2 = d\phi^2 + e^{2\phi}\,(dx^2 + dy^2) + 
   e^{4\phi}\,(dz - 2 x\, dy)^2\,,\label{ncCP2em}
\eea
where $\phi$ corresponds to the dilatonic scalar, and $x$, $y$ and $z$ correspond to the axionic scalars.  The metric (\ref{ncCP2em}) is related to the metric (\ref{ncCP2met}) by a somewhat involved coordinate transformation.  Further details showing how the transformation arises can be found in \cite{popscu}. 

\section{\label{conclusionssec}Conclusions}

In this paper, we have given a detailed discussion of some of the key features of the five-dimensional compact coset space $SU(3)/SO(3)_{\rm max}$,
which is known as the Wu manifold.  This is a very interesting manifold in that it is the simplest of the relatively rare explicit examples of
manifolds that admit neither an ordinary spin structure nor a spin$^c$ structure, but which do admit a spin$^h$ structure.  Our interest ultimately
is in employing the Wu manifold as the internal space in some dimensional reduction of a supergravity or related theory, and this will be
addressed in a companion paper that is currently a work in progress \cite{gigulapo2}.  Our discussion has accordingly been rather more informal
than some of the discussions of the Wu manifold in the mathematical literature, and our approach to the discussion of its spin$^h$ structure has
rather closely paralleled the discussion of the spin$^c$ structure of the $\CP^2$ manifold in \cite{Hawking:1977ab}.  We also revisited the discussion of the $\CP^2$ manifold, expressing the holonomy calculations in a language closely analogous to that which we have employed when investigating the Wu manifold.  In particular, for $\CP^2$ we explicitly demonstrated that it can be endowed with a spin$^h$ structure instead of a spin$^c$ structure.  That is, one can choose to repair the $-1$ holonomy of uncharged spinors in $\CP^2$, which is an obstruction to defining ordinary spinors, by coupling them in half-integer isospin representations of the $SU(2)$ Yang-Mills connection rather than the usual $U(1)$ connection that is employed when defining the spin$^c$ structure in $\CP^2$.  (This spin$^h$ structure for $\CP^2$ was described in \cite{Hawking:1977ab}, although the terminology came only later.)  In both $\CP^2$ and in the Wu manifold one can construct gauge-covariantly constant spinors, and these can then be used in order to generate the complete spectrum of generalised spinor eigenfunctions, in terms of the complete spectrum for scalar eigenfunctions.  We also gave a very explicit construction of the scalar eigenfunction spectrum in the Wu manifold.  Many of these results will be employed in a follow-up work \cite{gigulapo2}, in which we investigated various supergravity and string-theory compactifications involving the Wu manifold.

\section*{Acknowledgments}

This work is supported in part by DOE grant DE-SC0010813.
GL was supported by endowment funds from the Mitchell Family Foundation during the initial stages of this project. 

\newpage

\appendix

\section{Simple compact group \texorpdfstring{$G$}{G} bundle over base space}\label{Gbundlesec}

   Let $\Sigma^i$ be the left-invariant 1-forms on the group $G$.  They obey
\bea
d\Sigma^i = -\ft12 f^i{}_{jk}\, \Sigma^j\wedge \Sigma^k\,,
\eea
where $f^i{}_{jk}$ are the structure constants of the Lie algebra of the
group $G$.  If the base space $M$ has the metric $ds^2$, then we can write 
a family of metrics $d\tilde s^2$ on the space $\hat M$ 
of the $G$ bundle over $M$, 
with
\bea
d\tilde s^2 = ds^2 + c^2\, (\Sigma^i-A^i)^2\,,\label{Gbundlemet}
\eea
where $c$ is a constant squashing parameter and $A^i$ is a Yang-Mills
connection on the $G$ fibre.  The curvature of the connection, that is, the 
Yang-Mills field strength, is given by
\bea
F^i = dA^i +\ft12 f^i{}_{jk}\, A^j\wedge A^k\,.
\eea

   Choosing the natural orthonormal basis $\tilde e^A$ for the metric,
where $A=(a,i)$, we shall have
\bea
\tilde e^a=e^a\,,\qquad \tilde e^i= c\,(\Sigma^i-A^i)\,,
\label{Gviel}
\eea
where $e^a$ is an orthonormal basis for the base metric $ds^2$.
   It is useful to record here the expressions for the inverse vielbein
$\wtd E_A$ on the total space.  We shall have
\bea
\wtd E_a = E_a + A^i_a\, E_i\,,\qquad
\wtd E_i = \fft1{c}\, E_i\,,\label{Ginvviel}
\eea
where $E_a$ is the inverse vielbein on the base space, obeying
$\langle E_a, e^b\rangle = \delta_a^b$, and $E_i$
is the inverse vielbein on the $G$ fibres, obeying
$\langle E_i, \Sigma^j\rangle =\delta_i^j$.  It can easily be verified
that $\langle \wtd E_A,\td e^B\rangle =\delta_A^B$. Note that since the
tangent-frame indices refer to an orthonormal frame, we can freely raise and
lower tangent-frame indices using Kronecker deltas.  Thus it is immaterial
whether such indices are written upstairs or downstairs.

After straightforward algebra, one finds that the torsion-free spin
connection is given by
\bea
\tilde \omega_{ab} &=& \omega_{ab} + \ft12 c\, 
            F^i_{ab}\, \td e^i\,,\nn\\
\tilde\omega_{ij}&=& -\fft1{2c}\, f^i{}_{jk}\,\td e^k -
     f^i{}_{jk} \, A^k\,,\nn\\
\td\omega_{a i}&=& \ft12 c\, F^i_{ab}\, \td e^b\,,
\label{Gbundleom}
\eea
where $\omega_{ab}$ is the spin connection in the base space.
The orthonormal components of the 
Riemann tensor $\wtd R_{ABCD}$ and Ricci tensor $\wtd R_{AB}$ 
are then given by (see, for example, \cite{pagpop})
\bea
\wtd R_{abcd} &=& R_{abcd}  -
  \ft14 c^2\, (F^i_{ac} \,F^i_{bd} -
  F^i_{ad}\, F^i_{bc} +
  2 F^i_{ab}\, F^i_{cd})\,,\nn\\
\wtd R_{abc i} &=& 
     \ft12 c\, \cD_c\, F^i_{ab}\,,\nn\\
\wtd R_{a i b j} &=& \ft14 c^2\, F^i_{bc}\, F^j_{ac}
 -\ft14 f_{ijk}\, F^k_{ab}\,,\nn\\
\wtd R_{ijk\ell} &=& \fft1{4c^2}\, f_{ijm}\, f_{k\ell m}\,
\eea
and 
\bea
\wtd R_{ab} &=& R_{ab} - \ft12 c^2\, F^i_{ac}\,
F^i_{bc}\,,\nn\\
\wtd R_{ij} &=& \ft14c^2\, F^i_{ab}\, F^j_{ab} + 
   \fft1{2c^2}\,f^i{}_{k\ell}\, f^j{}_{k\ell}\,,\nn\\
\wtd R_{a i} &=& \ft12 c\, {\cal D}_c\, F^i_{c a}\,,
\label{GRicci}
\eea
where ${\cal D}_c$ is the Yang-Mills covariant derivative.  In all
the cases of interest to us, the field strength $F^i_{ab}$ obeys the
Yang-Mills equation, and so $\wtd R_{a i}$ will vanish.

   Since we are principally interested in the case where the fibre group 
$G$ is $SU(2)$ or $SO(3)$, for which the structure constants $f^i{}_{jk}$
will be simply $\ep_{ijk}$, we give the Ricci curvature explicitly for
this case.  The expressions (\ref{GRicci}) become
\bea
\wtd R_{ab} &=& R_{ab} - \ft12 c^2\, F^i_{ac}\,
F^i_{bc}\,,\nn\\
\wtd R_{ij} &=& \ft14 c^2\,F^i_{ab}\, F^j_{ab} +
   \fft1{c^2}\,\delta_{ij}\,,\nn\\
\wtd R_{a i} &=& \ft12 c\, {\cal D}_c\, F^i_{c a}\,.
\label{SU2Ricci}
\eea

\section{\label{paramsec}Parameterising group elements}

\subsection{Parameterising $SU(3)$}
\label{ParametSec}

We are writing a group element of $SU(3)$ as
\begin{equation}
    \mathcal{O}_1 B \mathcal{O}_2^T \label{su3elem}
\end{equation}
where $\mathcal{O}_i$ are two $SO(3)$ matrices and $B$ is a member of the Cartan subalgebra $H=U(1)\times U(1)$ of $SU(3)$. The diagonal matrix $B$ is written as
\begin{equation}
    B=\text{diag}\left( e^{i\left(\mu+\frac{1}{3}\nu\right)},e^{i\left(-\mu+\frac{1}{3}\nu\right)},e^{-\frac{2i}{3}\nu}    \right)\,.
\end{equation}
Due to the form of $B$, we have the two independent identifications 
\bea 
(\mu,\nu)\sim(\mu+2\pi,\nu) \qquad \hbox{and}\qquad (\mu,\nu)\sim(\mu+\pi,\nu+3\pi)\,.
\eea
From these it also follows that $(\mu,\nu) \sim(\mu,\nu+6\pi)$. 

There are three types of redundancy in the parameterisation (\ref{su3elem}). The first type of redundancy is 
\begin{equation}
    (\mathcal{O}_1)(B)(\mathcal{O}_2^T)=(\mathcal{O}_1 \gamma^{-1})(\gamma B ) (\mathcal{O}_2^T) \enspace \text{for} \enspace \gamma \in SO(3) \cap H\,.
\end{equation}
Possible choices of $\gamma$ form the dihedral group of order $4$ we denote $D_4$.

The second type of redundancy is
\begin{equation}
(\mathcal{O}_1)(B)(\mathcal{O}_2^T)=(\mathcal{O}_1 \omega^{-1})(\omega B \omega^{-1})(\omega \mathcal{O}_2^T) \enspace \text{for} \enspace  \omega \notin H \enspace \text{such that} \enspace \omega B \omega^{-1} \in H \enspace \forall B \in H \,.
\end{equation}
Possible choices of $\omega$ form a group of order $24$ that is the direct product of $D_4$ and a group of order $6$ we denote $W$.

The third type of redundancy is
\begin{equation}
(\mathcal{O}_1)(B)(\mathcal{O}_2^T)=(\mathcal{O}_1 \gamma')( B )(\gamma'^{-1} \mathcal{O}_2^T) \enspace \text{for} \enspace \gamma'  \enspace \text{such that} \enspace \gamma' B \gamma'^{-1} \in SO(3) \enspace \forall B \in H\,.
\end{equation}
Possible choices of $\gamma'$ also form the dihedral group $D_4$. 

In the following subsections, we discuss each of these redundancies in more detail then describe how to uniquely parameterise an $SU(3)$ element.

\subsubsection{Action of the binary dihedral group on $H$}
\label{firstAct}
For the first redundancy
\begin{equation}
   \mathcal{O}_1 H \mathcal{O}_2^T=(\mathcal{O}_1\gamma^{-1})(\gamma H)(\mathcal{O}_2^T) \enspace \text{for} \enspace \gamma \in SO(3) \cap H\,,
\end{equation}
we need to consider the intersection between $H$ and $SO(3)$, which is dihedral group $D_4$ made up of four matrices
\begin{equation}
    \gamma_1=\text{diag}(1,1,1), \enspace \gamma_2=\text{diag}(1,-1,-1), \enspace \gamma_3=\text{diag}(-1,1,-1), \enspace \gamma_4=\text{diag}(-1,-1,1)\,. \label{dihedral} 
\end{equation}
The action $B\rightarrow \gamma B$ has the following effect on the central coordinates $(\mu,\nu)$:
\begin{equation}
    \begin{split}
        &\gamma_1: (\mu,\nu)\rightarrow (\mu,\nu), \quad \gamma_2:(\mu,\nu) \rightarrow \left(\mu-\frac{1}{2}\pi,\nu+\frac{3}{2}\pi\right),\\
        &\gamma_3: (\mu,\nu) \rightarrow \left(\mu+\frac{1}{2}\pi,\nu+\frac{3}{2}\pi\right), \quad \gamma_4:(\mu,\nu)\rightarrow (\mu+\pi,\nu)\,.
    \end{split} \label{fourtrans}
\end{equation}

\subsubsection{Weyl transformations}
\label{secondAct}

Consider the second redundancy
\begin{equation}
(\mathcal{O}_1)(B)(\mathcal{O}_2^T)=(\mathcal{O}_1 \omega^{-1})(\omega B \omega^{-1})(\omega \mathcal{O}_2^T) \enspace \text{for} \enspace  \omega \notin H \enspace \text{such that} \enspace \omega b \omega^{-1} \in H \enspace \forall b \in H\,.
\end{equation}

Since $H$ is formed of diagonal matrices and conjugation by $\omega$ preserves eigenvalues, conjugation by $\omega$ must just swap diagonal entries of $b$. Therefore, we must take the quotient by matrices with exactly one entry of $\pm 1 $ in each row and column with determinant $+1$, which is a set of $24$ matrices. These $24$ matrices are the direct product of $D_4$ (listed in eqn (\ref{dihedral}) and $W$, where $W$ consists of the six matrices
\begin{alignat}{6} \nn
     \omega_1 &\equiv  +&& \begin{pmatrix}
        1 & 0 & 0\\
        0 & 1 & 0\\
        0 & 0 & 1
    \end{pmatrix},\quad
    \enspace  \omega_2 &&\equiv 
   -  && \begin{pmatrix}
        0 & 1 & 0\\
        1 & 0 & 0\\
        0 & 0 & 1
    \end{pmatrix},\quad
    \enspace \omega_3 &&\equiv 
    - && \begin{pmatrix}
        0 & 0 & 1\\
        0 & 1 & 0\\
        1 & 0 & 0
    \end{pmatrix},\\ 
    \omega_4 &\equiv 
    -&&\begin{pmatrix}
        1 & 0 & 0\\
        0 & 0 & 1\\
        0 & 1 & 0
    \end{pmatrix},\quad \enspace 
    \omega_5 && \equiv 
    +&& \begin{pmatrix}
        0 & 0 & 1\\
        1 & 0 & 0\\
        0 & 1 & 0
    \end{pmatrix},\quad \enspace 
    \omega_6 && \equiv 
   +&& \begin{pmatrix}
        0 & 1 & 0\\
        0 & 0 & 1\\
        1 & 0 & 0
    \end{pmatrix}. \label{weylperms}
\end{alignat}

Conjugation by elements of $D_4$ acts trivially on $(\mu,\nu)$. Conjugation of $B \in H$ by each of $\omega_i$ acts on $(\mu,\nu)$ as
\begin{equation}
    \begin{split}
        &\omega_1:(\mu,\nu)\rightarrow(\mu,\nu)\, , \enspace \omega_2:(\mu,\nu)\rightarrow (-\mu,\nu)\, , \\
        &\omega_3: (\mu,\nu)\rightarrow\left(\frac{1}{2}\mu-\frac{1}{2}\nu,-\frac{3}{2}\mu-\frac{1}{2}\nu\right), \enspace 
        \omega_4: (\mu,\nu)\rightarrow \left(\frac{1}{2}\mu+\frac{1}{2}\nu,\frac{3}{2}\mu-\frac{1}{2}\nu\right),\\
        &\omega_5:(\mu,\nu)\rightarrow \left(-\frac{1}{2}\mu-\frac{1}{2}\nu,\frac{3}{2}\mu-\frac{1}{2}\nu\right), \enspace \omega_6:(\mu,\nu)\rightarrow\left(-\frac{1}{2}\mu+\frac{1}{2}\nu,-\frac{3}{2}\mu-\frac{1}{2}\nu\right).
    \end{split} \label{sixtrans}
\end{equation}

\begin{figure}[t]
    \centering
    \includegraphics[width=\textwidth]{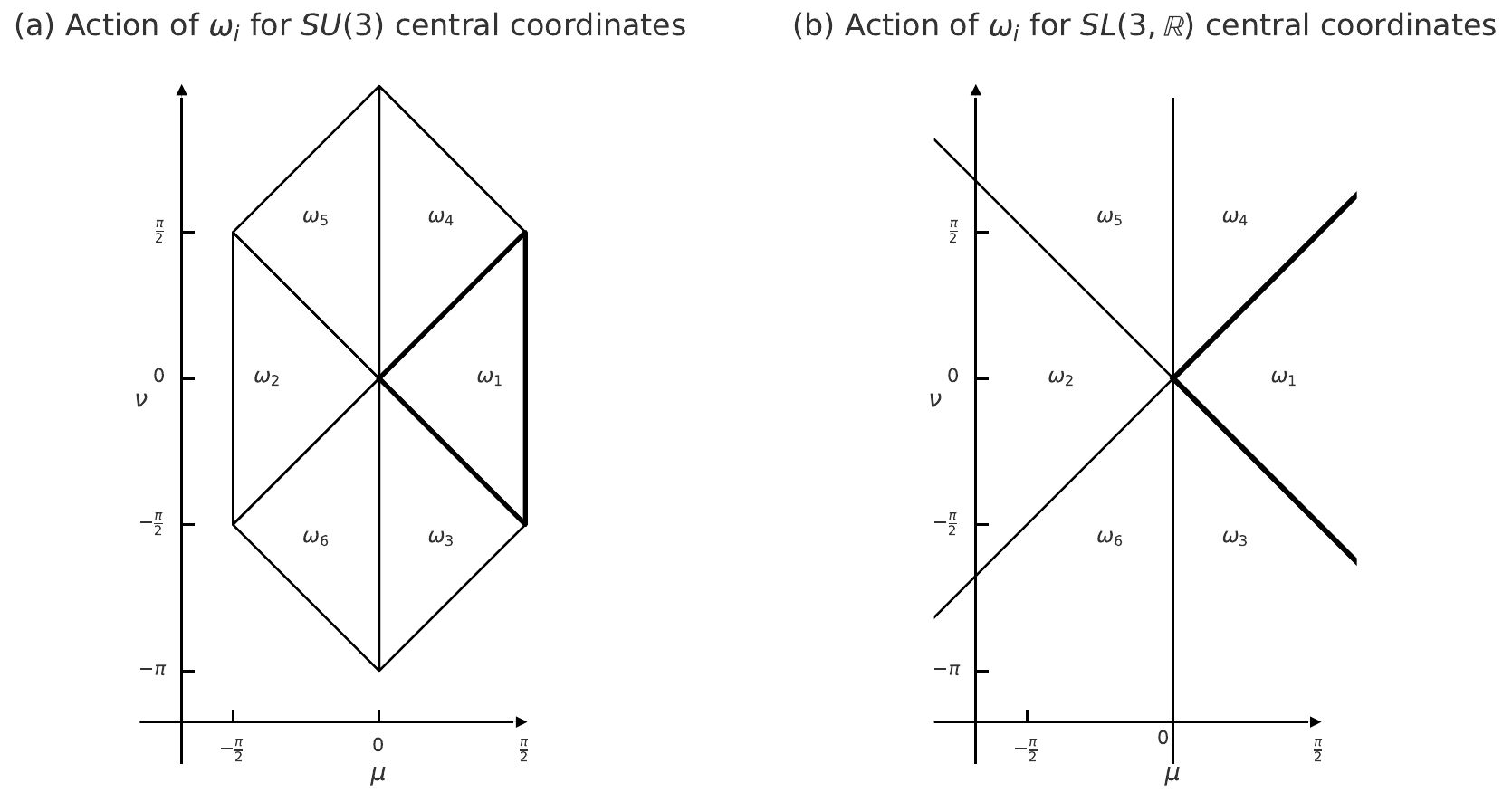}
    \caption{The fundamental domains in the $(\mu,\nu)$ plane for (a) $SU(3)$ and (b) $SL(3,\R)$ are highlighted in bold. The images of these domains under conjugation of a central element by $\omega_i$ are then plotted.}
    \label{munu}
\end{figure}

The effect of these actions is shown in Fig.~\ref{munu}.

\subsubsection{Action of the dihedral group $D_4$ on $SO(3)$}
\label{thirdAct}
Consider the third redundancy
\begin{equation}
\mathcal{O}_1 B \mathcal{O}_2^T=(\mathcal{O}_1\gamma'^{-1})B(\gamma' \mathcal{O}_2^T) \enspace \text{for} \enspace \gamma' \in D_4\,,
\end{equation}
where $D_4$ consists of the matrices in eqn (\ref{dihedral}).

$\mathcal{O}_2^T$ can be parameterised as
\begin{equation}
   \mathcal{O}_2^T= \left(
\begin{array}{ccc}
 \cos \theta  \cos \psi  \cos \phi -\sin \psi  \sin \phi  & \cos \theta  \cos \psi  \sin \phi +\sin \psi  \cos \phi  & -\sin \theta  \cos \psi  \\
 -\cos \theta  \sin \psi  \cos \phi -\cos \psi  \sin \phi  & \cos \psi  \cos \phi -\cos \theta  \sin \psi  \sin \phi  & \sin \theta  \sin \psi  \\
 \sin \theta  \cos \phi  & \sin \theta  \sin \phi  & \cos \theta  \\
\end{array}
\right).
\end{equation}

Quotienting by $D_4$ gives the identifications \begin{equation}
    (\theta,\phi,\psi) \sim (\theta,\phi,\psi+\pi)\sim(\pi-\theta,\phi+\pi,\pi-\psi) \sim (\pi-\theta,\phi+\pi,-\psi) \, , \label{dihedraliden}
\end{equation}
the last identification being the joint action of the first two.

\subsubsection{Uniquely parameterising an $SU(3)$ element}

Consider an $SU(3)$ element
\begin{equation}
    \mathcal{O}_1 B \mathcal{O}_2^T\,,
\end{equation}
with $\mathcal{O}_1,\mathcal{O}_2^T \in SO(3)$ and $B \in H$. 

We first use
\begin{equation}
(\mathcal{O}_1)(B)(\mathcal{O}_2^T)=(\mathcal{O}_1 \omega_i^{-1})(\omega_i B \omega_i^{-1})(\omega_i \mathcal{O}_2^T)\,,
\end{equation}
which quotients $(\mu,\nu)$ by the action of eqn (\ref{sixtrans}).

Secondly, we use
\begin{equation}
    (\mathcal{O}_1)(B)(\mathcal{O}_2^T)=(O_1 \gamma^{-1})(\gamma B ) (\mathcal{O}_2^T)\,,
\end{equation}
which quotients $(\mu,\nu)$ by the action of eqn (\ref{fourtrans}).

These two actions cut down the $(\mu,\nu)$ domain by a factor of $24$ to the fundamental domain defined by $0 \leq \mu \leq \frac{\pi}{2}$ and $-\mu \leq \nu \leq \mu$.

Finally, we use
\begin{equation}
(\mathcal{O}_1)(B)(\mathcal{O}_2^T)=(\mathcal{O}_1 \gamma'^{-1})(B)(\gamma' \mathcal{O}_2^T)\,,
\end{equation}
for $\gamma' \in D_4$ to take $\mathcal{O}_2^T$ from an element of $SO(3)$ to an element of $SO(3)/D_4$.

\subsection{Parameterising $SL(3,\R)$}

An element of $SL(3,\R)$ can be written as 
\begin{equation}
    \mathcal{O}_1 B \mathcal{O}_2^T
\end{equation}
where $\mathcal{O}_i$ are two $SO(3)$ matrices and $B$ is a member of the Cartan subalgebra $H$. $B$ is written as 
\begin{equation}
    B=\text{diag}\left(e^{-\mu-\frac{1}{3}\nu},e^{\mu-\frac{1}{3}\nu},e^{\frac{2}{3}\nu}\right).
\end{equation}

There are no longer periodic identifications arising from the form of $B$. Furthermore, since $H \cap SO(3)$ is now just the identity, there is no fourfold quotient of $(\mu,\nu)$.

However, it is still necessary to quotient $(\mu,\nu)$ by the action 
\begin{equation}
(\mathcal{O}_1)(B)(\mathcal{O}_2^T)=(\mathcal{O}_1 \omega^{-1})(\omega B \omega^{-1})(\omega \mathcal{O}_2^T) \enspace \text{for} \enspace  \omega \notin H \enspace \text{such that} \enspace \omega B \omega^{-1} \in H \enspace \forall B \in H\,.
\end{equation}
These are simply all matrices that swap diagonal entries of $B$ as they were for $SU(3)$. Their nontrivial action on $(\mu,\nu)$ was written down explicitly in eqns (\ref{sixtrans}), which is exactly the same for both $SU(3)$ and $SL(3,\R)$.

The fundamental $(\mu,\nu)$ domain can be chosen to be $-\mu \leq \nu \leq \mu $ and $\mu \geq 0$. Then the action of the six transformations in eqns (\ref{sixtrans}) give the following cones defined by the union of two bounding lines and an inequality which chooses which side of the origin is chosen
\begin{equation}
    \begin{split}
        \omega_1: \{(\mu=\nu) \cup (\mu=-\nu) : \mu \geq 0\}, \enspace \omega_2:\{(\mu=-\nu) \cup (\mu=\nu):\mu \leq 0\}\,,\\
        \omega_3:\{(\mu=0) \cup (\mu=-\nu):\mu \geq \nu\}, \enspace \omega_4:\{(\mu=\nu) \cup (\mu=0) :\mu  \geq -\nu\}\,,\\
        \omega_5: \{(\mu=0) \cup (\mu=\nu):\mu \leq -\nu\}, \enspace \omega_6:\{(\mu=0) \cup (\mu=-\nu): \mu \leq \nu\}\,.
    \end{split}
\end{equation}
These six regions are shown in Fig.~\ref{munu}.b.

Since 
\begin{equation}
    \mathcal{O}_1 B \mathcal{O}_2^T = (\mathcal{O}_1 \gamma)B(\gamma \mathcal{O}_2^T) \enspace \forall \gamma \in D_4\,,
\end{equation}
still holds in exactly the same way as for $SU(3)$, one factor of $SO(3)$ must be quotiented by $D_4$ as before in eqn (\ref{dihedraliden}).

\subsection{Action of the binary dihedral group $D^*_8$ on $SU(2)$}
\label{su2Act}

We can obtain the same identitifications as  eqn (\ref{dihedraliden}) by reinterpreting $SO(3)/D_4$ as $SU(2)/D_8^*$. The binary dihedral goup action on $S^3$ can be understood as follows. 
We can represent an $SU(2)$ element by a quaternion $q$ that obeys
\bea
\bar q \, q =1\,.
\eea
We can write
\bea
q= z_1 +z_2\,  \jm  \,,\qquad \hbox{where}\qquad z_1= x_1 + \im y_1\,,\quad
z_2= x_2 + \im y_2\,,\label{qzxy}
\eea
with $(\im,\jm,\km)$ being the three imaginary unit quaternions, obeying
\bea
\im^2 = \jm^2 = \km^2 =-1\,,\qquad \im \jm =-\jm \im = \km\,,\quad 
\hbox{etc.}
\eea
Note that in terms of the real coordinates, we have
\bea
q= x_1 + \im y_1 + \jm x_2 + \km y_2\,.
\eea
 
   The binary dihedral group $D^*_8$ is composed of the eight elements
\bea
\{1,-1,\im,-\im,\jm,-\jm,\km,-\km\}\,.\label{bindi}
\eea
To quotient $SU(2)$ by $D^*_8$, it suffices to impose the identifications
\bea
q\sim \im q\qquad \hbox{and} \qquad q\sim \jm q\,,
\label{ij}
\eea
since the other elements in (\ref{bindi}) will be produced by means of
repeated application of the two identifications in (\ref{ij}).

   We can parameterise the unit modulus $q$ as in eqns (\ref{qzxy}) in terms
of the two complex coordinates
\bea
z_1= \cos\ft12\theta\, e^{\ft{\im}{2} (\psi+\phi)}\,,\qquad
z_2= \sin\ft12\theta\, e^{\ft{\im}{2} (\psi-\phi)}\,,\label{eulerangles}
\eea
where $(\theta,\phi,\psi)$ are Euler angles.  In $SU(2)$ we have
the coordinates ranges
\bea
SU(2):&& 0\le\theta\le \pi\,,\qquad 0\le\phi<2\pi\,,\qquad
0\le\psi \le 4\pi\,.
\eea

Implementing the identification $q\sim \im q$ in eqns (\ref{ij}) we have
\bea
x_1+\im y_1 + \jm x_2 + \km y_2 \rightarrow \im x_1 - y_1 +\km x_2 -\jm y_2\,,
\eea
which implies
\bea
x_1 \rightarrow - y_1\,,\qquad y_1\rightarrow  x_1\,,\qquad
 x_2\rightarrow - y_2\,,\qquad y_2 \rightarrow x_2\,.
\eea
In terms of the complex coordinates this implies
\bea
z_1 \rightarrow \im\, z_1\,,\qquad z_2\rightarrow \im z_2\,.
\eea
In terms of the Euler angles in eqns (\ref{eulerangles}), this therefore
implies
\bea
q\rightarrow \im q:&& \psi\longrightarrow \psi + \pi\,.\label{itrans}
\eea
In a similar fashion, we can see that the $q\rightarrow \jm q$ identification
in eqns (\ref{ij}) implies
\bea
z_1\rightarrow -\bar z_2\,,\qquad z_2\rightarrow \bar z_1\,.
\eea
In terms of the Euler angles this therefore implies
\bea
q\rightarrow \jm q:&& \theta\rightarrow \pi-\theta\,,\quad 
\phi\rightarrow \phi+\pi\,,\quad \psi\rightarrow -\psi + \pi\,.\label{jtrans}
\eea

  Thus, in summary, the quotient $SU(2)/D^*_8$ is obtained by imposing the
identifications
\bea
SU(2)/D^*_8: && \quad (1)\qquad \psi\longrightarrow \psi + \pi\,,\label{imod}\\
&&\quad(2)\qquad \theta\rightarrow \pi-\theta\,,\quad
\phi\rightarrow \phi+\pi\,,\quad \psi\rightarrow -\psi + \pi\,,\label{jmod}
\eea
on the $(\theta,\phi,\psi)$ coordinates on $SU(2)$. 


\subsection{Volume of group manifolds}

Whereas the
left-invariant 1-forms on $SU(2)$ would give a volume
\bea
V(S^3) = \int_{S^3} \sigma_1\wedge\sigma_2\wedge\sigma_3 =16\pi^2\,,
\eea
($\sigma_1^2+\sigma_2^2+\sigma_3^2$ would be the metric on a round $S^3$ of
radius 2), on the quotient space $S^3/D_8^*$ the periods of the $\psi$ and $\phi$ coordinates must be such that 
\bea
V(S^3/D_8^*) = \int_{S^3/D_8^*} \sigma_1\wedge\sigma_2\wedge\sigma_3
=\ft18 \int_{S^3} \sigma_1\wedge\sigma_2\wedge\sigma_3  =2\pi^2\,.
\eea

With $\int\sigma_1\wedge\sigma_2\wedge\sigma_3$ equalling $2\pi^2$ we then have (choosing a positive orientation for the volume)
\be
\bald
{\rm Vol}(SU(3)/SO(3)_{\rm max})&= \int_0^{\frac{\pi}{2}} \frac{d\mu}{\sqrt{3}}\,
\int_{-\mu}^{\mu} d\nu \, \sin2\mu \, \sin(\mu-\nu)\sin(\mu+\nu)\, 
\int\sigma_1\wedge\sigma_2\wedge\sigma_3\nn\,,\\
&=\fft{\sqrt{3}\pi^3}{8}\,  ,
\label{CosetIntegral}
\eald
\ee
where the limits of the integration come from the fundamental region
of $(\mu,\nu)$ given in Fig.~\ref{munu}.
With the $SO(3)$ fibres in the description of $SU(3)$ as an $SO(3)$ bundle
over $SU(3)/SO(3)_{\rm max}$ (as in eqn (\ref{su3met})) contributing a 
factor $\int\Sigma_1\wedge\Sigma_2\wedge\Sigma_3=8\pi^2$, this gives the volume
of $SU(3)$ to be 
\bea
{\rm Vol}(SU(3))= {\rm Vol}(SU(3)/SO(3)_{\rm max})\, \int\Sigma_1\wedge\Sigma_2\wedge\Sigma_3  = \sqrt3\, \pi^5\,,
\eea
which is in agreement with results that can be found in the literature.  (Note, from eqn (\ref{RicSU3}), that we are considering the bi-invariant metric on $SU(3)$ that is normalised such that $R_{ab}= 3 g_{ab}$.)

\section{Totally geodesic submanifolds}
\label{SubmanSec}

In this appendix, we list the totally geodesic submanifolds of Wu manifold with the metric \eq{WuMetric}, matching those obtained using representation theory by Klein \cite{klein2009}. 

\subsection{Using connection 1-forms}

For $l^a \equiv e^a {}_\mu \dfft{dx^\mu}{ds}$, the geodesic equation is
\begin{equation}
\label{geoeq}
    l^b \nabla _b \,l^a =0 \iff l^b(\partial_b \,l^a + \omega_{b} {}^{a} {}_c\, l^c  )=0 \, .
\end{equation}
Choose local coordinates $\mu:1,... ,n$ such that they split into tangential directions $\bar a:1,...,m$ and normal directions $A:m+1,...,n$ with respect to a submanifold.
Splitting the geodesic equation using these indices, we get 
\begin{equation}
    \frac{dl^A}{ds}+l^{\bar b} \omega_{\bar b} {}^{ A} {}_{\bar c} \,l^{\bar c}+
      l^{\bar b} \omega_{\bar b} {}^{ A} {}_C \,l^C +
      l^B \omega_{B \phan \bar c}^{\phan A} \,l^{\bar c} + l^B \omega_{B \phan C}^{\phan A}\, l^C=0 \, .
\end{equation}
Therefore, if $l^A(s)=0$ at $s=0$ and if
\begin{equation} 
    \omega_{\bar b \phan \bar c}^{\phan A}=0 \enspace \text{for all } A \in (m+1,...,n) \enspace \text{and} \enspace \bar b,\bar c \in {1,...m} \, ,
\end{equation}
then this implies 
\begin{equation}
\frac{dl^A}{ds}(s) =0 \enspace \forall s \implies l^A(s)=0 \enspace \forall s \, .
\end{equation}

Therefore, under geodesic flow, any tangent vector to a submanifold will remain tangent to this submanifold if the mixed connection coefficients $\omega_{\bar b \phan \bar c}^{\phan A}$ are all zero.

\subsection{$(\mu,\nu)$ domain}
There is a subtlety that arises when the coordinates $\mu$ and $\nu$ are both varied in the metric. $\mu$ and $\nu$ have a fundamental domain $0 \leq \mu \leq \frac{\pi}{2}$ and $-\mu \leq \nu \leq \mu$. This was after taking discrete quotients (which cut down the fundamental domain by a factor of $24$). However, when varying both $\mu$ and $\nu$, it will be more convenient to consider twice this domain, defined by $0\leq \mu \leq \frac{\pi}{2}$ and $-\mu \leq \nu \leq -\mu+\pi$. Then, we just have to keep in mind that there is a $Z_2$ quotient acting on $(\mu,\nu)$. 

Specifically, this $Z_2$ quotient is one that arises on the level of writing an $SU(3)$ element as $o_1 h o_2$. It comes from an element $\omega \in SO(3), \omega \notin H$ such that $\omega h \omega^{-1} \in H$.
\subsubsection{$\mathbb{Z}_2$ quotient}
Even more specifically, consider a general central element
\begin{equation}
    h\equiv \begin{pmatrix}
        e^{i(\mu+\frac{\nu}{3})} & 0 & 0 \\
        0 & e^{i(-\mu+\frac{\nu}{3})} & 0 \\
        0 & 0 & e^{-2i\frac{\nu}{3}}
    \end{pmatrix}.
\end{equation}
Before any identifications are made, $(\mu,\nu) \sim (\mu+2\pi,\nu)\sim(\mu,\nu+6\pi)\sim(\mu+\pi,\nu+3\pi)$. Using these identifications and conjugating $h$ by 
\begin{equation}
    \omega \equiv - \begin{pmatrix}
        1 & 0 & 0\\
        0 & 0 & 1\\
        0 & 1 & 0
    \end{pmatrix}
\end{equation}
is equivalent to mapping the triangle with vertices $(0,0)$, $(0,\pi)$ and $(\frac{\pi}{2},\frac{\pi}{2})$ to our usual fundamental domain, which is the triangle with vertices $(0,0)$, $(\frac{\pi}{2},\frac{\pi}{2})$ and $(\frac{\pi}{2},-\frac{\pi}{2})$.

\subsection{$S^2$}
\label{S2Subsec}
Setting $\nu=0$ and $\theta,\phi$ both constant, to get
\begin{equation}
    ds^2=\frac{1}{4}(d(2\mu)^2+\sin^2(2\mu)d(2\psi)^2)\,.
\end{equation}

$\psi$ has periodicity $\pi$ which matches $S^2$ and $2\mu \in[0,\pi]$.

Orbits look like
\begin{equation}
    \begin{pmatrix}
        e^{i\mu}\cos \psi & e^{i\mu} \sin \psi & 0 \\
        -e^{-i\mu }\sin \psi & e^{-i\mu}\sin \psi & 0\\
        0 & 0 &1
    \end{pmatrix}\,,
\end{equation}
which is of the form
\begin{equation}
    \begin{pmatrix}
        B & & 0\\
         & & 0\\
         0 & 0 & 1
    \end{pmatrix}, \quad B \in SU(2)\,,
\end{equation}
The $SO(3)$ quotient factors out the action of block diagonal matrices $\text{diag}(R,1)\in SO(3)$ such that $R \in SO(2)$.
Then the orbit is $SU(2)/SO(2) = S^2$.

\subsection{$(S^1\times S^1)/\mathbb{Z}_2$}
Only varying $\mu$ and $\nu$ gives the metric
\begin{equation}
    ds^2=d\mu^2+\frac{1}{3}d\nu^2\,.
\end{equation}
$\mu$ and $\nu$ are defined in the doubled domain $0 \leq \mu \leq \frac{\pi}{2}$ and $-\mu \leq \nu \leq -\mu+\pi$. The $\mathbb{Z}_2$ action is given by $(\mu,\nu)\rightarrow (\frac{1}{2}\mu+\frac{1}{2}\nu,\frac{3}{2}\mu-\frac{1}{2}\nu)$.

Note that this space contains totally geodesic $S^1$'s or $\mathbb{R}$'s, depending on whether the $1D$ paths are periodic or injective and therefore dense.

\subsection{$(S^2 \times S^1)/\mathbb{Z}_2$}
If we vary only $(\mu,\nu,\psi)$, we get the metric
\begin{equation}
    ds^2=\frac{1}{3}d\nu^2+\frac{1}{4}\left(d(2\mu)^2+\sin^2(2\mu)d(2\psi)^2\right)\,.
\end{equation}
$2\psi$ has period $2\pi$ and the $\mathbb{Z}_2$ acts on $(\mu,\nu)$ as in the case of $(S^1\times S^1)/\mathbb{Z}_2$.

\subsection{$\RP^2$} 
Fixing $\mu=\frac{\pi}{2}$ and $\nu=0$ gives
\begin{equation}
ds^2=d\theta^2+\sin^2\theta d\phi^2 \,.
\end{equation}
$\phi \in [0,2\pi)$ and $(\phi,\theta)\sim (\phi+\pi,\pi-\theta)$ from the $D_2$ quotient of $SO(3)$ indeed give us $\RP^2$.

\section{\label{2compsec}Gauge-covariantly constant spinor using 2-component spinors}

  Here, we rewrite the gauge-covariantly constant spinor in the Wu space in
terms of two-component notation to describe the spin-$\ft32$ representation
of the gauge group.

  To begin, we define
\bea
A=\fft{1}{2\im}\, A^i\, \tau_i\,.
\eea
As can be verified from the
algebra $[\tau_i,\tau_j]=2\im\epsilon_{ijk}\, \tau_k$
of the Pauli matrices, we can write the Yang-Mills field strength as
\bea
F= dA + A\wedge A\,,\qquad \hbox{where}\qquad
F= \fft{1}{2\im}\, F^i\, \tau_i\,.
\eea
Using indices $\alpha$, $\beta$, etc., to label the components of the
$2\times2$ Pauli matrics, $(\tau_i)^\alpha{}_\beta$, we have, when acting on a scalar field carrying a
doublet representation of the gauge group,
\bea
D\phi = d\phi + A\,\phi\,,\qquad \hbox{i.e.}\qquad
  D\phi^\alpha = d\phi^\alpha + A^\alpha{}_\beta\, \phi^\beta\,.
\eea
A scalar field in the spin-$\ft32$ representation of the gauge group
is described in this
notation by $\phi^{\alpha\beta\gamma}$, totally-symmetric in the
three indices, with $D\phi^{\alpha\beta\gamma} =
d\phi^{\alpha\beta\gamma} +
        3 A^{(\alpha}{}_\lambda\, \phi^{\beta\gamma)\lambda}$.

  We now consider a spinor field $\psi^{\alpha\beta\gamma}$
in the spin-$\ft32$ representation of the gauge group. If this is
gauge-covariantly constant then it satisfies
\bea
D\psi^{\alpha\beta\gamma}\equiv d \psi^{\alpha\beta\gamma} +
  \ft14\omega_{ab}\, \Gamma^{ab}\, \psi^{\alpha\beta\gamma} +
  3 A^{(\alpha}{}_\lambda\, \psi^{\beta\gamma)\lambda} =0\,.
\eea
The integrability condition $D^2\psi^{\alpha\beta\gamma}=0$ then
implies
\bea
\fft14\Theta_{ab}\,\Gamma^{ab}\, \psi^{\alpha\beta\gamma}
  + 3 F^{(\alpha}{}_\lambda\, \psi^{\beta\gamma)\lambda} =0\,,
\eea
which in components implies
\bea
\fft14 R_{abcd}\,\Gamma^{cd}\, \psi^{\alpha\beta\gamma}
  + 3 F_{ab}{}^{(\alpha}{}_\lambda\, \psi^{\beta\gamma)\lambda} =0\,,
\label{2compriemid}
\eea
Using eqn (\ref{RiemFF}), and multiplying by $F^j_{ab}$ gives
\bea
\fft14 F^j_{cd}\,\Gamma^{cd}\, \psi^{\alpha\beta\gamma} +
\fft{3}{2\im}(\tau_j)^{(\alpha}{}_\lambda\, \psi^{\beta\gamma)\lambda}
  =0\,.
\eea
Multiplying by $(\tau_j)^\mu{}_\nu$ and using the identity
\bea
(\tau_j)^\mu{}_\nu\, (\tau_j)^\alpha{}_\lambda=
2 \delta^\mu_\lambda\, \delta^\alpha_\nu-
\delta^\mu_\nu\,\delta^\alpha_\lambda\,,
\eea
then gives
\bea
\Gamma^{ab}\, F_{ab}{}^\mu{}_\nu\, \psi^{\alpha\beta\gamma}=
  6\delta^{(\alpha}_\nu\, \psi^{\beta\gamma)\mu} -
     3 \delta^\mu_\nu\, \psi^{\alpha\beta\gamma}\,.\label{strong}
\eea
On the other hand, multiplying eqn (\ref{2compriemid}) by $\Gamma^{ab}$
and using that $R=30$ gives
\bea
\Gamma^{ab}\, F_{ab}{}^{(\alpha}{}_\lambda\, \psi^{\beta\gamma)\lambda}=
   5\psi^{\alpha\beta\gamma}\,.\label{weak}
\eea
Setting $\nu=\gamma$ in eqn (\ref{strong}) can be seen to imply eqn
(\ref{weak}); in other words, the two equations are self-consistent.

   If we make the choice
\bea
\Gamma^1 &=& \tau_2\otimes\tau_1\,,\qquad
\Gamma^2= \tau_2\otimes\tau_2\,,\qquad
\Gamma^3= \tau_2\otimes\tau_3\,,\nn\\
\Gamma^4 &=& -\tau_1\otimes\oneone_2\,,\qquad
\Gamma^5= \tau_3\otimes\oneone_2\,,\label{d5Gamma}
\eea
for the Dirac matrices in the five-dimensional base space (the Wu space),
eqn (\ref{strong}) can be solved explicitly, giving the unique (up to
overall scale) solution
\bea
\psi^{111}= \begin{pmatrix}\sqrt3\cr0\cr 0\cr0\end{pmatrix}\,,\quad
\psi^{112}= \begin{pmatrix}0\cr0\cr0\cr-1\end{pmatrix}\,,\quad
\psi^{122}= \begin{pmatrix}0\cr0\cr\ 1\cr 0\end{pmatrix}\,,\quad
\psi^{222}= \begin{pmatrix}0\cr-\sqrt3\cr\ 0\cr 0\end{pmatrix}\,.
\label{psires}
\eea

  Using eqn (\ref{omFA}), which can be rewritten in 2-component notation as
\bea
\omega_{ab}= -2 F_{ab}{}^\mu{}_\nu\, A^\nu{}_\mu\,,
\eea
it can be seen from eqn (\ref{strong}) that
\bea
D\psi^{\alpha\beta\gamma} &=& d \psi^{\alpha\beta\gamma}
  -3  A^{(\alpha}{}_\lambda\, \psi^{\beta\gamma)\lambda} +
  3  A^{(\alpha}{}_\lambda\, \psi^{\beta\gamma)\lambda}  \nn\\
&=& d \psi^{\alpha\beta\gamma}\,,
\eea
and that therefore the spinor defined in eqns (\ref{psires}) (\ie with
{\it constant} components),
satisfies the equation $D\psi^{\alpha\beta\gamma}=0$.

\section{\label{eigenvaluessec}Eigenvalues and completeness of the scalar eigenfunctions on the Wu manifold}

 In this appendix, we give the details of the calculation of the eigenvalues of the scalar Laplacian on $SU(3)/SO(3)$ that are found in section \ref{ConsScalEig}. We will use the following Fierz-like identities that are satisfied by the Gell-Mann matrices repeatedly:
\bea
\lambda_i^\alpha{}_\beta\,\,\lambda_i^\gamma{}_\delta &=& 
   2\delta^\alpha_\delta\, \delta^\beta_\gamma - \ft23 \delta^\alpha_\beta\,
\delta^\gamma_\delta\,,\qquad
\bar\lambda_i^\alpha{}_\beta\,\,\bar\lambda_i^\gamma{}_\delta =
   2\delta^\alpha_\delta\, \delta^\beta_\gamma - \ft23 \delta^\alpha_\beta\,
\delta^\gamma_\delta\,,\nn\\
\lambda_i^\alpha{}_\beta\,\,\bar\lambda_i^\gamma{}_\delta &=&
   2\delta^\alpha_\gamma\, \delta^\delta_\beta - \ft23 \delta^\alpha_\beta\,
\delta^\delta_\gamma\,,\label{su3ids}
\eea
where the bar denotes complex conjugation.  (Note that complex conjugation
changes the location of indices from up to down and down to up.)
It can be seen from the first two equations in (\ref{su3ids}) that
\bea
\lambda_i^\alpha{}_\beta\, \lambda_i^\beta{}_\gamma=\fft{16}{3}\, 
\delta^\alpha_\gamma\,,\qquad
\bar\lambda_i^\alpha{}_\beta\, \bar\lambda_i^\beta{}_\gamma=\fft{16}{3}\,
\delta^\alpha_\gamma\,.\label{casimir}
\eea

\subsection{Calculation of the eigenvalues}
\label{CalEigDeg}

 Acting on $f_{p,q}$ in eqn \eq{fpqdef} with $\nabla_i$ gives
\bea
\nabla_i\, f_{p,q} &=& p\, S^{\alpha^\phan_1\cdots\alpha^\phan_{2p}}{}_{\beta_1\cdots\beta_{2q}}\,
  (\nabla_i\,G_{\alpha^\phan_1\alpha^\phan_2})\cdots
      G_{\alpha^\phan_{2p-1}\alpha^\phan_{2p}}\,
\bar G^{\beta_1\beta_2}\cdots \bar G^{\beta_{2q-1}\beta_{2q}} \nn\\
&&+ q\, S^{\alpha^\phan_1\cdots\alpha^\phan_{2p}}{}_{\beta_1\cdots\beta_{2q}}\,
  G_{\alpha^\phan_1\alpha^\phan_2}\cdots
      G_{\alpha^\phan_{2p-1}\alpha^\phan_{2p}}\,
(\nabla_i\, \bar G^{\beta_1\beta_2})\cdots \bar G^{\beta_{2q-1}\beta_{2q}}\,,
\label{nabfpq}
\eea
after taking into account the symmetries of 
$S^{\alpha^\phan_1\cdots\alpha^\phan_{2p}}{}_{\beta_1\cdots\beta_{2q}}$.
Acting now with $\nabla^i$ gives
\bea
\square\,f_{p,q}&=&
p\,                                                      S^{\alpha^\phan_1\cdots\alpha^\phan_{2p}}{}_{\beta_1\cdots\beta_{2q}}\,
  (\square\,G_{\alpha^\phan_1\alpha^\phan_2})\cdots
      G_{\alpha^\phan_{2p-1}\alpha^\phan_{2p}}\,
\bar G^{\beta_1\beta_2}\cdots \bar G^{\beta_{2q-1}\beta_{2q}} \nn\\
&&+ q\, S^{\alpha^\phan_1\cdots\alpha^\phan_{2p}}{}_{\beta_1\cdots\beta_{2q}}\,
  G_{\alpha^\phan_1\alpha^\phan_2}\cdots
      G_{\alpha^\phan_{2p-1}\alpha^\phan_{2p}}\,
(\square\, \bar G^{\beta_1\beta_2})\cdots \bar G^{\beta_{2q-1}\beta_{2q}}\,,
\nn\\
&&+p(p-1)\,                                                      S^{\alpha^\phan_1\cdots\alpha^\phan_{2p}}{}_{\beta_1\cdots\beta_{2q}}\,
  (\nabla_i\,G_{\alpha^\phan_1\alpha^\phan_2})
(\nabla^i\, G_{{\alpha^\phan_3}{\alpha^\phan_4}})\cdots
      G_{\alpha^\phan_{2p-1}\alpha^\phan_{2p}}\,
\bar G^{\beta_1\beta_2}\cdots \bar G^{\beta_{2q-1}\beta_{2q}} \nn\\
&&+ q(q-1)\, S^{\alpha^\phan_1\cdots\alpha^\phan_{2p}}{}_{\beta_1\cdots\beta_{2q}}\,
  G_{\alpha^\phan_1\alpha^\phan_2}\cdots
      G_{\alpha^\phan_{2p-1}\alpha^\phan_{2p}}\,
(\nabla_i\, \bar G^{\beta_1\beta_2})(\nabla^i\, \bar G^{\beta_3\beta_4})
\cdots \bar G^{\beta_{2q-1}\beta_{2q}}\,,\nn\\
&& + 2 p\, q\, S^{\alpha^\phan_1\cdots\alpha^\phan_{2p}}{}_{\beta_1\cdots\beta_{2q}}\,
  (\nabla^i\,G_{\alpha^\phan_1\alpha^\phan_2})\cdots
      G_{\alpha^\phan_{2p-1}\alpha^\phan_{2p}}\,
(\nabla_i\, \bar G^{\beta_1\beta_2})\cdots \bar G^{\beta_{2q-1}\beta_{2q}}\,,
\label{boxfpq}
\eea
where $\square=\nabla^i\, \nabla_i$ is the scalar Laplacian.

   The various terms in eqn (\ref{boxfpq}) can be evaluated as follows.  
First, we note that
\bea
\square\, g^\alpha{}_\beta = - g^\alpha{}_\delta \lambda_i^\delta{}_\gamma\,
  \lambda_i^\gamma{}_\beta= -\fft{16}{3}\,
g^\alpha{}_\beta\,,\label{boxg}
\eea
and similarly, $\square\, \bar g^\alpha{}_\beta = -\dfft{16}{3}\, \bar          g^\alpha{}_\beta$.
In the first line of eqn (\ref{boxfpq}) we then have
\bea
\square\, G_{\alpha\beta}&=& (\square\, g^\gamma{}_\alpha)\, g^\gamma{}_\beta
 + g^\gamma{}_\alpha\, (\square\, g^\gamma{}_\beta) + 
  2(\nabla^i\, g^\gamma{}_\alpha)\,(\nabla_i\, g^\gamma{}_\beta) \,,\nn\\
&=& -\fft{32}{3}\, G_{\alpha\beta} -g^\gamma{}_\delta\, 
\lambda_i^\delta{}_\alpha\, g^\gamma{}_\sigma\, \lambda_i^\sigma{}_\beta\,,\nn\\
&=& -\fft{32}{3}\, G_{\alpha\beta} -
 (2\delta^\delta_\beta\, \delta^\sigma_\alpha - \fft23 
   \delta^\delta_\alpha\, \delta^\sigma_\beta) \,G_{\delta \sigma}\,,\nn\\
&=& \Big(-\fft{32}{3} - 2 + \fft23\Big)\, G_{\alpha\beta}\nn\\
&=& -\fft{40}{3}\, G_{\alpha\beta}\,.\label{boxG}
\eea
Thus in the first line of eqn (\ref{boxfpq}) we have $\square\, 
G_{\alpha_1\alpha_2}= -\dfft{40}{3}\, G_{\alpha_1\alpha_2}$.  An 
exactly analogous calculation gives for the second line that
$\square\, 
\bar G^{\beta_1\beta_2}= -\dfft{40}{3}\, \bar G^{\beta_1\beta_2}$.

  For the third line of (\ref{boxfpq}) we note that it suffices to calculate
\bea
&&S^{\alpha^\phan_1 \alpha^\phan_2\alpha^\phan_3\alpha^\phan_4}\,
   (\nabla_i\,G_{\alpha^\phan_1\alpha^\phan_2})
(\nabla^i\, G_{{\alpha^\phan_3}{\alpha^\phan_4}})=4 S^{\alpha^\phan_1 \alpha^\phan_2\alpha^\phan_3\alpha^\phan_4}\,
(\nabla_i\, g^{\gamma_1}{}_{\alpha^\phan_1})\, g^{\gamma_1}{}_{\alpha^\phan_2}\,
 (\nabla^i \, g^{\gamma_3}{}_{\alpha^\phan_3})\, 
   g^{\gamma_3}{}_{\alpha^\phan_4}\,,\nn\\
&=& -4 S^{\alpha^\phan_1 \alpha^\phan_2\alpha^\phan_3\alpha^\phan_4}\,
  g^{\gamma_1}{}_{\gamma_2}\, \lambda_i^{\gamma_2}{}_{\alpha^\phan_1}\, g^{\gamma_1}{}_{\alpha^\phan_2}\,
  g^{\gamma_3}{}_{\gamma_4}\, \lambda_i^{\gamma_4}{}_{\alpha^\phan_3}\,
  g^{\gamma_3}{}_{\alpha^\phan_4}\,,\nn\\
&=& -4 S^{\alpha^\phan_1 \alpha^\phan_2\alpha^\phan_3\alpha^\phan_4}\,
\Big(2\delta^{\gamma_2}_{\alpha_3}\, \delta^{\gamma_4}_{\alpha_1}-
   \fft23 \delta^{\gamma_2}_{\alpha_1}\, \delta^{\gamma_4}_{\alpha_3}\Big)\,
   g^{\gamma_1}{}_{\gamma^\phan_2}\,         g^{\gamma_1}{}_{\alpha^\phan_2}\,
 g^{\gamma_3}{}_{\gamma^\phan_4}\,
  g^{\gamma_3}{}_{\alpha^\phan_4}\,,\nn\\
&=& -4 S^{\alpha^\phan_1 \alpha^\phan_2\alpha^\phan_3\alpha^\phan_4}\,
\big( 2 G_{\alpha^\phan_3\alpha^\phan_2} G_{\alpha^\phan_1\alpha^\phan_4} -\fft23
       G_{\alpha^\phan_1 \alpha^\phan_2}\, G_{\alpha^\phan_3 \alpha^\phan_4}
  \big)\,,\nn\\
&=& -\fft{16}{3}\, S^{\alpha^\phan_1 \alpha^\phan_2\alpha^\phan_3\alpha^\phan_4}
\, 
G_{\alpha^\phan_1 \alpha^\phan_2}\, G_{\alpha^\phan_3 \alpha^\phan_4}\,,
\eea
where the final line follows because of the total symmetry of
$S^{\alpha^\phan_1 \alpha^\phan_2\alpha^\phan_3\alpha^\phan_4}$.  An
exactly analogous calculation shows that 
\bea
S_{\beta_1\beta_2\beta_3\beta_4}\, 
(\nabla_i\, \bar G^{\beta_1\beta_2})(\nabla^i\, \bar G^{\beta_3\beta_4}) =
-\fft{16}{3}\,S_{\beta_1\beta_2\beta_3\beta_4}\, 
\bar G^{\beta_1\beta_2}\, \bar G^{\beta_3\beta_4}\,,
\eea
which is the result needed for the fourth line in eqn (\ref{boxfpq}).

   Finally, for the fifth line in eqn (\ref{boxfpq}) it suffices to calculate
\bea
&&S^{\alpha^\phan_1 \alpha^\phan_2}{}_{\beta_1\beta_2}\,
(\nabla^i\,G_{\alpha^\phan_1\alpha^\phan_2})\, 
(\nabla_i\, \bar G^{\beta_1\beta_2})=
4S^{\alpha^\phan_1 \alpha^\phan_2}{}_{\beta_1\beta_2}\,
(\nabla^i\, g^{\gamma_1}{}_{\alpha^\phan_1})\, 
g^{\gamma_1}{}_{\alpha^\phan_2}\,
(\nabla_i\, \bar g^{\delta_1}{}_{\beta_1})\, \bar g^{\delta_1}{}_{\beta_2}
\nn\\
&=& 4 S^{\alpha^\phan_1 \alpha^\phan_2}{}_{\beta_1\beta_2}\,
g^{\gamma_1}{}_{\gamma_2}\, \lambda_i^{\gamma_2}{}_{\alpha^\phan_1}\,
 g^{\gamma_1}{}_{\alpha^\phan_2}\,
 \bar g^{\delta_1}{}_{\delta_2}\, \bar \lambda_i^{\delta_2}{}_{\beta_1}\,
\bar g^{\delta_1}{}_{\beta_2}\,,\nn\\
&=& 4 S^{\alpha^\phan_1 \alpha^\phan_2}{}_{\beta_1\beta_2}\,
\Big(2\delta^{\gamma_2}_{\delta_2}\, \delta^{\beta_1}_{\alpha_1} -
   \fft23 \delta^{\gamma_2}_{\alpha_1}\, \delta^{\beta_1}_{\delta_2}\,\Big)\,
   g^{\gamma_1}{}_{\gamma^\phan_2}\,g^{\gamma_1}{}_{\alpha^\phan_2}\,
\bar g^{\delta_1}{}_{\delta_2}\, 
\bar g^{\delta_1}{}_{\beta_2}\,,\nn\\
&=& 4 S^{\alpha^\phan_1 \alpha^\phan_2}{}_{\beta_1\beta_2}\,
\Big( -\fft23 
G_{\alpha^\phan_1 \alpha^\phan_2}\, \bar G^{\beta_1\beta_2}\Big)\,,\nn\\
&=& -\fft83 S^{\alpha^\phan_1 \alpha^\phan_2}{}_{\beta_1\beta_2}\,
 G_{\alpha^\phan_1 \alpha^\phan_2}\, \bar G^{\beta_1\beta_2}\,,
\eea
where the step in getting to the fourth line follows 
by virtue of the tracelessness properties of 
$S^{\alpha^\phan_1 \alpha^\phan_2}{}_{\beta_1\beta_2}$.

  We now have all the results needed to read off the contributions 
from each line in eqn (\ref{boxfpq}). 
   Putting all of the above together, we finally conclude that the
function $f_{p,q}$ defined in eqn (\ref{fpqdef}) is an eigenfunction
of the scalar Laplacian on the coset $SU(3)/SO(3)$, obeying
$-\square\, f_{p,q}= \lambda_{p,q}\, f_{p,q}$ with
\bea
\lambda_{p,q}= \fft83\, \Big( 2p^2 + 2 q^2 + 3p + 3q + 2 p\, q\Big)\,.
\label{lampq}
\eea

\subsection{$SU(3)$ transformations of the eigenfunctions}

  To verify that the construction described above does indeed generate
all of the scalar eigenfunctions on $SU(3)/SO(3)$, we can examine
how the eigenfunctions $f_{p,q}$ transform under the action of the
$SU(3)$ isometry group of the coset manifold.  

   We write the general $SU(3)$ group element as
\bea
g= \cO_1\, B\, \cO_2^T\,,
\eea
where $\cO_1$ is an orthogonal $3\times 3$ matrix parameterised by
the fibre coordinates $(\Theta,\Phi,\Psi)$.  Since this describes
the $SU(3)/SO(3)$ manifold as a right coset, it follows that the
$SU(3)$ Killing vectors on the coset are constructed from 
the left-invariant 1-forms $g^\dagger\, dg$.  Thus we may define
$K^\alpha{}_\beta = \ft14 (g^\dagger\, dg)^\alpha{}_\beta$, and hence
as vectors
\bea
K^\alpha{}_\beta = \ft{\im}{4}\, \bar g^\gamma{}_\alpha\, 
   g^\gamma{}_\delta\, \lambda_i^\delta{}_\beta\, \nabla^i\,.
\label{KVdef}
\eea
As a check, calculating the commutator of Killing vectors on a scalar
field, one can verify after some algebra, using eqns (\ref{nabg3}) and
the identities (\ref{su3ids}),
that
\bea
[K^\alpha{}_\beta, K^\gamma{}_\delta] = 
\delta^\gamma_\beta\,K^\alpha{}_\delta -
  \delta^\alpha_\delta\, K^\gamma{}_\beta\,,\label{KVcom}
\eea
which is the expected result for $SU(3)$ Killing vectors.

  Acting with the Killing vectors on the quantities $G_{\alpha\beta}=
g^\gamma{}_\alpha\, g^\gamma{}_\beta$ and $\bar G^{\alpha\beta}=
 \bar g^\gamma{}_\alpha\, \bar g^\gamma{}_\beta$ that we defined earlier,
it can be seen by using eqns (\ref{nabg3}) and (\ref{su3ids}) that
\bea
K^\alpha{}_\beta(G_{\gamma\delta}) &=& -\delta^\alpha_{(\gamma}\,
   G_{\delta)\beta} +\ft13 \delta^\alpha_\beta\, G_{\gamma\delta}
   \,,\label{KG}\\
K^\alpha{}_\beta(\bar G^{\gamma\delta}) &=&
  \delta^{(\gamma}_\beta\,\bar G^{\delta)\alpha}
  -\ft13 \delta^\alpha_\beta\, \bar G^{\gamma\delta}\,.\label{KbG}
\eea
If we introduce the constants 
$\epsilon_\alpha{}^\beta$ (with $\epsilon_\alpha{}^\alpha=0$)
as the parameters of infinitesimal $SU(3)$ symmetry transformations then
we see acting on the eigenfunctions $f_{p,q}$ defined in eqn
(\ref{fpqdef}) with $\epsilon_\alpha{}^\beta\, K^\alpha{}_\beta$
gives
\bea
\delta f_{p,q}&=& \epsilon_\alpha{}^\beta\, K^\alpha{}_\beta(f_{p,q})\nn\\
&=& 
-p\,\epsilon_{\alpha^\phan_1}{}^\gamma\, 
S^{\alpha^\phan_1\cdots\alpha^\phan_{2p}}{}
 _{\beta_1\cdots\beta_{2q}}\, 
 G_{\gamma\alpha^\phan_2}\, G_{\alpha^\phan_3\alpha^\phan_4}\cdots
G_{\alpha^\phan_{2p-1} \alpha^\phan_{2p}} \bar G^{\beta_1\beta_2}\cdots
\bar G^{\beta_{2q-1}\beta_{2q}}\nn\\
&&+q\, \epsilon_\gamma{}^{\beta_1}\, S^{\alpha^\phan_1\cdots\alpha^\phan_{2p}}{}
 _{\beta_1\cdots\beta_{2q}}\, G_{\alpha^\phan_1\alpha^\phan_2}\cdots
G_{\alpha^\phan_{2p-1} \alpha^\phan_{2p}} \,\bar G^{\gamma\beta_2} \, 
\bar G^{\beta_3\beta_4}\cdots
\bar G^{\beta_{2q-1}\beta_{2q}}
\eea
It follows that the $SU(3)$ transformations can be viewed as inducing the
following transformation on the tensor 
$S^{\alpha^\phan_1\cdots\alpha^\phan_{2p}}{}_{\beta_1\cdots\beta_{2q}}$:
\bea
\delta 
S^{\alpha^\phan_1\cdots\alpha^\phan_{2p}}{}_{\beta_1\cdots\beta_{2q}}=
-p\, \epsilon_\gamma{}^{(\alpha^\phan_1}\, 
S^{\alpha^\phan_2\cdots\alpha^\phan_{2p}) \gamma}{}_{\beta_1\cdots \beta_{2q}}
+q\, \epsilon_{(\beta_1}{}^\gamma\, 
S^{\alpha^\phan_1\cdots\alpha^\phan_{2p}}{}
 _{\beta_2\cdots\beta_{2q})\gamma}\,.
\eea
This shows that the tensor 
$S^{\alpha^\phan_1\cdots\alpha^\phan_{2p}}{}_{\beta_1\cdots\beta_{2q}}$
indeed acquires an induced transformation as the $(2p,2q)$ irreducible
representation of $SU(3)$, and that therefore the eigenfunctions
$f_{p,q}$ form the $(2p,2q)$ irreducible representation of $SU(3)$.

\subsection{Eigenfunctions obeying the conformality condition}
\label{ConfScal}
  A subset of the scalar eigenfunctions on the Wu manifold
$SU(3)/SO(3)$ obey what is known as the {\it conformality condtion},
namely
\bea
(\nabla^i f)\, (\nabla_i f) = -\kappa\, f^2\,,\label{conf1}
\eea
where $\kappa$ is a constant.  A  scalar eigenfunction $f$ obeying
$-\square\, f= \lambda\, f$ and also eqn (\ref{conf1}) has the
property that any power of $f$ will also be a scalar eigenfunction.  Thus
if $h\equiv f^n$ then
\bea
-\square h &=& -n\, (\square\, f)\, f^{n-1} -n(n-1)\, 
              (\nabla^i f)\,(\nabla_i f)\, f^{n-2}\,,\nn\\
&=& n\, \lambda\, h + n(n-1)\,\kappa\, h\,,
\eea
and so $h$ obeys
\bea
-\square\, h = \hat\lambda\, h\,,\qquad 
 \hat\lambda= n\, \lambda + n(n-1)\, \kappa\,.\label{hev}
\eea

Consider first the eigenfunctions
$f_{p,q}$ with $p=1$ and $q=0$; these are in the $(2,0)$ 6-dimensional 
representation of $SU(3)$, and they have the form
\bea
f= S^{\alpha\beta}\, g^\gamma{}_\alpha\, g^\gamma{}_\beta\,.
\eea
From eqn (\ref{lampq}) 
they have eigenvalue $\lambda = \lambda_{1,0}= \dfft{40}{3}$.  
Using eqns (\ref{nabg2}), (\ref{nabg3}) and (\ref{su3ids}), 
it can be seen that
\bea
(\nabla^i f)\, (\nabla_i f) &=& -4 S^{\alpha\beta}\,
 S^{\gamma\delta}\, g^\mu{}_\nu\, \lambda_i^\nu{}_\alpha\, 
g^\mu{}_\beta\, g^\rho{}_\sigma\, \lambda_i^\sigma{}_\gamma\,
g^\rho{}_\delta\,,\nn\\
&=& - 8 S^{\alpha\beta}\, S^{\gamma\delta}\, g^\mu{}_\gamma\, 
g^\mu{}_\beta\, g^\rho{}_\alpha\, g^\rho{}_\delta +
\fft83 S^{\alpha\beta}\, S^{\gamma\delta}\, g^\mu{}_\alpha\,
g^\mu{}_\beta\, g^\rho{}_\gamma\, g^\rho{}_\delta\,,\nn\\
&=& - 8 S^{\alpha\beta}\, S^{\gamma\delta}\, g^\mu{}_\gamma\, 
g^\mu{}_\beta\, g^\rho{}_\alpha\, g^\rho{}_\delta  + \fft83 f^2\,.
\eea
The conditions on $S^{\alpha\beta}$ in order that the 
right-hand side should give purely a constant times
$f^2$ are therefore that
\bea
S^{\alpha\beta}\, S^{\gamma\delta}\,  g^\mu{}_\gamma\, 
g^\mu{}_\beta\, g^\rho{}_\alpha\, g^\rho{}_\delta  = c\, 
S^{\alpha\beta}\, S^{\gamma\delta}\, g^\mu{}_\alpha\,
g^\mu{}_\beta\, g^\rho{}_\gamma\, g^\rho{}_\delta\,,\label{Scon}
\eea
where $c$ is a constant.  By substituting in the form of the general
$SU(3)$ element $g$, it is straightforward to see that for (\ref{Scon}) to
be satisfied, it must be that
\bea
S^{\alpha\beta}= a^\alpha\, a^\beta\,,\qquad \hbox{and}\qquad c=1\,,
\eea
where $a^\alpha$ is an arbitrary constant vector in ${\mathbb C}^3$.  With 
$S^{\alpha\beta}$ of this form, $f$ then satisfies the conformality 
condition (\ref{conf1}) with
\bea
\kappa= \fft{16}{3}\,.
\eea

   From eqn (\ref{hev}), it follows that $h=f^n$ is then a scalar eigenfunction 
satisfying $-\square\, h= \hat\lambda h$ with
\bea
\hat\lambda= \fft{8n\, (2n+3)}{3}\,,
\eea
which is exactly of the form for $\lambda_{p,q}$ in eqn (\ref{lampq}) 
with $p=n$ and $q=0$.

\section{Pinors on $\RP^2$ }
\label{sec:pin}
\subsection{Pin group}
The Clifford algebra $Cl_2$ is generated by 
\begin{equation}
    \oneone,\, e_1,\, e_2,\, e_1e_2, \enspace \text{such that} \enspace e_1^2=e_2^2=\oneone, \,e_1e_2=-e_2e_1\,.
\end{equation}

The group Pin$(n)$ is generated by elements of the unit sphere $S^{n-1}$ embedded in $\mathbb{R}^n$ written as $(a,b,...,c)$ such that $a^2+b^2+...+c^2=1$. Every element of Spin$(n)$ can be written as the product of an even number of elements $(a_1,b_1,...,c_1)$, $(a_2,b_2,...,c_2)$,$...$ of $S^{n-1}$ expanded in a basis of $e_i$. For $n=2$, 
\begin{equation}
    \text{Spin}(2)=\prod^{2k}_{i=1} (a_i e_1+b_i e_2), \qquad a_i ^2+b_i^2=1 \enspace \forall i\,.
\end{equation}
The product of two elements gives
\begin{equation}
    (a_1 e_1+b_1e_2)(a_2e_1+b_2e_2)=(a_1a_2+b_1b_2)+(a_1b_2-a_2b_1)e_1e_2\,.
\end{equation}
Since $(a_1a_2+b_1b_2)^2+(a_1b_2-a_2b_1)^2=(a_1^2+b_1^2)(a_2^2+b_2^2)=1$, the product of two elements can be written as
\begin{equation}
    A\,\oneone +Be_1e_2: \quad A^2+B^2=1\,. \label{spin2}
\end{equation}
The product of two elements with another two elements is 
\begin{equation}
(A_1\, \oneone+B_1e_1e_2)(A_2 \,\oneone + B_2e_1e_2)=(A_1A_2-B_1B_2)+(A_1B_2+A_2B_1)e_1e_2\,.
\end{equation}
Since $(A_1A_2-B_1B_2)^2+(A_1B_2+A_2B_1)^2=(A_1^2+B_1^2)(A_2^2+B_2^2)=1$, by induction, all elements of Spin$(2)$ can be written in the form of eqn (\eq{spin2}). Therefore,
\begin{equation}
\label{spin22}
    \text{Spin}(2)=\{a\, \oneone+be_1e_2:\ a^2+b^2=1\} \sim U(1)\,.
\end{equation}

All products of odd numbers of elements can then be written as
\begin{equation}
    (a_1e_1+b_1e_2)(A_1 \,\oneone+B_1e_1e_2)=(a_1A_1-b_1B_1)e_1+(b_1A_1+a_1B_1)e_2\,.
\end{equation}
Since $(a_1A_2-b_1B_2)^2+(b_1A_2+a_1B_2)^2=(a_1^2+b_1^2)(A_2^2+B_2^2)=1$, the product of an odd number of elements can always be written as
\begin{equation}
\label{pin2pre}
C e_1+De_2:\ C^2+D^2=1\,.
\end{equation}
Since Pin group elements are products of even and odd number of elements, with \eq{spin22} and \eq{pin2pre}, we get  
\begin{equation}
\text{Pin}(2)=\{a\oneone+be_1e_2:\ a^2+b^2=1\} \cup \{ce_1+de_2:\ c^2+d^2=1\}\sim U(1) + U(1)\,.
\end{equation}

\subsection{Pin structure on \texorpdfstring{$\RP^2$}{RP2}}

Starting with the sphere $S^2$, viewed as the unit sphere with
metric $ds^2=d\theta^2+\sin^2\theta\, d\phi^2$, define
the antipodal map $P$:
\bea
P:\qquad \{\theta\longrightarrow \pi-\theta\,, \quad\phi\longrightarrow \phi+\pi\}\,.
\label{Pdef}
\eea
$\RP^2$ is obtained by quotienting $S^2$ by $P$.

We can use the following procedure to try to construct spinors on $\RP^2$.  We
begin by constructing a complete set of spinor eigenfunctions on $S^2$.  With
the Dirac operator being
\bea
\im\nabslash = \im\, E^\mu_i\, \tau^i\, \nabla_\mu\,,
\eea
where $\tau^i=\tau_i$ are the Pauli matrices and $E^i_\mu$ is the inverse of
the zweibein $e^i_\mu$ on $S^2$, which we take to be $e^1=d\theta$ and
$e^2=\sin\theta\, d\phi$, one finds
\bea
\im\nabslash = \im \tau_1(\del_\theta + \ft12\cot\theta) +
   \im\, \tau_2\, \csc\theta\, \del_\phi\,.
\eea
It is straightforward then to establish (see, for example, \cite{abrik}) that
the complete sets spinor eigenfunctions $\psi_\pm$ obeying
\bea
\im\nabslash\psi^\pm_{\ell,m} = \pm\, \lambda_\ell\, \psi^\pm_{\ell,m}\,,\qquad
\lambda_\ell=\ell+\ft12\,,
\eea
are given by
\bea
\psi^\pm_{\ell,m} = e^{\im m\phi}\,
 \begin{pmatrix} \sin^{m_-}(\ft12\theta)\, \cos^{m_+}(\ft12\theta)\,
           P_{\ell-m}^{m_+,m_-}(\cos\theta) \cr
           \mp \im\sin^{m_+}(\ft12\theta)\, \cos^{m_-}(\ft12\theta)\,
           P_{\ell-m}^{m_-,m_+}(\cos\theta)\end{pmatrix}\,,
\label{psipm}
\eea
where
\bea
m_\pm\equiv m \pm \ft12\,,\qquad \ell=\ft12\,,\ft32\,,\ft52\,,\cdots\,,\qquad
\hbox{and}\quad m=\pm\ft12\,,\pm\ft32\,,\cdots \,,\pm \ell\,.
\eea
The functions $P_n^{\alpha,\beta}(x)$ are the Jacobi polynomials, which
can be defined by the generalised Rodrigues' formula
\bea
P_n^{\alpha,\beta}(x) =\fft{(-1)^n}{2^n\, n!}\,
(1-x)^{-\alpha}\, (1+x)^{-\beta}\, \fft{d^n}{dx^n}\, \Big[
  (1-x)^{\alpha+n}\, (1+x)^{\beta+n}\Big]\,.
\eea
Note that the eigenvalue spectrum is symmetric between positive and negative,
with
\bea
\lambda= \pm1\,,\pm2\,,\pm3\,,\pm4\,,\cdots\,.
\eea

  It is now straightforward to see that none of the Dirac eigenspinors on
$S^2$ are invariant under the action of the antipodal map $P$ defined
in (\ref{Pdef}).  Therefore, the subset of the Dirac eigenspinors
on $S^2$ that survive the projection $S^2/P\longrightarrow \RP^2$ comprises
just the empty set.  Since the Dirac eigenspinors on $S^2$ formed a
complete set, this means that there are no spinors at all on $\RP^2$; i.e., it does not admit a spin structure.

  As an example, consider the $\ell=\ft12$ Dirac eigenspinors on
$S^2$.  We have
\bea
\lambda=+1:\qquad \psi^+_{\ft12,\ft12}&=& e^{\ft{\im}{2}\phi}\,
\begin{pmatrix} \cos\ft12\theta\cr -\im \sin\ft12\theta\end{pmatrix}\,,\qquad
\psi^+_{\ft12,-\ft12}= e^{-\ft{\im}{2}\phi}\,
\begin{pmatrix} -\sin\ft12\theta\cr -\im \cos\ft12\theta\end{pmatrix}\,,\\
&&\nn\\
\lambda=-1:\qquad \psi^-_{\ft12,\ft12}&=& e^{\ft{\im}{2}\phi}\,
\begin{pmatrix} \cos\ft12\theta\cr \im \sin\ft12\theta\end{pmatrix}\,,\qquad
\psi^-_{\ft12,-\ft12}= e^{-\ft{\im}{2}\phi}\,
\begin{pmatrix} -\sin\ft12\theta\cr \im \cos\ft12\theta\end{pmatrix}\,.
\eea
One can easily explicitly see that none of these eigenspinors is invariant
under the action of the antipodal identification $P$ defined in eqn
(\ref{Pdef}). 

$\RP^2$ does, however, admit two $\text{Pin}^-$ structures. To see this, consider the charge conjugation matrix $C$, which in our
basis for the Dirac matrices will be given by
\bea
C \equiv  \im \tau_2 = \begin{pmatrix}0&1\cr -1&0\end{pmatrix}\, \in \text{Pin}(2).\label{Cdef}
\eea
The $S^2$ spinor eigenfunctions $\psi^\pm_{\ell,m}$
defined in eqns (\ref{psipm}) obey
\bea
P(\psi^\pm_{\ell,m}) = \pm (-1)^{\ell+\ft12}\, C \psi^\pm_{\ell,m}\,.
\eea

   It can now be observed that if instead of requiring $P(\psi)=\psi$ we
require
\bea
P(\psi)= C\psi\,,\label{Cid}
\eea
then we find that this condition is satisfied by
\bea
&&\psi^+_{\ell,m}\,,\qquad \hbox{for}\qquad \ell=\ft32\,,\ft72\,,\ft{11}{2}\,,
  \cdots\,,\\
&&\psi^-_{\ell,m}\,,\qquad \hbox{for}\qquad \ell=\ft12\,,\ft52\,,\ft{9}{2}\,,
  \cdots\,.
\eea
Thus we have an asymmetric set of eigenspinors that survive the identification
in (\ref{Cid}), with the survivors having Dirac eigenvalues
\bea
\lambda=2\,,4\,,6\,,8\,,\cdots\,,\qquad \hbox{and}\qquad
\lambda=-1\,,-3\,,-5\,,-7\,,\cdots\,.
\eea
This is exactly of the form seen in section 8.2 of Trautman's paper
\cite{traut}.

Note that if we had required $P(\psi)=-C\psi$, instead of (\ref{Cid}), then
the roles of $+$ and $-$ would be exchanged, and the spectrum would instead be
\bea
\lambda=1\,,3\,,5\,,7\,,\cdots\,,\qquad \hbox{and}\qquad
\lambda=-2\,,-4\,,-6\,,-8\,,\cdots\,.\label{Cid2}
\eea

These two cases are the two $\text{Pin}^-$ structures on $\RP^2$, with the $-$ denoting that $C^2=-1$, 
and the choice of two given by whether $P(\psi)=C\psi $ or $P(\psi)=-C\psi$.

\subsection{$U(1)$ bundle over $\RP^2$}

There are two inequivalent line bundles on $\RP^2$ corresponding to $H^2(\RP^2,\mathbb{Z})=\mathbb{Z}_2$. After lifting to the double cover of $\RP^2$ which is $S^2$, they correspond to a choice of sign in 
\begin{equation}
    A(\pi-\theta,\phi+\pi)=\pm A(\theta,\phi)\,. \label{ident}
\end{equation}
 The two choices of sign correspond exactly to two different holonomies $\pm 1$ respectively around the nontrivial orientation-reversing loop within $\RP^2$ \cite{Kim2013}.

This is important since on the totally geodesic $\RP^2$ in $SU(3)/SO(3)$, $A^1=A^2=0$ and $A^3=-\cos \theta d\phi$. Since $A^3(\pi-\theta,\phi+\pi)=-A^3(\theta,\phi)$, this selects the minus sign in eqn (\eq{ident}).

\newpage

\bibliographystyle{utphys} 
\bibliography{bibliography}

\end{document}